\documentclass{iopjournal}

\usepackage{lmodern}

\usepackage{hyperref}
\hypersetup{colorlinks=true, citecolor=blue, urlcolor=teal, linkcolor=blue}

\usepackage{amssymb, amsmath}
\usepackage[english]{babel}
\usepackage{bm}
\usepackage{graphicx}
\usepackage[utf8]{inputenc}
\usepackage{bbold}
\usepackage{multirow}

\usepackage{csquotes}
\usepackage[
    sorting = none,
    style = phys,
    biblabel=brackets
]{biblatex}
\usepackage[dvipsnames]{xcolor}
\newcommand{\red}[1]{{ #1}}

\newcommand{\opvec}[1]{\hat{\mathbf{#1}}}
\newcommand{\qnm}[2]{\tilde{\mathbf{#1}}_{#2}}

\begin{document}

\articletype{Paper} 

\title{Quasinormal mode quantization of bound and propagating photons in complex lightguiding nanostructures for integrated devices}

\author{Robert Meiners Fuchs$^{1*}$\orcid{0000-0002-6280-4261}, and Marten Richter$^{1\dagger}$\orcid{0000-0003-4160-1008}}

\affil{$^{1}$Institut für Physik und Astronomie, Nichtlineare Optik und
Quantenelektronik, Technische Universität Berlin, Hardenbergstr. 36, EW 7-1, 10623
Berlin, Germany}

\affil{$^{*}$ now at 
DTU Electro, Technical University of Denmark, rmefu@dtu.dk
}

\affil{$^{\dagger}$Author to whom any correspondence should be addressed.}

\email{marten.richter@tu-berlin.de}


\begin{abstract}
Open optical or plasmonic resonators are placed on and connected through surfaces or via waveguides, forming complex lightguiding nanostructures, e.g. for integrated photonic quantum devices. We derive general boundary conditions for quasinormal modes that account for the structure's specific geometry. We then present a general quantization scheme for multiple, interacting quasinormal-mode cavities coupled to quantum emitters and to a non-bosonic bath of propagating photons on waveguides or a surface. We derive a system-bath Hamiltonian \red{for coupled-cavity structures, where the photonic Green's function can be expressed in terms of bound quasinormal modes for the resonators and a Green's function for propagating photons for the connecting devices.
Coupling elements are defined in terms of the quasinormal mode functions, and} can be computed using Maxwell solvers, including light-matter coupling between the electromagnetic field and quantum emitters.
We define system-bath correlation functions for an effective, bath-mediated, and time-delayed interaction between the quasinormal modes and quantum emitters, which is a main ingredient commonly used to simulate open quantum system dynamics. \red{We illustrate our approach on the simple model of 2D waveguide-coupled cavities.}
\end{abstract}

\section{Introduction}
\begin{figure}
    \centering
    \includegraphics[width=0.9\linewidth]{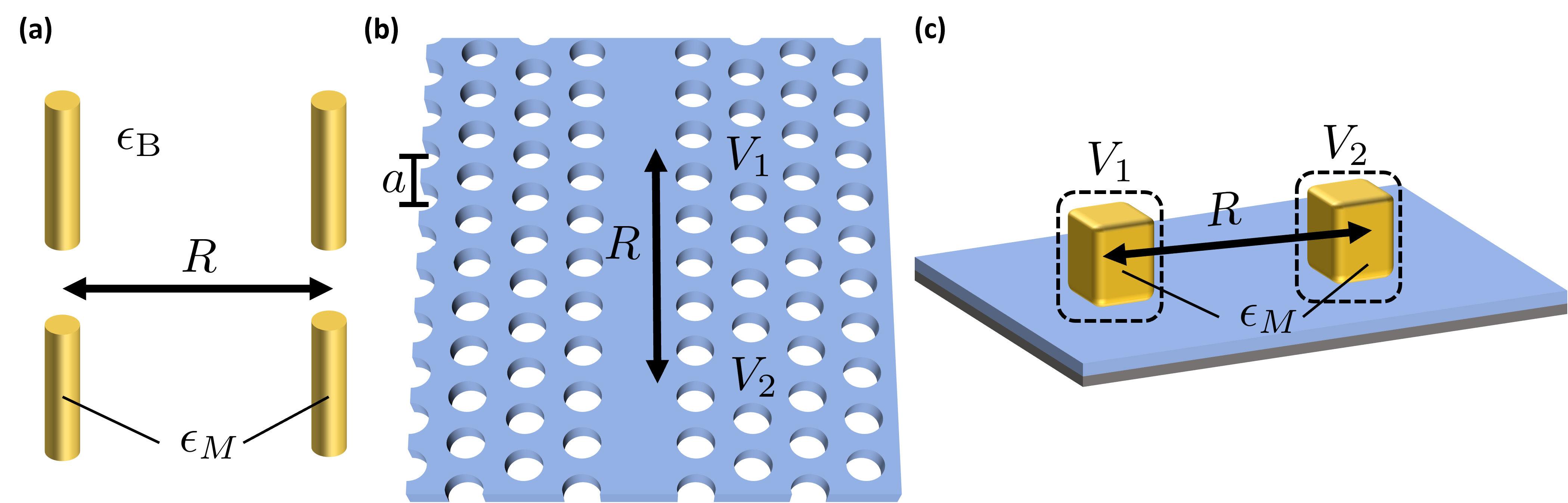}
    \caption{Examples of different types of coupled QNM cavities separated by distance \(R\) in various types of background structures. (a) Metal dimers described by a Drude permittivity \(\epsilon_M\) in a 3D homogeneous background medium \(\epsilon_B\) \cite{fuchs2026greens, fuchs2026quantum}. (b) Cavities \(V_1, V_2\) on a photonic crystal with lattice constant \(a\), side-coupled to the same periodic photonic crystal waveguide \cite{hughes2007coupled, yang2009all, kristensen2017theory}. (c) Plasmonic nanoparticles \(V_1, V_2\) described by the Drude permittivity \(\epsilon_M\) near a planar surface \cite{yan2018rigorous, binkowski2022computation}.}
    \label{fig:examples}
\end{figure}

Quantum emitters such as atoms, molecules, or quantum dots, coupled to the modes of optical cavities or plasmonic resonators, are fundamental building blocks of many quantum technologies \cite{pellizzari1995decoherence, posani2006nanoscale, noginov2009demonstration, suh2012plasmonic, pichler2016photonic, benito2019optimized, borjans2020resonant, lee2021quantum}, and a must-have for integrated photonic circuits. While setups using resonators in homogeneous background media have limited applications in quantum information technologies, e.g., the transmission of entangled states between satellites \cite{yin2017satellite, yin2017satellite1200}, many quantum technologies use multiple cavities in structured environments, such as waveguide-coupled cavities \cite{cirac1997quantum, pellizzari1997quantum, pichler2016photonic}, for efficient intercavity coupling. See Fig.~\ref{fig:examples} for example structures of coupled cavities in different environments.

Theoretical treatments often start with a Hamiltonian or Lindbladian, where the cavity modes are assumed to be the orthonormal modes of closed resonators, while losses and coupling between different cavities are added heuristically \cite{cirac1997quantum, blais2004cavity, cirac1991two}. Decay rates and coupling elements for realistic systems \red{have to be guessed or fitted to experimental data.}

In contrast, quasinormal modes (QNMs) are the solutions to the source-free Helmholtz equation with a complex permittivity (or dielectric function) under outgoing wave boundary conditions \cite{muljarov2011brillouin, kristensen2012generalized, sauvan2013theory}. The QNMs have complex eigenfrequencies leading to temporal decay and spatial divergence outside the resonator caused by causality \cite{wu2025reflections}. This divergence can be addressed in a few-mode approximation by replacing the QNMs outside the resonator with regularized, frequency-dependent QNM fields, obtained from a Dyson scattering equation \cite{ge2014quasinormal, fuchs2026greens}.
Quasinormal modes have been used to study a variety of setups in optics and photonics, and yield accurate results, even for systems with significant losses \cite{sauvan2013theory, weiss2016from, yan2020shape, primo2020quasinormal, gustin2025what, muljarov2026rigorous}.

The construction of operators for quantized, quasi-bound QNMs has been performed for individual cavities \cite{franke2019quantization} as well as coupled-cavity systems \cite{franke2022quantized}, including multi-cavity systems with significant retardation delays \cite{fuchs2024quantization}. These quantum theories yield rigorously defined, numerically calculable coupling parameters that allow for the implementation of quantized QNMs into established methods for complex quantum dynamics \cite{franke2020quantized, fuchs2026quantum}, such as the time-convolutionless (TCL) method \cite{breuer2002theory}, the hierarchical equations of motion (HEOM) \cite{kubo1989time, tanimura2020numerically, fuchs2023hierarchical}, or path integral methods \cite{caldeira1983path, grabert1988quantum, chernyak1996collective, garraway1997nonperturbative}.
However, previous work on quantized QNMs has focused on cavities in vacuum or homogeneous background media and is not immediately applicable to other environments. A rigorous quantized QNM theory for multiple coupled cavities in structured environments, such as waveguides or surfaces, is therefore needed.

We present here a generalization of the quantization schemes for QNMs from Refs.~\cite{franke2019quantization, fuchs2024quantization, fuchs2026quantum} to structured background media, including waveguide-coupled cavities or cavities near surfaces. In Sec.~\ref {sec:BCs}, we derive general boundary conditions for the electric field, which also apply to the QNMs. From these general boundary conditions, we derive the specific boundary conditions for closed systems, systems in 1D, 2D, and 3D homogeneous background media (i.e., the Silver-Müller radiation condition \cite{muller1948grundzuge, silver1984microwave}), waveguide-coupled cavities \cite{kristensen2017theory}, or cavities on surfaces. We also discuss the case of combined background structures, where different boundary conditions apply in different directions.

In Sec.~\ref{sec:QNMs}, we discuss the classical theory of QNMs, their definition and their regularization (for use outside the cavity regions), and then introduce the adapted quantization scheme for cavities in different background. We discuss the overlap between the quasi-bound modes of separated cavities, that decreases exponentially with the \textit{cavity separation parameter}, which acts as a measure for the separation between the cavities \cite{fuchs2024quantization}. The QNM quantization naturally yields a set of non-bosonic bath operators that primarily describe photons propagating through the background medium between the cavities. We introduce the QNM system-bath Hamiltonian with rigorously defined eigenenergies and coupling parameters between the system and bath, as well as coupling to quantum emitters. As an example, we consider two-level systems (TLSs) that couple to the electric field via dipole coupling. We define coupling elements between the TLSs and the QNMs, respectively, to bath photons. The extension to multi-level emitters is straightforward.

Finally, in Sec.~\ref{sec:corrfunc}, we present the dynamics of the non-bosonic bath operators and define correlation functions for bath-mediated, time-delayed QNM-QNM, QNM-TLS, and TLS-TLS interactions, including coupling elements that are rigorously defined in terms of QNM parameters. These correlation functions can be implemented in many standard open quantum system dynamics schemes
\cite{tanimura2020numerically, fuchs2023hierarchical, vagov2011real, strathearn2017efficient, richter2022enhanced}. 

\begin{table}
    \centering
    \caption{Overview of the types of boundary conditions (BCs) in different dimensions and different kinds of lightguiding background structures.}
    \begin{tabular}{c c c}
       \hline\hline BC & Section & Equation\\ \hline
       General & \ref{sec:genBC} & \eqref{eq:BCintegral}\\
       Closed cavity & \ref{sec:closedcavity} & \eqref{eq:closedcavityBC} \\
         Free space 1D & \ref{sec:BChomogeneous} & \eqref{eq:1DBC}\\
         Free space 2D & \ref{sec:BChomogeneous} & \eqref{eq:2DBC}\\
         Free space 3D & \ref{sec:BChomogeneous} & \eqref{eq:silvermuller}\\
         Waveguide & \ref{sec:wavBC} & \eqref{eq:waveguideBC}\\
         Surfaces & \ref{sec:surfaceBC} & \eqref{eq:surfaceBCfree},\eqref{eq:surfaceBCz0}\\
         Multiple structures & \ref{sec:combinedwavs} & \eqref{eq:BCintegralseparate}\\
         \hline\hline
    \end{tabular}
    \label{tab:summary}
\end{table}

\section{Boundary conditions for cavities in structured media} \label{sec:BCs}
\subsection{Cavity modes in different background environments}
We consider an absorptive, isotropic, inhomogeneous, and non-magnetic medium. The electric field in this medium solves the Helmholtz equation
\begin{align}\label{eq:ehelm}
    \Big[\nabla\times\nabla\times-\frac{\omega^2}{c^2}\epsilon(\mathbf{r},\omega)\Big]\mathbf{E}(\mathbf{r},\omega) = i\omega\mu_0\mathbf{J}(\mathbf{r},\omega),
\end{align}
where \(\epsilon(\mathbf{r},\omega) \) is the electric permittivity (or complex dielectric constant), and \(\mathbf{J}(\mathbf{r},\omega)\) is the source current-density. Extensions to anisotropic or magnetic media are possible, see for example Ref.~\cite{kristensen2020modeling}. 
The field from Eq.~\eqref{eq:ehelm} must also obey the correct boundary conditions (BCs) to match the specific geometry of the medium. The goal of this section is to derive BCs for the field in different kinds of environments. 

If the system is an optical cavity or nanoplasmonic resonator in a lightguiding background medium (e.g., waveguides, surfaces, or free space), the field from Eq.~\eqref{eq:ehelm} will be dominated by the resonances of the cavity. These resonances are described by the resonator modes, which solve the source-free Helmholtz equation at a specific eigenfrequency \(\tilde{\omega}_{\mu}\):
\begin{align}\label{eq:helmquasi}
    \nabla\times\nabla\times\qnm{f}{\mu}(\mathbf{r}) - \frac{\tilde{\omega}_{\mu}^2}{c^2}\epsilon(\mathbf{r},\tilde{\omega}_{\mu})\qnm{f}{\mu}(\mathbf{r}) = 0. 
\end{align}

In an isolated system without absorption, the modes from Eq.~\eqref{eq:helmquasi} are the Hermitian normal modes, while for open boundary conditions, they are quasinormal modes (QNMs) with complex eigenfrequencies \cite{leung1994completeness, muljarov2011brillouin, kristensen2012generalized, sauvan2013theory}.  
For non-degenerate modes, the electric field near a specific resonance takes the form \cite{bai_efficient_2013-1, sauvan2022normalization}
\begin{align}\label{eq:fieldqnmexpand}
    \lim_{\omega\to \tilde{\omega}_{\mu}} (\omega-\tilde{\omega}_{\mu})\mathbf{E}(\mathbf{r},\omega) = \lim_{\omega\to \tilde{\omega}_{\mu}}a_{\mu}(\omega)\qnm{f}{\mu}(\mathbf{r}),
\end{align}
where \(a_{\mu}(\omega) = (1/2i\epsilon_0)\int\mathrm{d}^3r \qnm{f}{\mu}(\mathbf{r})\cdot\mathbf{J}(\mathbf{r},\omega)\) is an analytic expansion coefficient \cite{kristensen2020modeling}.
The boundary conditions derived below can be represented via a linear operator \(\hat{L}(\mathbf{r},\omega)\) which generates the BCs for the electric field on a piecewise smooth surface \(\mathcal{S}\) [e.g., Eq.~\eqref{eq:silvermuller}]. In this case,
\begin{align}\label{eq:QNMBC}
    \Big[\hat{L}(\mathbf{r},\omega)\mathbf{E}(\mathbf{r},\omega)\Big]_{\mathbf{r}\in \mathcal{S}} = 0 \to \Big[\hat{L}(\mathbf{r},\tilde{\omega}_{\mu})\qnm{f}{\mu}(\mathbf{r})\Big]_{\mathbf{r}\in \mathcal{S}} = 0
\end{align}
relates the BCs for the electric field to the BCs of the QNMs.

\subsection{General boundary condition}\label{sec:genBC}
For a general formulation of the boundary conditions for the electric field from Eq.~\eqref{eq:ehelm}, we assume a volume \(V\) bounded by a piecewise smooth surface \(\mathcal{S}\). The electromagnetic Green's function (or dyad) solves the Helmholtz equation~\eqref{eq:ehelm} for a point-source, 
\begin{align}\label{eq:greenhelm}
    \Big[\nabla\times\nabla\times-\frac{\omega^2}{c^2}\epsilon(\mathbf{r},\omega)\Big]\mathbf{G}(\mathbf{r},\mathbf{r}',\omega) = \frac{\omega^2}{c^2}\mathbb{1}\delta(\mathbf{r}-\mathbf{r}'),
\end{align}
where \(\mathbb{1}\) is the unit dyad, and \(\delta(\mathbf{r}-\mathbf{r}')\) is the Dirac delta function.
Towards the goal of deriving the boundary conditions on \(\mathcal{S}\), we use Green's second identity \cite{franke2020fluctuation}:
\begin{align}\label{eq:secondgreen}
    \int_V\mathrm{d}^3r'\Big\{\mathbf{G}^T(\mathbf{r}',\mathbf{r},\omega)\cdot\big[\nabla_{r'}\times\nabla_{r'}\times\mathbf{E}(\mathbf{r}',\omega)\big]-\mathbf{E}(\mathbf{r}',\omega)\cdot\big[\nabla_{r'}\times\nabla_{r'}\times\mathbf{G}(\mathbf{r}',\mathbf{r},\omega)\big]^T\Big\}\nonumber\\
    = \oint_{\mathcal{S}}\mathrm{d}A_s \opvec{n}_s\cdot\Big\{\mathbf{E}(\mathbf{s},\omega)\times\big[\nabla_s\times\mathbf{G}(\mathbf{s},\mathbf{r},\omega)\big]^T-\mathbf{G}^T(\mathbf{s},\mathbf{r},\omega)\times\big[\nabla_s\times\mathbf{E}(\mathbf{s},\omega)\big]\Big\},
\end{align}
where \(\opvec{n}_s\) is the normal vector on \(\mathcal{S}\), which points outwards with respect to \(V\).
We insert the Helmholtz equations for the electric field and Green's function on the left-hand-side, and rearrange the terms to obtain the following expression for the full electric field in \(V\):
\begin{align}
    \mathbf{E}(\mathbf{r},\omega) &= \frac{i}{\omega\epsilon_0}\int_V\mathrm{d}^3r' \mathbf{G}(\mathbf{r},\mathbf{r}',\omega)\cdot\mathbf{J}(\mathbf{r}',\omega)\nonumber\\
    &-\frac{c^2}{\omega^2}\oint_{\mathcal{S}}\mathrm{d}A_s\opvec{n}_s\cdot\Big\{\mathbf{E}(\mathbf{s},\omega)\times\big[\nabla_s\times\mathbf{G}(\mathbf{s},\mathbf{r},\omega)\big]^T-\mathbf{G}^T(\mathbf{s},\mathbf{r},\omega)\times\big[\nabla_s\times\mathbf{E}(\mathbf{s},\omega)\big]\Big\}.
\end{align}

The first term on the right-hand side represents the field that is generated by sources within the volume, while the second term is the net flow of radiation through the surface \(\mathcal{S}\), e.g., due to emitters or absorbers outside of \(V\).  If there is no flow through \(\mathcal{S}\), e.g., for an isolated system or if \(V\) covers the entire space, the integral over the outer surface \(\mathcal{S}_{\rm out}\) of \(V\) must vanish, yielding boundary conditions (BCs) for \(\mathbf{E}(\mathbf{r},\omega)\) via
\begin{align} \label{eq:BCintegral}
    \oint_{\mathcal{S}_{\rm out}}\mathrm{d}A_s \opvec{n}_s\cdot\Big\{\mathbf{E}(\mathbf{s},\omega)\times\big[\nabla_s\times\mathbf{G}(\mathbf{s},\mathbf{r},\omega)\big]^T-\mathbf{G}^T(\mathbf{s},\mathbf{r},\omega)\times\big[\nabla_s\times\mathbf{E}(\mathbf{s},\omega)\big]\Big\} = 0.
\end{align}
In what follows, we derive several well-known BCs from this integral. An overview of these BCs is given in Tab.~\ref{tab:summary}. These boundary conditions can be used to numerically calculate the fields of resonators coupled to various lightguiding background structures. Using the relation from Eq.~\eqref{eq:fieldqnmexpand}, they can be used to derive analytic BCs for the QNMs \(\qnm{f}{\mu}(\mathbf{r})\) in various backgrounds. 
\red{
\subsection{Boundary conditions for a closed cavity}\label{sec:closedcavity}
To show that the requirement from Eq.~\eqref{eq:BCintegral} recovers well-known boundary conditions, we consider a closed system inside a finite volume \(V\) with no transmissive surfaces such that no energy can enter or leave \(V\) through the closed surface \(\mathcal{S}_{\rm closed}\) surrounding \(V\). A sketch of such a system is given in Fig.~\ref{fig:silvbound}(a) for a scattering medium inside a closed resonator. We make no assumption about the permittivity distribution inside \(V\), and hence no assumptions about the electric field or Green's function can be made, so the two terms under the integral in Eq.~\eqref{eq:BCintegral} generally do not cancel. Instead, to hold for  any permittivity distribution inside the closed resonator, the two terms must vanish independently, i.e.,}
\begin{align}\label{eq:closedcavityBC}
    \opvec{n}_s\times\mathbf{E}(\mathbf{s},\omega)\big|_{s\in\mathcal{S}_{\rm closed}} = 0,\nonumber\\
    \opvec{n}_s\times\big[\nabla\times\mathbf{E}(\mathbf{s},\omega)\big]\big|_{s\in\mathcal{S}_{\rm closed}} = 0,
\end{align}
which are \red{the well-known} boundary conditions on a \red{closed} surface with no external surface currents.

\subsection{Homogeneous background media}\label{sec:BChomogeneous}
\begin{figure}
    \centering
    \includegraphics[width=0.8\linewidth]{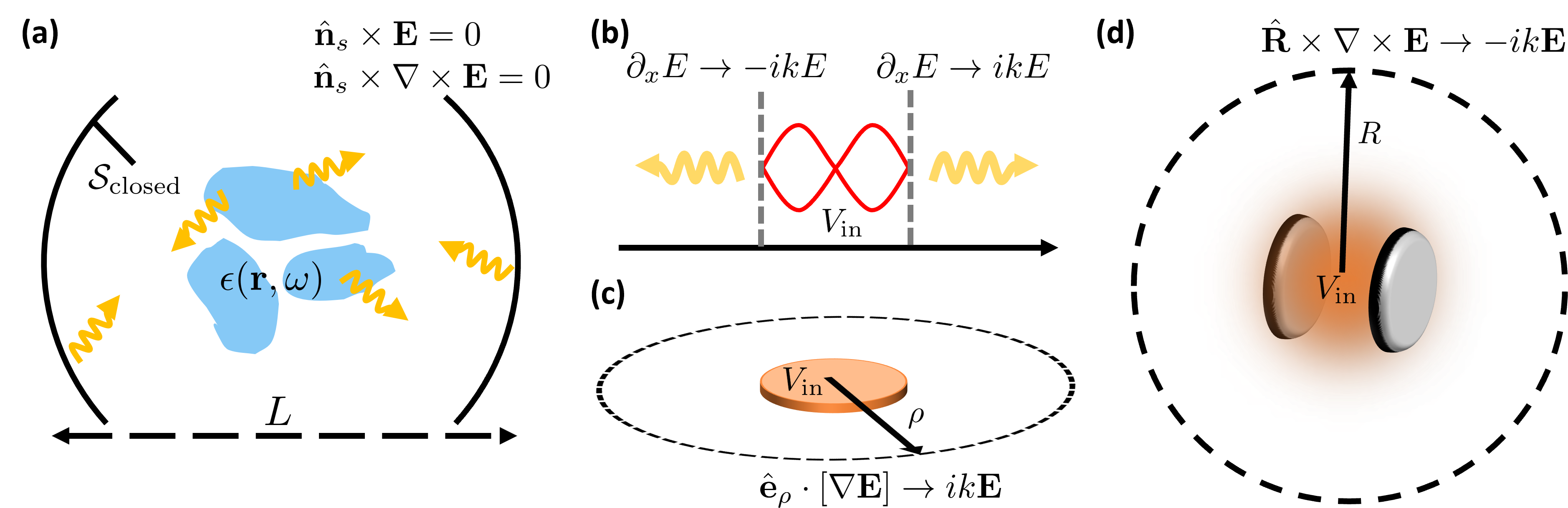}
    \caption{(a) Closed cavity system. A scattering medium described by permittivity \(\epsilon(\mathbf{r},\omega)\) is placed inside a closed resonator of length \(L\), so that waves will scatter back from the resonator surface, described by the boundary conditions in Eq.~\eqref{eq:closedcavityBC}.
    (b)-(d) Cavity in a homogeneous background medium for 1D, 2D, and 3D, respectively. In 1D, the condition from Eq.~\eqref{eq:BCintegral1D} yields outgoing boundary conditions at the cavity surface [Eq.~\eqref{eq:1DBC}]. For 2D (3D), we choose a circular (spherical) far-field surface for the integral from Eq.~\eqref{eq:BCintegral2D} [Eq.~\eqref{eq:BCintegral}], which yields the outgoing wave boundary conditions from Eq.~\eqref{eq:2DBC} [Eq.~\eqref{eq:silvermuller}].}
    \label{fig:silvbound}
\end{figure}

Next, we consider a finite scattering geometry inside a volume \(V_{\rm in}\) embedded in a homogeneous background medium \(V_{\rm out}\), where \(\epsilon(\mathbf{r},\omega)|_{r\in V_{\rm out}} = \epsilon_B\). We discuss this setup in one, two, and three dimensions (1D, 2D, 3D). We illustrate example structures together with the boundary conditions for all three cases in Fig.~\ref{fig:silvbound}(b)-(d). In three dimensions, the outgoing boundary condition is the Silver-Müller radiation condition \cite{muller1948grundzuge, silver1984microwave}, which is the typical choice for open cavity quasinormal modes in homogeneous media \cite{kristensen2020modeling, franke2019quantization, fuchs2024quantization}.

\textit{One Dimensional System.} First, we consider a 1D structure embedded in a homogeneous background medium [Fig.~\ref{fig:silvbound}(b)]. Such a setup can be achieved, for example, by a dielectric slab that is infinite in extent in the $y$-$z$ plane, so that equations reduce to their 1D equivalents along the $x$-axis. The 1D equivalent of Eq.~\eqref{eq:BCintegral} reads (Appendix~\ref{appsec:1D2D})
\begin{align}\label{eq:BCintegral1D}
    [E(x,\omega) \partial_{x}G(x,x',\omega)-G(x',x,\omega)\partial_{x}E(x,\omega)]_{x\to\pm\infty} = 0,
\end{align}
where \(x'\) lies within the scattering volume, and \(x\) lies in the very far field. This yields two boundary conditions, one to the left-hand side \(x\to -\infty\), and one to the right-hand side \(x\to +\infty\). We derive the boundary condition for \(x\to +\infty\) here, the case for \(x\to -\infty\) is obtained accordingly. In the far-field limit \(x\gg x'\), the Green's function in Eq.~\eqref{eq:BCintegral1D} reduces to the Green's function of a point source in a lossless homogeneous background medium, i.e.,  \(G(x,x',\omega)\to ik\exp[ik(x-x')]/2\) [cf.~Eq.~\eqref{eq:greenhelm}], where \(k=n_B\omega/c\) is the wavenumber in the homogeneous background medium with refractive index \(n_B = \sqrt{\epsilon_B}\). Hence, we obtain from Eq.~\eqref{eq:BCintegral1D},
\begin{align}
    \lim_{x\to\infty}\frac{ik}{2}\mathrm{e}^{ik(x-x')}[ikE(x,\omega)-\partial_x E(x,\omega)] = 0. 
\end{align}

For a cavity with a clearly defined surface, such as a dielectric slab or open Fabry-Perot cavity, this far-field boundary condition is equivalent to the boundary condition at the cavity surface [cf.~Fig.~\ref{fig:silvbound}(b)]
\begin{align}\label{eq:1DBC}
    [\partial_x E(x,\omega) -ikE(x,\omega)]_{x\searrow x_R} = 0,
\end{align}
where \(x_R\) is the right-hand side surface of the cavity \cite{lalanne2018light, kristensen2020modeling}. A similar boundary condition is obtained from the left-hand side limit \(x\to -\infty\) and reads [cf.~Fig.~\ref{fig:silvbound}(b)],
\begin{align*}
    [\partial_x E(x,\omega) +ikE(x,\omega)]_{x\nearrow x_L} = 0,
\end{align*}
where \(x_L\) is the left-hand side surface of the cavity, which yields waves propagating away from the cavity in the other direction.

\textit{Two Dimensional System.} Next, we consider a two-dimensional example, such as a 2D microdisk resonator \cite{wiersig2003boundary, franke2022quantized, ren2022connecting} 
as sketched in Fig.~\ref{fig:silvbound}(c). Here, we consider purely 2D phenomena in free space, while phenomena at a planar surface in 3D are considered in Sec.~\ref{sec:surfaceBC}. The 2D equivalent of Eq.~\eqref{eq:BCintegral} reads (Appendix~\ref{appsec:1D2D}),
\begin{align} \label{eq:BCintegral2D}
    &\lim_{\rho\to\infty}\oint_{\mathcal{C}(\rho)}\mathrm{d}s\,\opvec{e}_{\rho}\cdot\Big\{\big[\nabla_{\bm{\rho}}\mathbf{G}(\bm{\rho},\mathbf{r},\omega)\big]^T\cdot\mathbf{E}(\bm{\rho},\omega)-\mathbf{G}^T(\bm{\rho},\mathbf{r},\omega)\cdot\big[\nabla_{\bm{\rho}}\mathbf{E}(\bm{\rho},\omega)\big]\Big\} = 0,
\end{align}
where \(\mathbf{r}\in V_{\rm in}\) lies in the scattering volume, while \(\bm{\rho}=\rho\opvec{e}_{\rho}\) runs to the surrounding circular boundary \(\mathcal{C}(\rho)\) with radius \(\rho\) and outward pointing normal vector \(\opvec{e}_{\rho}\) [Fig.~\ref{fig:silvbound}(c)]. Furthermore, \(\mathrm{d}s = \rho \,\mathrm{d}\varphi\) with \(\varphi\in [0,2\pi)\). 

Again, in the limit of a far-field boundary \(\rho\to\infty\), the Green's function reduces to the 2D free space Green's function \cite{morse1946methods}
\begin{align}
    \mathbf{G}(\bm{\rho},\mathbf{r},\omega) \to \frac{ik^2}{4\epsilon_B}H^{(1)}_0(ks)\mathbb{1},
\end{align}
where \(\mathbb{1}\) is the unit dyadic, again \(k=n_B\omega/c\), and \(H^{(1)}_0(ks)\) is the zero-th Hankel function of the first kind. Also, \(s = |\bm{\rho}-\mathbf{r}|\), since, in a homogeneous medium, the Green's function depends only on the distance between two points.
In the limit \(ks\to \infty\) (which follows from \(\rho\to\infty\)), the Hankel function tends to \(H^{(1)}_0(ks) \to\sqrt{2/(\pi ks)}\mathrm{e}^{iks-i\pi/4}\) \cite{morse1946methods}.
Since \(\rho\gg |\mathbf{r}|\), we approximate \(s \approx \rho\) everywhere except in the exponential.
Hence, Eq.~\eqref{eq:BCintegral2D} reduces to
\begin{align}
    0=\lim_{\rho\to\infty}&\int_0^{2\pi}\mathrm{d}\varphi \frac{i\sqrt{2k^3}\mathrm{e}^{iks-i\pi/4}}{4\epsilon_B}\sqrt{\rho}\Big\{\left(ik-\frac{1}{2\rho}\right)\mathbf{E}(\bm{\rho},\omega)-\opvec{e}_{\rho}\cdot\big[\nabla_{\bm{\rho}}\mathbf{E}(\bm{\rho},\omega)\big]\Big\} .
\end{align}

This expression vanishes if [cf.~Fig.~\ref{fig:silvbound}(c)]
\begin{align}\label{eq:2DBC}
    &\lim_{\rho\to\infty}\sqrt{\rho}\,\mathbf{E}(\bm{\rho},\omega)\quad \mathrm{finite},\quad {\rm and}\nonumber\\
    &\lim_{\rho\to\infty}\sqrt{\rho}\,\Big\{ik\mathbf{E}(\bm{\rho},\omega)-\opvec{e}_{\rho}\cdot\big[\nabla_{\bm{\rho}}\mathbf{E}(\bm{\rho},\omega)\big]\Big\} = 0,
\end{align}
which constitute the outgoing wave boundary conditions in 2D.

\textit{Three Dimensional System: Silver-Müller radiation condition.}
For a cavity in a three-dimensional homogeneous background medium, we use for the outer boundary a spherical surface \(\mathcal{S}(R)\) with radius \(R\), as sketched in Fig.~\ref{fig:silvbound}(d). Since, in the limit \(R\to\infty\)\, no sources are outside the volume, the condition from Eq.~\eqref{eq:BCintegral} applies. 
In the far-field limit, the scattering geometry behaves like a point source, so that the full Green's function becomes the transverse Green's function of the homogeneous background 
\begin{align}
    \lim_{R\to\infty}\mathbf{G}(\mathbf{r},\mathbf{R},\omega) \to \frac{\omega^2}{c^2}\frac{\mathrm{e}^{iks}}{4\pi s}\left(\mathbb{1}-\hat{\mathbf{s}}\otimes\hat{\mathbf{s}}\right),
\end{align}
where \(\mathbf{s} = \mathbf{r}-\mathbf{R}\), \(s = |\mathbf{s}|\), and \(\hat{\mathbf{s}}=\mathbf{s}/s\). Here, \(\mathbf{r}\in V_{\rm in}\), and \(\mathbf{R}\) runs to the outer surface \(\mathcal{S}(R)\). Together with Eq.~\eqref{eq:BCintegral}, we obtain
\begin{align}
    0 = \lim_{R\to\infty}\oint_{\mathcal{S}(R)}\mathrm{d}A\, \frac{\mathrm{e}^{iks}}{4\pi s}\Big\{\left(ik-\frac{1}{s}\right)\left[\mathbf{E}(\mathbf{R},\omega)\times(\mathbb{1}\times\hat{\mathbf{s}})\right]-\left(\mathbb{1}-\hat{\mathbf{s}}\otimes\hat{\mathbf{s}}\right)\times\left[\nabla\times\mathbf{E}(\mathbf{R},\omega)\right]\Big\}\cdot\hat{\mathbf{R}},
\end{align}
where \(\hat{\mathbf{R}}=\mathbf{R}/|\mathbf{R}|\) is the normal vector on \(\mathcal{S}(R)\). Now, we set \(\mathbf{s} \approx \mathbf{R}\), everywhere except in the exponential, and perform a far-field approximation, where \(|\mathbf{R}|\gg |\mathbf{r}|\). We furthermore rewrite the integral into an integral over the solid angle \(\Omega\) via \(\mathrm{d}A = R^2\mathrm{d}\Omega\). We then find
\begin{align}
    \lim_{R\to\infty}\int_{\Omega}\mathrm{d}\Omega \frac{\mathrm{
    e}^{iks}}{4\pi}\Big[R\big\{ik\mathbf{E}^T(\mathbf{R},\omega)+\hat{\mathbf{R}}\times\big[\nabla\times\mathbf{E}^T(\mathbf{R},\omega)\big]\big\}-\mathbf{E}^T(\mathbf{R},\omega)\Big] = 0.
\end{align}

The condition is satisfied if the relations are fulfilled [cf.~Fig.~\ref{fig:silvbound}(d)]
\begin{align*}
    &\lim_{R\to\infty} R \,\mathbf{E}^T(\mathbf{R},\omega) \,\,\mathrm{finite},\quad {\rm and}\\
    &\lim_{R\to\infty} R\left\{ik\mathbf{E}^T(\mathbf{R},\omega)+\hat{\mathbf{R}}\times\left[\nabla\times\mathbf{E}^T(\mathbf{R},\omega)\right]\right\} = 0,
\end{align*}
which constitute the \textit{Silver-Müller radiation condition} \cite{silver1984microwave, muller1948grundzuge}. They can be combined into the single expression \cite{kristensen2020modeling}
\begin{align}\label{eq:silvermuller}
    \hat{\mathbf{R}}\times\left[\nabla\times\mathbf{E}^T(\mathbf{R},\omega)\right] \to -
    ik\mathbf{E}^T(\mathbf{R},\omega),\, R\to\infty.
\end{align}
Note that the radiation condition applies to the transverse part of the electric field only. Since \(\nabla\times\mathbf{E}^L = 0\) holds, the longitudinal part \(\mathbf{E}^L\) reads [cf.~Eq.~\eqref{eq:ehelm}]
\begin{align}\label{eq:Elong}
    \mathbf{E}^L(\mathbf{r},\omega) = \frac{-i}{\omega\epsilon_0}\frac{\mathbf{J}^L(\mathbf{r},\omega)}{\epsilon(\mathbf{r},\omega)}.
\end{align}

\subsection{Waveguide radiation condition}\label{sec:wavBC}
\begin{figure}
    \centering
    \includegraphics[width=0.65\linewidth]{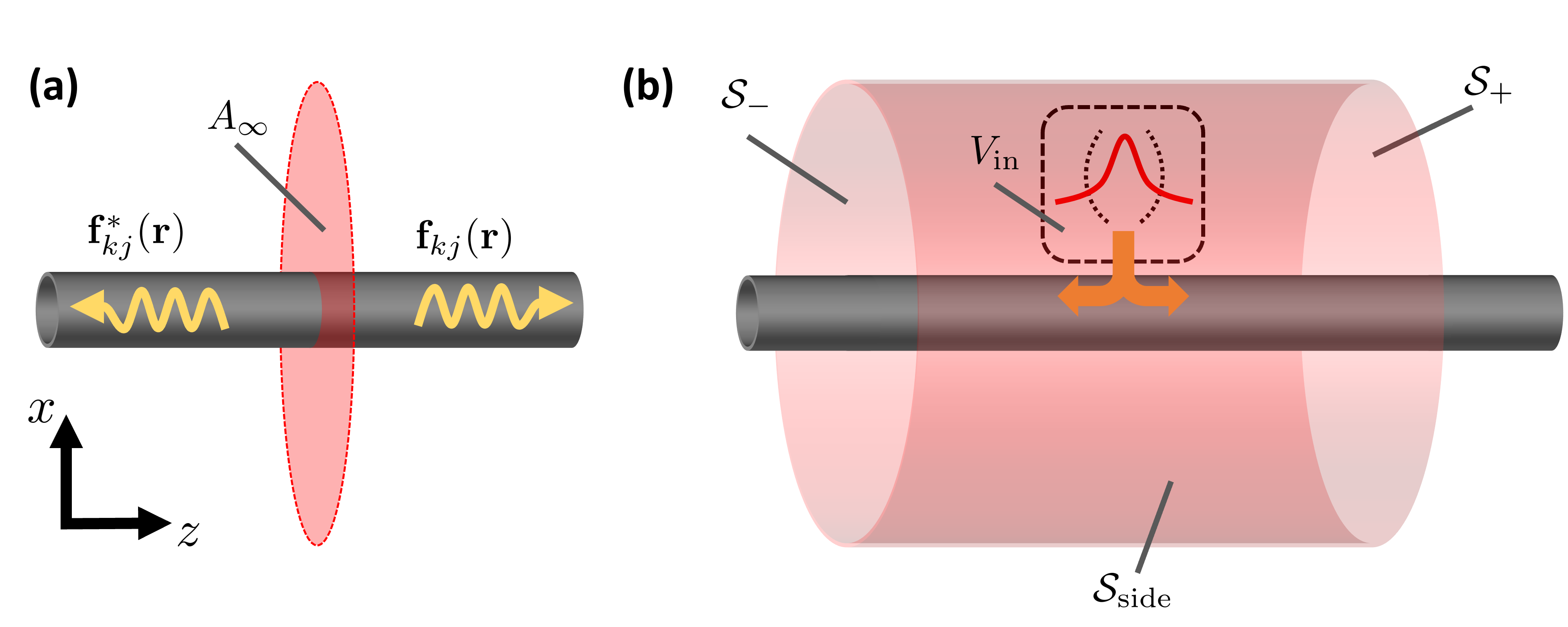}
    \caption{(a) Sketch of a single cylindrical waveguide with forward (backward) propagating waveguide modes \(\mathbf{f}^{(*)}_{kj}(\mathbf{r})\) and infinite cross-section \(A_{\infty}\) over which the normalization of the waveguide modes is carried out [Eq.~\eqref{eq:waveguidenorm}]. (b) Sketch of a cavity coupled to the waveguide from (a) as its only loss channel. The field of the open cavity extends into the waveguide. Equation~\eqref{eq:BCintegral} is evaluated over a far-field surface \(\mathcal{S}_{\rm side}\cup\mathcal{S}_-\cup\mathcal{S}_+\), yielding the waveguide boundary conditions from Eq.~\eqref{eq:waveguideBC}.}
    \label{fig:wavbound}
\end{figure}

In this section, we want to derive the boundary conditions that apply to the field of a cavity coupled to a lossless waveguide. (cf.~Fig.~\ref{fig:wavbound}) We assume that the waveguide constitutes the only loss channel to the cavity, i.e., no energy leaks into the surrounding medium. 
We consider a finite-size scattering geometry (e.g., an optical cavity) centered at \(z=0\) coupled to an infinite, non-absorptive waveguide as sketched in Fig.~\ref{fig:wavbound}(b), or a periodic photonic crystal waveguide as in Fig.~\ref{fig:examples}(b). The waveguide has modes \(\mathbf{f}_{kj}(\mathbf{r}) = \mathbf{e}_{kj}(\mathbf{r})\mathrm{e}^{ik_jz}\), where \(k_j\) is the \(j\)-th mode index (propagation constant) at wavenumber \(k=\omega/c\), which is obtained from the eigenvalue equation for the modes \cite{snyder1983optical}. 
The modes \(\mathbf{f}_{kj}\) propagate forwards or backward in the direction of the waveguide. We assume \(k_j>0\) so that \(\mathbf{f}_{kj}(\mathbf{r})\) is a forward-propagating mode and \(\mathbf{f}_{-kj}(\mathbf{r})\equiv\mathbf{f}^*_{kj}(\mathbf{r})\) is a backward-propagating mode.

Meanwhile, we assume that the mode functions \(\mathbf{e}_{kj}(\mathbf{r})\) decay exponentially perpendicular to the waveguide (or are zero outside if the waveguide cladding is perfectly reflecting). For a periodic waveguide (e.g., on a photonic crystal) with lattice constant \(a\), \(\mathbf{e}_{kj}(\mathbf{r}) = \mathbf{e}_{kj}(\mathbf{r}+a\opvec{e}_z)\) are Bloch modes, which are normalized across the unit cell \cite{hughes2005extrinsic, kristensen2017theory}.

Generally, the waveguide modes \(\mathbf{f}_{kj}(\mathbf{r})\) are normalized according to \cite{kristensen2017theory, snyder1983optical}
\begin{align} \label{eq:waveguidenorm}
   \int_{A_{\infty}}\mathrm{d}A \,\opvec{n}\cdot&\Big\{\mathbf{f}_{kj}(\mathbf{s})\times\big[\nabla\times\mathbf{f}^*_{k'j'}(\mathbf{s})\big]+\mathbf{f}^*_{k'j'}(\mathbf{s})\times\big[\nabla\times\mathbf{f}_{kj}(\mathbf{s})\big]\Big\}= \delta_{jj'}\delta_{kk'},
\end{align}
where \(A_{\infty}\) is the infinite cross-section of the waveguide [Fig.~\ref{fig:wavbound}(a)], and \(\opvec{n}\) is the normal surface vector on \(A_{\infty}\), parallel to the waveguide axis. Note that for an infinite waveguide, the position of the cross-section along the \(z\)-axis is arbitrary. 

Next, we consider a far-field cylindrical surface \(\mathcal{S}_{\infty} = \mathcal{S}_{\rm side}\cup\mathcal{S}_-\cup\mathcal{S}_+\) around the waveguide and cavity, where \(\mathcal{S}_{\rm side}\) is the side of the cylinder, while \(\mathcal{S}_-\) and \(\mathcal{S}_+\) are the bases of the cylinder in the negative and positive \(z\)-directions, respectively [see Fig.~\ref{fig:wavbound}(b) for a sketch of the far-field surface]. In the limit of an infinite cylinder, there are no sources outside the far-field surface \(\mathcal{S}_{\infty}\), and Eq.~\eqref{eq:BCintegral} applies. Similar to the derivation in homogeneous background media, the full Green's function in the far-field limit can be expanded in terms of the waveguide modes with propagation constants \(k_j\) at frequency \(\omega\) \cite{hughes2005extrinsic}
\begin{align}\label{eq:waveguidegreen}
    \mathbf{G}(\mathbf{r},\mathbf{r}',\omega)\to \sum_j A_{kj} &\Big[\mathbf{f}_{kj}(\mathbf{r})\mathbf{f}^*_{kj}(\mathbf{r}')\Theta(z-z')+\mathbf{f}^*_{kj}(\mathbf{r})\mathbf{f}_{kj}(\mathbf{r}')\Theta(z'-z)\Big],
\end{align}
where \(A_{kj} = ia\omega/(2k_j)\), where \(a\) is the lattice constant for a periodic waveguide, and \(a=1\) for a continuous waveguide. For practical calculations, an integration over the dispersion of the bound modes is often carried out, so that, instead of many modes \(j\) at wavenumber \(k\), Eq.~\eqref{eq:waveguidegreen} is expressed via a single mode \(k_{\omega}\). In this case, \(A_k =  ia\omega/(2v_g)\), where \(v_g\) is the group velocity \cite{hughes2005extrinsic, kristensen2017theory}.

We insert the Green's function in the far-field limit into Eq.~\eqref{eq:BCintegral}:
\begin{align}\label{eq:waveguideBC}
    0 = \sum_j A_{kj} \mathbf{f}_{kj}(\mathbf{r})\int_{\mathcal{S}_-}\mathrm{d}A \,\opvec{n}\cdot\Big\{\mathbf{E}(\mathbf{s},\omega)\times\big[\nabla\times\mathbf{f}_{kj}^*(\mathbf{s})\big]-\mathbf{f}_{kj}^*(\mathbf{s})\times\big[\nabla\times\mathbf{E}(\mathbf{s},\omega)\big]\Big\}\nonumber\\
    +\sum_j A_{kj} \mathbf{f}^*_{kj}(\mathbf{r})\int_{\mathcal{S}_+}\mathrm{d}A \,\opvec{n}\cdot\Big\{\mathbf{E}(\mathbf{s},\omega)\times\big[\nabla\times\mathbf{f}_{kj}(\mathbf{s})\big]-\mathbf{f}_{kj}(\mathbf{s})\times\big[\nabla\times\mathbf{E}(\mathbf{s},\omega)\big]\Big\}.
\end{align}

The integral over the lateral surface of the cylinder vanishes because the lossless waveguide modes are assumed to decay exponentially in that direction.
From the normalization condition of the waveguide modes in Eq.~\eqref{eq:waveguidenorm}, it follows that Eq.~\eqref{eq:waveguideBC} holds for
\begin{align} \label{eq:conds}
    \mathbf{E}(\mathbf{r},\omega) &\to \sum_j\sigma^{\rm backw}_{kj}\mathbf{f}^*_{kj}(\mathbf{r}),\, z\to -\infty,\nonumber\\
    \mathbf{E}(\mathbf{r},\omega) &\to \sum_j\sigma^{\rm forw}_{kj}\mathbf{f}_{kj}(\mathbf{r}),\, z\to\infty,
\end{align}
i.e., the field far away to the left of the scatterer is represented by the left-propagating waveguide modes only, and the field far away to the right is represented by the right-propagating modes only.
Here,
\begin{align}\label{eq:wavexpand}
    \sigma^{\rm backw}_{kj} = \int_{A_{\infty}}\mathrm{d}A \,\opvec{n}\cdot\Big\{\mathbf{E}(\mathbf{s},\omega)\times\big[\nabla\times\mathbf{f}_{kj}(\mathbf{s})\big]-\mathbf{f}_{kj}(\mathbf{s})\times\big[\nabla\times\mathbf{E}(\mathbf{s},\omega)\big]\Big\},\nonumber\\
    \sigma^{\rm forw}_{kj}=\int_{A_{\infty}}\mathrm{d}A \,\opvec{n}\cdot\Big\{\mathbf{E}(\mathbf{s},\omega)\times\big[\nabla\times\mathbf{f}_{kj}^*(\mathbf{s})\big]-\mathbf{f}_{kj}^*(\mathbf{s})\times\big[\nabla\times\mathbf{E}(\mathbf{s},\omega)\big]\Big\}
\end{align}
are the projections of the field onto the left-propagating and right-propagating waveguide modes, respectively. Note that, for \(z\to \pm\infty\) only waveguide modes with real \(k_j\) contribute, since the modes with \(\mathrm{Im}(k_j) >0\) are damped to zero. 

The equations in Eq.~\eqref{eq:conds} constitute the boundary conditions for the electric field of a cavity coupled to an infinite waveguide as its only loss channel and hold in this form also for periodic waveguides, e.g., on a photonic crystal \cite{kristensen2017theory}. The boundary conditions demand that the field far away from the scatterer can be represented by the waveguide modes traveling \textit{away} from the scatterer, thus fulfilling the condition that there is no backscattering from outside \(\mathcal{S}_{\infty}\).

\section{Boundary conditions for combinations of multiple lightguiding structures}\label{sec:combinedstructures}
After deriving specific boundary conditions for setups in which the cavity emission enters a single type of guiding background structure, we now turn to setups in which the boundary conditions vary with the direction of emission. As a first example, we discuss a cavity near the interface between layered media, where the far-field emission can appear as either spherical waves in free space or bound surface waves. We formulate the boundary condition integral from Eq.~\eqref{eq:BCintegral} for this case, which is dominated by the spherical waves or surface waves, depending on the direction of the emission. As our second example, we consider a cavity coupled to multiple waveguides simultaneously. Here, we formulate conditions under which the waveguides can be treated as separate background structures so that far-field emissions into one of the waveguides take the form of outward propagating waveguide modes of that waveguide only.

\subsection{Cavities at planar surfaces}\label{sec:surfaceBC}
\begin{figure}    \centering
    \includegraphics[width=0.9\linewidth]{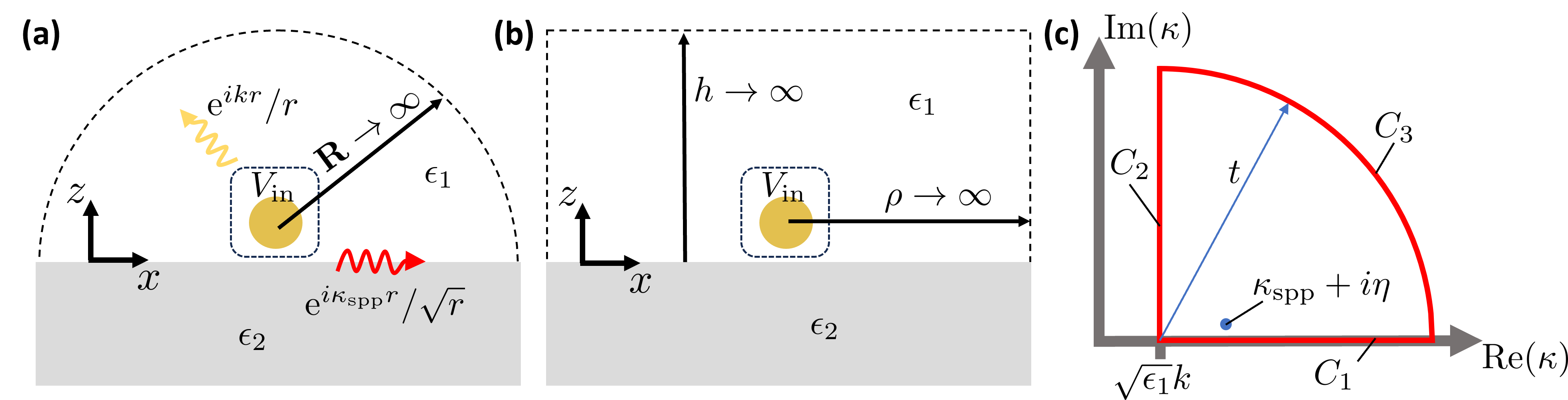}
    \caption{Sketch of a cavity (here, a plasmonic particle) located near the planar interface between two different media \(\epsilon_1\neq \epsilon_2\). (a) As our far-field surface for the boundary condition integral from Eq.~\eqref{eq:BCintegral}, we chose a half sphere with radius \(R\to \infty\). The far field takes different forms, depending on the direction in which the emission occurs. Far away from the surface, the waves behave like spherical waves the form \(\mathrm{e}^{i\kappa_{\rm spp} r}/\sqrt{r}\) (red arrow) appear, similar to the 2D problem from Sec.~\ref{sec:BChomogeneous}. Here, \(\kappa_{\rm spp}\) is the in-plane wavenumber of a surface plasmon polarition [cf.~Eq.~\eqref{eq:sppkappa}]. (b) Same as before but with a cylindrical far-field surface with height \(h\to\infty\) and radius \(\rho\to\infty\). (c) Integration path to calculate the integral from Eq.~\eqref{eq:Gsurderiv}. The integrals over the paths \(C_2\) and \(C_3\) vanish in the limit \(\rho\to\infty\), since there, \({\rm Im}(\kappa)>0\). Thus, the integration over the full path \(C_1+C_2+C_3\) recovers the integral from Eq.~\eqref{eq:Gsurderiv} in the limit \(t\to\infty\). Setting \(\eta\to 0\) after the integration yields the result from Eq.~\eqref{eq:Gsur}. }
    \label{fig:surfacecav}
\end{figure}

We start with a cavity near a planar interface between media, as sketched in Fig.~\ref{fig:surfacecav}. For a simple example, we consider the case with a single interface between homogeneous media, so that the permittivity reads \(\epsilon(z) = \Theta(z)\epsilon_1 + \Theta(-z)\epsilon_2\) with \(\epsilon_1\neq \epsilon_2\). Extensions to layered media are possible \cite{vial2014quasimodal, binkowski2022computation}. Such setups are common in nano-optics, where cavities or quantum emitters (atoms, molecules, or quantum dots) are located on substrate surfaces \cite{yan2018rigorous, sauvan2022normalization}. For a simple illustration, we consider a finite-size cavity above the interface centered at \(\mathbf{r}\approx 0\). We also focus only on the region \(z\geq 0\) and do not discuss the transmitted fields below the interface, which can be treated similarly, accounting for the different material properties for \(z<0\). 

Theoretical treatments of such systems are usually based on one of two approaches: Numerical treatments, including the calculation of QNMs for these setups, use perfectly matched layers that surround the cavity and substrate to use fictitious far-field losses to simulate open boundary conditions \cite{yan2018rigorous, binkowski2022computation, wu2025reflections}. Meanwhile, treatments based on analytic approaches use a partial  spatial Fourier transform of the Green's function and fields in the plane parallel to the interface and then describe the reflections at the interface in terms of Fresnel coefficients \cite{sipe1987new, paulus2000accurate, novotny2012principles}. 
These analytic expressions of the Green's function take the form of Sommerfeld integrals \cite{sipe1987new, novotny2012principles, paulus2000accurate},
\begin{align}\label{eq:sommerfeldG}
    \mathbf{G}(\rho,z,\omega)\big|_{z>0} = k^2\int_0^{\infty}\mathrm{d}\kappa \frac{\kappa}{q_1} \mathbf{K}(\rho,\kappa,\omega)\mathrm{e}^{iq_1 z},
\end{align}
where \(q_j = \sqrt{\epsilon_jk^2-\kappa^2}\) is the z-component of the wave vector in the j-th medium, which is an implicit function of \(k = \omega/c\) and \(\kappa\). Furthermore, the dyadic kernel \(\mathbf{K}\) (see Appendix~\ref{appsec:greensurface} for the definition) depends on Bessel functions \(J_n(\kappa\rho)\), which account for the lateral wave part, and the Fresnel coefficients \cite{novotny2012principles}
\begin{align}\label{eq:fresnelcoeffs}
    r_s(\kappa) = \frac{q_1-q_2}{q_1+q_2},\quad r_p(\kappa) = \frac{\epsilon_2 q_1-\epsilon_1 q_2}{\epsilon_2 q_1+\epsilon_1 q_2}. 
\end{align}

Inserting Eq.~\eqref{eq:sommerfeldG} into the boundary integral from  Eq.~\eqref{eq:BCintegral} does not immediately yield an analytic expression of the far-field boundary conditions for the electric field. For such an expression, the Green's function in real space is required. However, for the boundary conditions, we only require the asymptotic Green's function in the limit \(r = \sqrt{\rho^2+z^2}\to \infty\). 

The Sommerfeld integral from Eq.~\eqref{eq:sommerfeldG} contains a \red{singularity} at \(\kappa = \sqrt{\epsilon_1}k\) from the \(1/q_1\)-term. 
It can be formally separated into propagating contributions \(\kappa\leq \sqrt{\epsilon_1} k\) (and therefore \(q_1 \geq 0\)), and evanescent contributions \(\kappa> \sqrt{\epsilon_1} k\), where \(q_1\) is imaginary with positive imaginary part, leading to exponential decay. For \(z\to \infty\), the evanescent contributions vanish, and we obtain a purely propagating solution in free space in terms of spherical waves \cite{novotny2012principles}
\begin{align}\label{eq:Gspace}
    \mathbf{G}(\rho,z,\omega)\big|_{z\to \infty} \approx \mathbf{G}_{\rm free}(\rho,z,\omega) = \frac{k^2e^{ikr}}{4\pi r}\big[(\mathbb{1}-\opvec{e}_r\opvec{e}_r)+\mathbf{A}^{\rm ref}(\theta)\big],
\end{align}
where the \(\mathbb{1}-\opvec{e}_r\opvec{e}_r\) yields the transverse free space Green's function (sec.~\ref{sec:BChomogeneous}), and \(\mathbf{A}^{\rm ref}(\theta)\) (see Appendix~\ref{appsec:greensurface} for the analytic form) yields the reflected contributions. The reflections depend on the incident angle \(\theta\) between \(\mathbf{r}\) and the \(z\)-axis, and hence \(\sin(\theta) = \rho/r,\, \cos(\theta) = z/r\). In the limit \(z\to 0\) (\(\theta \to \pi/2\)), \(\mathbf{A}^{\rm ref}(\pi/2) \to -(\mathbb{1}-\opvec{e}_r\opvec{e}_r)\) so that \(\mathbf{G}_{\rm free}(\rho,z,\omega)|_{z\to 0} \approx 0\). Physically, this vanishing of the free space contribution to the Green's function accounts for destructive interference between the free space emission and reflections as the wave propagates almost parallel to the interface. 

For finite \(z\) (and especially \(z\to 0\)), the evanescent contributions to Eq.~\eqref{eq:sommerfeldG} cannot be neglected. \red{In this regime, the Green's function is dominated by the singularities of the \(\kappa\)-integral. One of these singularities is the branch point at \(\kappa = \sqrt{\epsilon_1} k\) discussed above. We discuss this singularity in Appendix~\ref{appsec:greensurface}, and show that it leads to a contribution to the asymptotic Green's function that scales as \(1/\rho\). The other singularity is a pole in}
the Fresnel coefficient \(r_p(\kappa)\) if \(q_1 = -(\epsilon_1/\epsilon_2)q_2\). This can be achieved for \(\epsilon_2 < 0 <\epsilon_1 < |\epsilon_2|\) if \cite{novotny2012principles}
\begin{align}\label{eq:sppkappa}
    \kappa^2= \kappa_{\rm spp}^2 = k^2 \frac{\epsilon_1\epsilon_2}{\epsilon_1+\epsilon_2},
\end{align}
which is the lateral wavenumber of a surface plasmon polariton \red{(SPP)}.
Since \(\kappa_{\rm spp} > k\), \(q_1(\kappa_{\rm spp}) \equiv iq_{\rm spp}\) is imaginary, so that this surface wave decays exponentially perpendicular to the interface and vanishes for \(z\to \infty\). For our discussion here, we assume that the setup from Fig.~\ref{fig:surfacecav} supports surface plasmon polaritons, which \red{scale as \(1/\sqrt{\rho}\) and therefore} dominate in the far field near the surface over \red{the branch point contribution or} other contributions\red{, such as lateral waves \cite{king2005lateral, park2017characterization, emelyanenko2017surface}}.

\red{To derive a boundary condition from the SPP contribution, we consider} the evanescent part of the Sommerfeld integral from Eq.~\eqref{eq:sommerfeldG} for the limit \(\rho\to \infty, z\to 0\). We only consider the contributions to the dyadic \(\mathbf{K}\) which contain the Fresnel coefficient \(r_p(\kappa)\) which has a pole at \(\kappa = \kappa_{\rm spp}\). The Bessel functions \(J_n(\kappa\rho)\) in \(\mathbf{K}\) in the limit \(\rho\to\infty\) scale as
\begin{align}
    J_n(\kappa\rho)\big|_{\rho\to \infty} \approx \sqrt{\frac{1}{2\pi\kappa\rho}}\left(\mathrm{e}^{i\kappa\rho-in\pi/2-i\pi/4}+\mathrm{e}^{-i\kappa\rho+in\pi/2+i\pi/4} \right).
\end{align}

We only include the retarded contribution, yielding an overall scaling of \(\mathrm{e}^{i\kappa\rho}/\sqrt{\kappa\rho}\) from the Bessel function. Hence, the Green's function approximately reads,
\begin{align}\label{eq:Gsurderiv}
    \mathbf{G}(\rho,z,\omega)\big|_{\rho\to\infty, z\to 0} \approx k^2\int_{\sqrt{\epsilon_1}k}^{\infty}\mathrm{d}\kappa \frac{\kappa}{q_1}\tilde{\mathbf{K}}^p(\kappa,\omega) \frac{\mathrm{e}^{i\kappa\rho}}{\sqrt{\kappa\rho}}\frac{\mathrm{e}^{iq_1 z}}{\epsilon_2 q_1+\epsilon_1 q_2}.
\end{align}

Here, the dyadic \(\tilde{\mathbf{K}}^p(\kappa,\omega) \) contains additional \(\kappa\)-dependent contributions as well as prefactors from the Bessel functions, which we leave general here, since they do not factor into the boundary conditions below. Assuming that all contributions (except the pole at \(\kappa = \kappa_{\rm spp}\)) are analytic, we can solve the integral via contour integral methods. Towards this, we shift the pole to \(\kappa = \kappa_{\rm spp}+i\eta\), with \(\eta>0\), since we are interested in the retarded solution. In the limit \(\eta\to 0\), the original pole is recovered. Then, we integrate over the path shown in Fig.~\ref{fig:surfacecav}(c), where the integrals over the paths \(C_2\) and \(C_3\) vanish, since the exponential \(\mathrm{e}^{i\kappa\rho}\) goes to zero in the limit \(\rho \to\infty\) for any \(\kappa\) with positive imaginary part. Thus, the red path \(C_1+C_2+C_3\) in Fig.~\ref{fig:surfacecav}(c) for \(t\to \infty\) recovers the integral from Eq.~\eqref{eq:Gsurderiv}.
Evaluating the expression at the pole and then setting \(\eta\to 0\) yields
\begin{align}\label{eq:Gsur}
     \mathbf{G}(\rho,z,\omega)\big|_{\rho\to\infty, z\to 0} \approx  \mathbf{G}_{\rm sur}(\rho,z,\omega) = \mathbf{B}(\kappa_{\rm spp},\omega) \frac{\mathrm{e}^{i\kappa_{\rm spp}\rho}}{\sqrt{\rho}}\mathrm{e}^{-q_{\rm spp} z},
\end{align}
where again, we left the dyadic \(\mathbf{B}(\kappa_{\rm spp},\omega)\) general, since it does not factor into the boundary conditions below. We merely note that the shape of \(\mathbf{B}(\kappa_{\rm spp},\omega)\) is such that the Green's function remains transverse: \(\nabla\cdot\mathbf{G}_{\rm sur} = 0\).

Hence, this setup presents a combination of two different cases discussed above in Sec.~\ref{sec:BChomogeneous}: 
For \(z\to \infty\), the freely propagating and reflected fields take the form of spherical waves, similar to the 3D free space case. Meanwhile, depending on the ratio between \(\epsilon_1\) and \(\epsilon_2\), bound surface states may appear for \(\rho\to \infty\) and finite \(z\), which behave like the 2D waves with an additional exponential decay perpendicular to the interface. Assuming that all other contributions decay faster that \(1/r\), we can write the asymptotic Green's function as a simple sum of these two contributions:
\begin{align}\label{eq:surfaceG}
    \mathbf{G}(\rho,z,\omega)|_{r\to\infty} \approx \mathbf{G}_{\rm free}(\rho,z,\omega)+\mathbf{G}_{\rm sur}(\rho,z,\omega),
\end{align}
where \(\mathbf{G}_{\rm free}\) vanishes as we approach the surface, while \(\mathbf{G}_{\rm sur}\) vanishes as we move way from the surface. Thus, the two contributions represent two different regimes depending on the direction of the emission from the cavity. We note that the neglected higher order terms could in principle also be included in the Green's function and may even play a crucial role for cases without bound surface states, where they constitute the dominant contribution to lateral emissions from the cavity. 

Inserting Eq.~\eqref{eq:surfaceG} into the boundary integral from Eq.~\eqref{eq:BCintegral}, we obtain, for \(z>0\),
\begin{align} \label{eq:surfaceBCintegral}
    \oint_{\mathcal{S}_{\rm out}}\mathrm{d}A_s \opvec{n}_s\cdot\Big[&\mathbf{E}(\mathbf{s},\omega)\times\big\{\nabla_s\times\big[\mathbf{G}_{\rm free}(\rho,z,\omega)+\mathbf{G}_{\rm sur}(\rho,z,\omega)\big]\big\}^T\nonumber\\
    &-\big[\mathbf{G}_{\rm free}^T(\rho,z,\omega)+\mathbf{G}_{\rm sur}^T(\rho,z,\omega)\big]\times\big\{\nabla_s\times\mathbf{E}(\mathbf{s},\omega)\big\}\Big] = 0,
\end{align}
where we set the position of the source \(\mathbf{r}\approx 0\), and \(\rho,z\) are the in-plane and out-of-plane components of \(\mathbf{s}\), respectively. Since the free space spherical waves and surface waves are different phenomena, the Green's functions \(\mathbf{G}_{\rm free}\) and \(\mathbf{G}_{\rm sur}\) are linearly independent, and the two contributions to Eq.~\eqref{eq:surfaceBCintegral} must vanish independently. For the far-field outer surface \(\mathcal{S}_{\rm out}\), we may choose any appropriate surface that matches the geometry of the waves. Indeed, we can even choose different far-field surfaces for each of the two contributions to simplify the derivation.

For the free space part, we choose a semi-spherical surface with radius \(R\to\infty\) in the upper half space \(z>0\) [cf.~Fig.~\ref{fig:surfacecav}(a)]. For this surface, \(\mathrm{d}A_s = R^2 \mathrm{d}\Omega\), where \(\Omega\) is the solid angle of the half sphere. Thus, the resulting boundary conditions take a similar shape as the Silver-Müller radiation condition from Eq.~\eqref{eq:silvermuller}, except that we must distinguish between the polarization directions to account for the reflection (see Appendix~\ref{appsec:greensurface} for the derivation):
\begin{align}\label{eq:surfaceBCfree}
    &\lim_{R\to\infty} R\opvec{e}_{\varphi}\cdot \left\{\opvec{e}_r\times[\nabla\times\mathbf{E}(\mathbf{R},\omega)]-\opvec{e}_r\times\Big[\frac{\nabla r_s(\theta)}{1+r_s(\theta)}\times\mathbf{E}(\mathbf{R},\omega)\Big]+ik\mathbf{E}(\mathbf{R},\omega) \right\} = 0,\quad {\rm and}\nonumber\\
    &\lim_{R\to\infty} R\opvec{e}_{\theta}\cdot  \left\{\opvec{e}_r\times[\nabla\times\mathbf{E}(\mathbf{R},\omega)]+\mathbf{E}(\mathbf{R},\omega)\left[ik-\opvec{e}_{\theta}\cdot\frac{\nabla \sin(2\theta)r_p(\theta)}{1-\cos(2\theta)r_p(\theta)}\right] \right\} = 0,
\end{align}
where \(\opvec{e}_{r},\opvec{e}_{\theta},\opvec{e}_{\varphi}\) are the unit vectors in spherical coordinates, and \(r_s(\theta),r_p(\theta)\) are the Fresnel coefficients from Eq.~\eqref{eq:fresnelcoeffs} in real space, which depend on the incident angle \(\theta\) between \(\mathbf{R} = R\opvec{e}_r\) and the \(z\)-axis. Without reflection, Eq.~\eqref{eq:surfaceBCfree} recovers the Silver-Müller radiation condition from Eq.~\eqref{eq:silvermuller}. 

For the surface plasmon part, we instead use a cylindrical far-field surface with radius \(\rho\to \infty\) and height \(h\to \infty\) [Fig.~\ref{fig:surfacecav}(b)]. The integral from Eq.~\eqref{eq:surfaceBCintegral} for the surface wave part thus separates into an integral over the top of the cylinder at \(z=h\to \infty\), where the Green's function \(\mathbf{G}_{\rm sur}\) is zero, and an integral over the side of the cylinder, where \(\mathrm{d}A_s = \rho \mathrm{d}\varphi\mathrm{d}z\). Thus, we obtain (cf.~Appendix~\ref{appsec:greensurface}) for the surface waves:
\begin{align}\label{eq:surfaceBCz0}
    &\lim_{\rho\to\infty}\sqrt{\rho}\mathbf{E}(\mathbf{r},\omega) \, {\rm finite},\quad {\rm and},\nonumber\\
    &\lim_{\rho\to\infty}\sqrt{\rho}\left\{i\mathbf{k}_{\rm spp}\times[\opvec{e}_{\rho}\times\mathbf{E}(\mathbf{r},\omega)]-\opvec{e}_{\rho}\times[\nabla\times\mathbf{E}(\mathbf{r},\omega)]\right\} = 0,
\end{align}
where \(\mathbf{k}_{\rm spp} = \kappa_{\rm spp}\opvec{e}_{\rho}+iq_{\rm spp}\opvec{e}_z\) is the wavevector of the surface plasmon polariton, so that the exponential decay of the field in the \(z\)-direction follows from the boundary conditions. If we restrict ourselves to the regime immediately at the surface \(z \to 0\), we find the familiar Silver-Müller form [cf.~Eq.~\eqref{eq:silvermuller}]: \(\lim_{\rho\to\infty}\sqrt{\rho}[i\kappa_{\rm spp}\mathbf{E}+\opvec{e}_{\rho}\times(\nabla\times\mathbf{E})] = 0\).

Thus, the far-field behavior of the electric field depends on whether the emission occurs into free space or along the interface. For a setup of multiple cavities on a substrate that supports surface waves [cf.~Fig.~\ref{fig:examples}(c)], propagation along the surface offers a more efficient transfer of energy between the cavities than propagation through free space, since the surface waves decay as \(1/\sqrt{r}\) instead of \(1/r\). 

\subsection{Combination of multiple lightguiding structures}\label{sec:combinedwavs}
\begin{figure}
    \centering
    \includegraphics[width=0.4\linewidth]{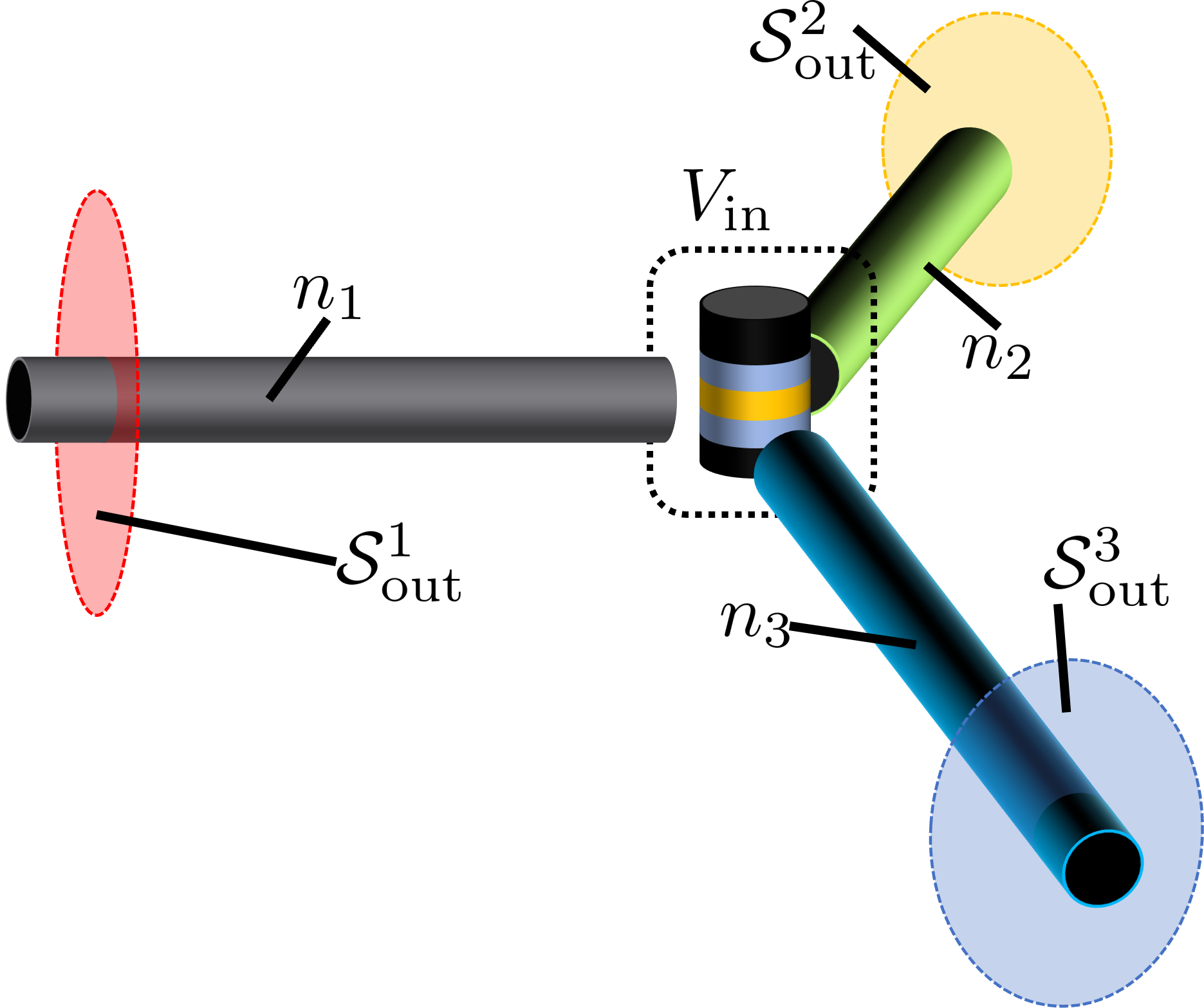}
    \caption{Example for a system with separate background structures. A cavity is coupled to three semi-finite waveguides with different refractive indices \(n_{1/2/3}\). The waveguides can be made from different materials or have different cross-section profiles, thereby supporting very different waveguide modes (indicated by the different colors). If the waveguide modes do not extend into the outside medium [\(\Delta\mathbf{G}\approx 0\), cf.~Eq.~\eqref{eq:DeltaG}], the three waveguides can be treated separately and the cavity emission inside one of the waveguides obeys the boundary conditions for that waveguide alone [Eq.~\eqref{eq:waveguideBC}]. Here, \(\mathcal{S}^{1/2/3}_{\rm out}\) are the far-field infinite cross-sections through the different waveguides, used for obtaining the boundary conditions, as shown in Sec.~\ref{sec:wavBC}.}
    \label{fig:multiwaveguide}
\end{figure}

As our second example, we consider a cavity coupled to multiple, non-intersecting waveguides in different directions, as sketched in Fig.~\ref{fig:multiwaveguide}. Our goal is to derive a measure for when the different background structures (here, the separate waveguides) can be treated separately when deriving the boundary conditions. 
Before discussing the example system below, we introduce a generalized formulation. We assume that the background volume \(V_{\rm out} = \cup_{\alpha} V_{\rm out}^{\alpha}\) separates into a finite number of continuous volumes \(V_{\rm out}^{\alpha}\) which each contain a part of the background structure, e.g., a waveguide, surface, or free space. Then, the outer surface surrounding the whole structure (scatterer+background) \(\mathcal{S}_{\rm out} = \cup_{\alpha}\mathcal{S}_{\rm out}^{\alpha}\) separates into a finite number of piecewise smooth surfaces, where \(\mathcal{S}_{\rm out}^{\alpha} = \mathcal{S}_{\rm out}\cap V_{\rm out}^{\alpha}\) (e.g., the different waveguide cross-sections in Fig.~\ref{fig:multiwaveguide}). Thus, we separate the integral from Eq.~\eqref{eq:BCintegral} into a sum of integrals that account for losses through the different loss channels:
\begin{align}\label{eq:BCintegralseparate}
    0 = \sum_{\alpha} \int_{\mathcal{S}_{\rm out}^{\alpha}}\mathrm{d}A_s& \opvec{n}_s\cdot\Big\{\mathbf{E}(\mathbf{s},\omega)\times\big[\nabla_s\times\mathbf{G}(\mathbf{s},\mathbf{r},\omega)\big]^T-\mathbf{G}(\mathbf{r},\mathbf{s},\omega)\times\big[\nabla_s\times\mathbf{E}(\mathbf{s},\omega)\big]\Big\}.
\end{align}

\red{Generally, the different integrals in Eq.~\eqref{eq:BCintegralseparate} will not vanish independently. For example, if the waveguides couple evanescently or intersect with one another, or if a leaky waveguide couples to the free space modes in the outside medium, a separate treatment of the different background substructures is not possible, and a joint treatment is required. Methods such as mode-matching or coupled-mode theory are established approaches to obtain the fields of coupled waveguides \cite{hardy1985coupled, eleftheriades1994some, huang1994coupled, jiang2002mode} However, calculating the full background Green's function is costly for complex devices. For some setups, a separate treatment of the different \(\mathcal{S}^{\alpha}_{\rm out}\)-integrals is instead possible to ease the computation. For this, we use} a Dyson scattering equation similar to the one derived in Appendix~\ref{appsec:regularization} (see also Refs.~\cite{ge2014quasinormal, fuchs2026greens}) for \(\mathbf{s}\in \mathcal{S}^{\alpha}_{\rm out}\):
\begin{align}
    \mathbf{G}(\mathbf{r},\mathbf{s},\omega) = \mathbf{G}^{\alpha}(\mathbf{r},\mathbf{s},\omega) +\Delta\mathbf{G}(\mathbf{r},\mathbf{s},\omega),
\end{align}
where \(\mathbf{G}^{\alpha}(\mathbf{r},\mathbf{s},\omega)\) is the Green's function for the background structure \(\alpha\) alone (e.g., just a single waveguide, surface, or free space) and 
\begin{align}\label{eq:DeltaG}
    \Delta\mathbf{G}(\mathbf{r},\mathbf{s},\omega) = \int_{\overline{V}^{\alpha}_{\rm out}}\mathrm{d}^3r' [\epsilon(\mathbf{r},\omega)-\epsilon^{\alpha}(\mathbf{r},\omega)]\mathbf{G}(\mathbf{r},\mathbf{r}',\omega)\cdot\mathbf{G}^{\alpha}(\mathbf{r}',\mathbf{s},\omega)
\end{align}
is the difference between the full Green's function and the contribution from background structure \(\alpha\) only. Here, \(\epsilon^{\alpha}(\mathbf{r},\omega)\) is the permittivity for background structure \(\alpha\) only, and \(\overline{V}^{\alpha}_{\rm out}\) is the complement of the background volume \(V^{\alpha}_{\rm out}\). Hence, it follows from Eq.~\eqref{eq:DeltaG} that this difference is small if the Green's function \(\mathbf{G}^{\alpha}(\mathbf{r}',\mathbf{s},\omega)\) is small for positions \(\mathbf{r}'\notin V_{\rm out}^{\alpha}\). For example, if \(\alpha\) denotes a single waveguide and the relevant modes of the waveguide used to expand the Green's function \(\mathbf{G}^{\alpha}\) decay fast perpendicular to the waveguide, \(\Delta\mathbf{G}\) will be small, and the full Green's function in the \(\mathcal{S}^{\alpha}_{\rm out}\)-integral can be approximated via the Green's function for that part of the background structure only.
\red{If this is possible for all the \(\mathcal{S}^{\alpha}_{\rm out}\)-integrals in Eq.~\eqref{eq:BCintegralseparate}, then the integrals vanish \textit{independently}, and different boundary conditions are applied to the cavity field depending on the direction. If not, a separation is not possible, and one has to use a different treatment, such as coupled-mode theory \cite{huang1994coupled} for the joint background structure.}

For a simple example, we consider the structure sketched in Fig.~\ref{fig:multiwaveguide} of a cavity coupled to three different semi-finite waveguides. The waveguides can be different in shape, material, or core refractive index, and hence support different waveguide modes. For illustrative purposes, we assume that no emission from the cavity into the surrounding free space occurs; instead, all emissions enter one of the three waveguides, depending on the direction of emission. We also assume lossless waveguides with modes that decay exponentially in the direction perpendicular to the waveguide axis. Hence, in Eq.~\eqref{eq:BCintegralseparate}, \(\alpha= 1,2,3\), denoting the three different waveguides, and \(\mathcal{S}^{1/2/3}_{\rm out}\) are the far-field infinite cross-sections of the waveguides. Since the waveguide modes decay exponentially perpendicularly to the waveguide axis, \(\Delta\mathbf{G}\approx 0\) in Eq.~\eqref{eq:DeltaG}, and, for the integral over each of the cross-sections, we can (approximately) expand the Green's function from Eq.~\eqref{eq:BCintegralseparate} in terms of the waveguide modes of that waveguide only. Consequently, three different boundary conditions apply to the cavity field, depending on the direction of emission. The emitted field far away from the cavity takes the shape of the outward propagating waveguide modes of only that waveguide into which the emission occurs [cf.~Eq.~\eqref{eq:conds}]
\begin{align}
    \mathbf{E}(\mathbf{r},\omega)\big|_{\mathbf{r}\in V_{\rm out}^{\alpha}} &\to \sum_j\sigma^{\alpha}_{kj}\mathbf{f}^{\alpha}_{kj}(\mathbf{r}),\, \mathbf{r}\to \infty,
\end{align}
where \(\mathbf{f}^{\alpha}_{kj}(\mathbf{r})\) is the waveguide mode of the \(\alpha\)-th waveguide and \(\sigma^{\alpha}_{kj}\) is the expansion coefficient from Eq.~\eqref{eq:wavexpand}. 

Thus, combinations of different types of background structures can be computed approximately, if \red{the background structures are separable} (\(\Delta\mathbf{G}\approx 0\)), via separate boundary integrals for the different kinds of lightguiding background media. This way, knowledge of the full background Green's function is not necessary to obtain an approximation of the far-field behavior of the cavity field.

\section{\red{Classical} quasinormal modes in the presence of \red{lightguiding} structures}\label{sec:QNMs}

Having discussed the derivation of boundary conditions for the field of a resonator embedded in a specific type of background structure (or combination of background structures), we now move on to the \red{theory of classical quasinormal modes, including multi-cavity structures.}
We consider multiple optical cavities or nanoplasmonic resonators in a general background medium. The full permittivity reads, 
\begin{align}
    \epsilon(\mathbf{r},\omega) = \epsilon_{\rm back}(\mathbf{r},\omega) + \Delta\epsilon(\mathbf{r},\omega)\sum_i \chi_{V_i}(\mathbf{r}),
\end{align}
where \(\epsilon_{\rm back}(\mathbf{r},\omega)\) is the permittivity of the background medium, and \(\Delta\epsilon(\mathbf{r},\omega) = \epsilon(\mathbf{r},\omega)-\epsilon_{\rm back}(\mathbf{r},\omega)\) is the perturbation of the permittivity from the background value due to the presence of the cavity. Furthermore, \(\chi_{V_i}(\mathbf{r})\) is unity inside the cavity volume \(\mathbf{r}\in V_i\) and zero elsewhere. For generality, we assume here a frequency-dependent background permittivity, which (in principle) can include absorption and dispersion. In practice, to enable the definition of QNMs, the absorption and dispersion must be weak enough so that there is negligible backscattering from the background to the cavity \cite{ching1998quasinormal}.

Instead of calculating joint QNMs for the entire multi-cavity structure (which is generally not feasible), we calculate the \textit{single-cavity} QNMs \(\qnm{f}{i_{\mu}}(\mathbf{r})\) from the source-free Helmholtz equation~\eqref{eq:helmquasi} using only the single-cavity permittivity \(\epsilon^{(i)}(\mathbf{r},\omega) = \epsilon_{\rm back}(\mathbf{r},\omega) + \Delta\epsilon(\mathbf{r},\omega) \chi_{V_i}(\mathbf{r})\), where the permittivity is set to the background value at all other cavities.
Multi-cavity properties such as the full electromagnetic Green's function are then constructed from the single-cavity QNMs via a set of scattering equations, as discussed in Ref.~\cite{fuchs2026greens}. The boundary conditions are obtained as described in the previous sections (cf.~Tab.~\ref{tab:summary}).

\subsection{QNM regularization}
The QNMs have complex eigenfrequencies \(\Tilde{\omega}_{i_{\mu}} = \omega_{i_{\mu}}-i\gamma_{i_{\mu}},\, \gamma_{i_{\mu}}>0\) so that the temporal decay of the cavity field is inherent to the mode. However, this causes the QNMs to diverge in the far field \cite{muljarov2011brillouin, sauvan2013theory, kristensen2020modeling}. This is a straightforward consequence of causality, since the field at \(r\to\infty\) must have been emitted at \(t\to -\infty\), which, for a temporally decaying mode, leads to spatial divergence. 

We address this divergence by adapting the QNM regularization approach to general background media. The approach yields frequency-dependent, non-diverging regularized QNM fields which account for propagation away from the resonator. These fields can be calculated via a Dyson scattering equation \cite{ge2014quasinormal}, or alternatively, via the field-equivalence principle (see Appendix~\ref{appsec:regularization} for a derivation) \cite{franke2020quantized, ren2020near}:
\begin{align}\label{eq:surfacereg}
    \qnm{F}{i_{\mu}}(\mathbf{r},\omega) = i\tilde{\omega}_{i_{\mu}}\mu_0\oint_{\mathcal{S}_i}\mathrm{d}A_s \mathbf{G}_{\rm back}(\mathbf{r},\mathbf{s},\omega)\cdot\qnm{J}{i_{\mu}}(\mathbf{s})-\oint_{\mathcal{S}_i}\mathrm{d}A_s\qnm{M}{i_{\mu}}(\mathbf{s})\cdot\Big[\nabla\times\mathbf{G}_{\rm back}(\mathbf{s},\mathbf{r},\omega)\Big],
\end{align}
where \cite{ren2020near}
\begin{align}
    \qnm{J}{i_{\mu}}(\mathbf{s}) = \opvec{n}_s\times\nabla\times\qnm{f}{i_{\mu}}(\mathbf{s})/(i\tilde{\omega}_{i_{\mu}}\mu_0),\quad \qnm{M}{i_{\mu}}(\mathbf{s}) = -\opvec{n}_s\times\qnm{f}{i_{\mu}}(\mathbf{s}),
\end{align}
are the electric and magnetic surface currents generated by the QNM \(i_{\mu}\) on the cavity surface \(\mathcal{S}_i\) of \(V_i\), and \(\opvec{n}_s\) is the surface normal vector, pointing outwards with respect to \(V_i\). 

The presence of the background Green's function \(\mathbf{G}_{\rm back}\) in Eq.~\eqref{eq:surfacereg} ensures that the shape of the outward propagating regularized QNM field matches the specific geometry of the background, e.g., propagation through a waveguide or along a surface. For combined background guiding structures as discussed in Sec.~\ref{sec:combinedwavs}, The shape of the regularized QNM is different within the different background media. Similar to Sec.~\ref{sec:combinedwavs}, we separate the outside volume into different parts containing the different lightguiding background structures \(V_{\rm out} = \cup_{\alpha}V_{\rm out}^{\alpha}\). We furthermore denote by \(\mathcal{S}_i^{\alpha}\) that part of the cavity surface \(\mathcal{S}_i\) through which energy is radiated into the background structure \(V_{\rm out}^{\alpha}\). Then, for \(\mathbf{r}\in V_{\rm out}^{\alpha}\) (and sufficiently far away from the cavity), the regularized QNM field can be approximately written as:
\begin{align}\label{eq:combinedsurfacereg}
    \qnm{F}{i_{\mu}}(\mathbf{r},\omega)\big|_{\mathbf{r}\in V_{\rm out}^{\alpha}} \approx i\tilde{\omega}_{i_{\mu}}\mu_0\oint_{\mathcal{S}^{\alpha}_i}\mathrm{d}A_s \mathbf{G}^{\alpha}(\mathbf{r},\mathbf{s},\omega)\cdot\qnm{J}{i_{\mu}}(\mathbf{s})-\oint_{\mathcal{S}^{\alpha}_i}\mathrm{d}A_s\qnm{M}{i_{\mu}}(\mathbf{s})\cdot\Big[\nabla\times\mathbf{G}^{\alpha}(\mathbf{s},\mathbf{r},\omega)\Big],
\end{align}
where \(\mathbf{G}^{\alpha}\) is again the background Green's function for the type of lightguiding structure in \(V_{\rm out}^{\alpha}\) only. Equation~\eqref{eq:combinedsurfacereg} allows for an efficient approximation of the regularized fields in complex environments.

\begin{figure}
    \centering
    \includegraphics[width=0.95\linewidth]{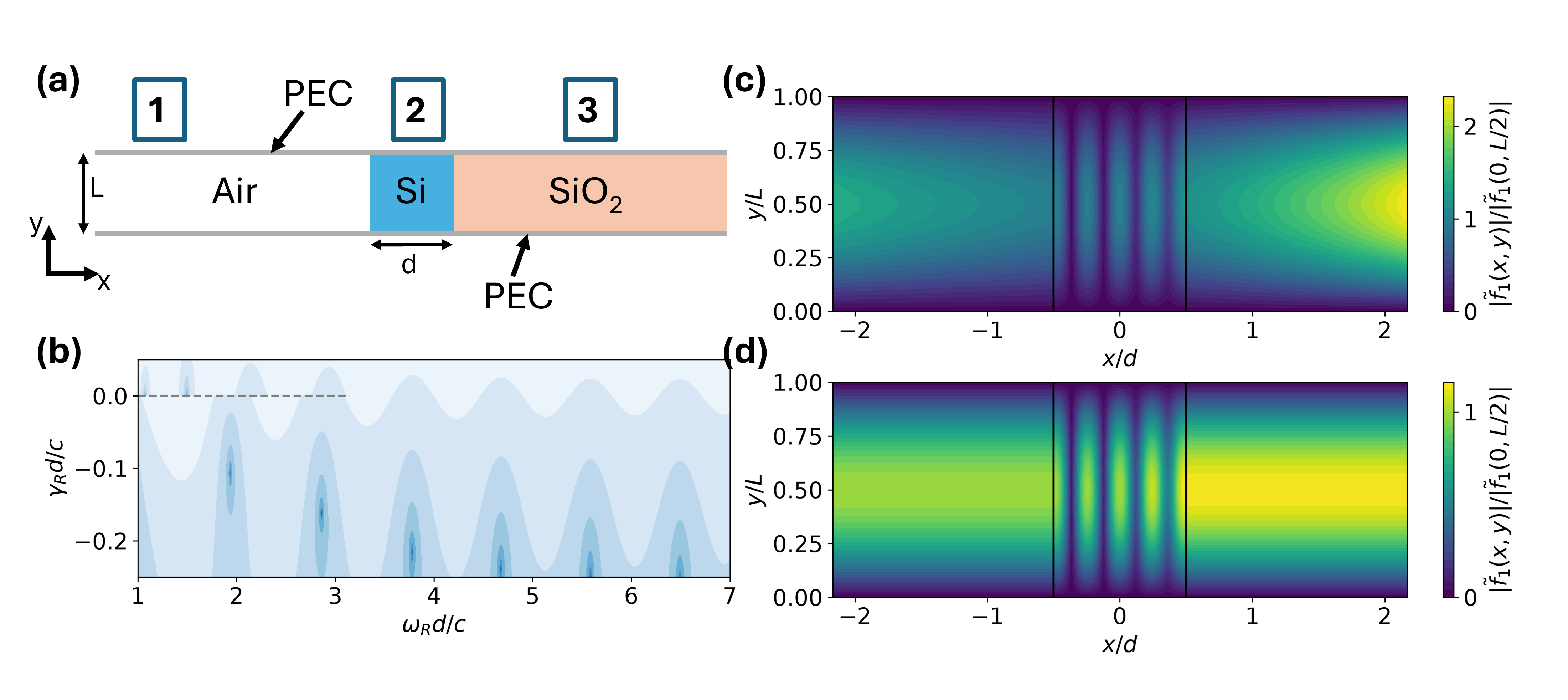}
    \caption{\red{(a) Sketch of the 2D example structure. The structure is comprised of three domains, a semiinfinite air waveguide (domain 1) with permittivity \(\epsilon_1 \approx 1\), a silicon cavity (domain 2) with permittivity \(\epsilon_2 \approx 11.7\), and a semiinfinite silicon dioxide waveguide (domain 3) with permittivity \(\epsilon_3 \approx 3.75\). The structure is bound at the top and bottom with perfect electric conductors (PEC), so that it supports standing waves in the \(y\)-direction, and propagating waves in the waveguides in the \(x\)-direction. The height of the structure is \(L = 300\, {\rm nm}\), and the width of the cavity is \(d = 300\,{\rm nm}\). (b) Resonance spectrum of the silicon cavity for the fundamental waveguide mode \(m=1\) [cf.~Eq.~\eqref{eq:characteristicequation}]. The dark spots mark the QNM resonance frequencies. The branch cut is highlighted. The cutoff frequency is determined by the air waveguide and has the value \(\omega^{\rm air}_{\rm cutoff} = \pi d/c\). Modes below this point cannot propagate through both waveguides. The silicon dioxide waveguide has a lower cutoff of \(\omega^{\rm SiO2}_{\rm cutoff} =\omega^{\rm air}_{\rm cutoff}/\sqrt{\epsilon_3} \). Hence, the QNMs visible below the branch cut are related to loss via the silicon dioxide waveguide only. (c) Lowest order QNM \(\mu = c\) [cf.~Eq.~\eqref{eq:2DQNM}] above the cutoff of the air waveguide with a resonance frequency of \(\tilde{\omega}_c d/c \approx 3.776-0.215i\), corresponding to a Q-factor of \(Q\approx 8.777\). The spatial divergence of the QNM is visible at the edge of the plot, and it is more pronounced towards the right, since the contrast between the silicon cavity and silicon dioxide waveguide is less than between the cavity and air waveguide. (d) Regularized lowest order QNM [cf.~Eq.~\eqref{eq:2Dreg}] at \(\omega = \omega_c\). The regularized QNMs do not diverge and instead propagate through the waveguide at a real frequency. }}
    \label{fig:figure}
\end{figure}

\red{
\subsection{Example: Classical quasinormal modes in a 2D waveguide-coupled cavity}\label{sec:classexample}
To illustrate our approach so far, we consider the example model sketched in Fig.~\ref{fig:figure}(a) of a 2D square silicon cavity with side lengths \(d = L = 300\,{\rm nm}\), coupled to two semiinfinite waveguides with different core materials (air and silicon dioxide, respectively). The structure is bound by perfect electric conductors (PEC) at \(y = 0\) and \(y=L\), leading to standing waves in the \(y\)-direction, and propagating waves in the waveguides in the \(x\)-direction. This structure is the simplest non-trivial example of a cavity in a structured environment, and therefore instructive to our approach. We stress however, that our goal is not to give a comprehensive study of this device, but merely point out how such a device would be approached theoretically. Here, we derive the single-cavity QNMs of a cavity coupled to two different waveguides, and consider the case of two coupled cavities in a similar device in the context of QNM correlation functions below (cf.~Fig.~\ref{fig:2QNMcorr}). Examples and applications to more realistic and complex multi-cavity structures are subject to future work.

We assume that the electric field (and electric field modes) is polarized in the \(z\)-direction, allowing us to restrict the discussion in the following to scalar functions of \(x\) and \(y\). The waveguides support modes of the form
\begin{align}\label{eq:2Dwaveguidemodes}
    f_{mj}(x,y, \omega) = \sin(k_y y)\mathrm{e}^{ik_{jx} x},
\end{align}
where \(k_y = m \pi/L\) is the wavenumber in the \(y\)-direction which determines the mode index \(m\) to satisfy the boundary conditions \(f_{mj}(x,0,\omega) = f_{mj}(x,L,\omega) = 0\) imposed by the perfect electric conductor cladding. Furthermore, \(k_{jx} = \sqrt{\epsilon_j(\omega^2/c^2)-k_y^2}\) is the wavenumber in the \(x\)-direction, where \(j\) is the index of the domain, so \(j=1\) for the air waveguide, and \(j=3\) for the silicon dioxide waveguide. The waveguide modes experience a branch point at \(\omega_{\rm cutoff} = c k_y/\sqrt{\epsilon_j}\). Below this frequency, the wavenumber in the \(x\)-direction is imaginary, and the waves become evanescent, i.e., no propagation through the waveguides is possible. Since the waveguides have different refractive indices, their cutoff frequencies are different. In the following, we will only focus on the regime above the cutoff frequency of the air waveguide, which lies higher than the cutoff for the silicon dioxide waveguide [cf.~Fig.~\ref{fig:figure}(b)]. 
We leave the dependence of \(k_y\) and \(k_{jx}\) on the mode index \(m\) and frequency \(\omega\) implicit for brevity, since in the following we will focus on a single dominant mode.}

\red{To construct the QNMs, we require boundary conditions. Following the discussion from Sec.~\ref{sec:wavBC} for cavities coupled to waveguides, the QNMs in the far field must take the shape of outward propagating waveguide modes. Since our waveguides to the left and right of the cavity have different core materials, we must impose different boundary conditions to the left and right of the silicon cavity. The most general Ansatz reads:
\begin{align}\label{eq:2DQNM}
    \tilde{f}_{\mu}(x,y)\Big|_{x\in V_j} = A_{\mu,j}\sin(k_y y)\mathrm{e}^{i k_{jx}^{\mu} x} + B_{\mu,j} \sin(k_y y) \mathrm{e}^{-ik_{jx}^{\mu} x},
\end{align}
where \(j \in \{1,2,3\}\). To satisfy the outgoing-wave boundary conditions, we must choose \(A_{\mu,1} = B_{\mu,3} = 0\). We can furthermore set \(A_{\mu,2} = 1\) due to our freedom of choosing the norm. The wavenumber \(k_y\) is identical across all domains to satisfy the boundary conditions in the \(y\)-direction, while \(k_{jx}^{\mu} = \sqrt{\epsilon_j(\tilde{\omega}_{\mu}^2/c^2)-k_y^2}\)  depends on \(j\) due to the different permittivities \(\epsilon_j\) in the three domains. It also implicitly depends on the waveguide mode index \(m\). In the following, we restrict ourselves to the mode \(m=1\). 

To obtain the eigenfrequency \(\tilde{\omega}_{\mu}\), we make use of the condition of continuity at the cavity boundary, \(\lim_{x\searrow \pm d/2} \tilde{f}_{\mu} = \lim_{x\nearrow \pm d/2} \tilde{f}_{\mu}\) and \(\lim_{x\searrow \pm d/2} \partial_x\tilde{f}_{\mu}=\lim_{x\nearrow \pm d/2} \partial_x\tilde{f}_{\mu}\) (cf.~Ref.~\cite{kristensen2020modeling} for a discussion of the QNMs of a 1D dielectric slab, which has the same continuity conditions). We can reorder these equations to obtain the characteristic function:
\begin{align}\label{eq:characteristicequation}
    f(\tilde{\omega}_{\mu})=(k_{2x}^{\mu}+k_{3x}^{\mu})(k_{2x}^{\mu}+k_{1x}^{\mu})\mathrm{e}^{-2ik_{2x}^{\mu} d} -(k_{2x}^{\mu}-k_{1x}^{\mu})(k_{2x}^{\mu}-k_{3x}^{\mu}),
\end{align}
where the zeros of \(f(\tilde{\omega}_{\mu})\) are the QNM eigenfrequencies. Plotting \(1-\ln[f(\tilde{\omega}_{\mu})]\) as a function of the real and imaginary parts of \(\tilde{\omega}_\mu\), we obtain the resonance spectrum of our cavity shown in Fig.~\ref{fig:figure}(b), where the dark spots mark the QNM eigenfrequencies. The branch cut for frequencies below the air waveguide cutoff frequency is highlighted. There are QNM resonances visible below the branch cut in Fig.~\ref{fig:figure}(b). These modes are related to coupling to the silicon dioxide waveguide only, and cannot propagate through the air waveguide.
We note that branch cuts in the spectrum can cause problems for the expansion in terms of QNMs, and must be treated carefully \cite{kristensen2020modeling, sauvan2022normalization}. We therefore stress again that our goal here is not to provide a comprehensive study of the present device, but only to illustrate our approach. 

We focus on the regime close to the first QNM \(\mu = c\) above the branch point at \(\tilde{\omega}_c d/c \approx 3.776-0.215i\), and discuss only the regime above the air waveguide cutoff frequency. 
In Fig.~\ref{fig:figure}(c), we plot the absolute value of this QNM as a function of \(x\) and \(y\). The standing waves in the \(y\)-direction are visible, as well as the spatial divergence of the QNM in the waveguides. This divergence is larger in the silicon dioxide waveguide, since the contrast between silicon and silicon dioxide is less strong than between silicon and air, meaning that the cavity loses more energy to the right than to the left. 

Since the QNM is a simple standing wave in the \(y\)-direction, the regularization takes the same form as for the 1D dielectric slab in Ref.~\cite{franke2023impact}, and we obtain:
\begin{align}\label{eq:2Dreg}
    \tilde{F}_c(x,y,\omega)\Big|_{x<-d/2} &= \tilde{f}_c(-d/2,y)\mathrm{e}^{-ik_{1x} (x+d/2)},\nonumber\\
    \tilde{F}_c(x,y,\omega)\Big|_{x>d/2} &= \tilde{f}_c(d/2,y)\mathrm{e}^{ik_{3x} (x-d/2)},
\end{align}
which preserves the continuity of the field at \(x=\pm d/2\) by construction, and \(k_{1x}\) and \(k_{3x}\) now depend on the real frequency \(\omega\) as well as the waveguide mode index \(m=1\). The field of the regularized QNMs is plotted in Fig.~\ref{fig:figure}(d) for \(\omega = \omega_c = \mathrm{Re}(\tilde{\omega}_c)\), where now, there is no divergence in the waveguides, but instead the cavity field takes the shape of outward propagating waveguide modes. 

Below, after a general discussion of the quantization of QNMs in structured environments, we will return to this example of a 2D waveguide-coupled cavity and show how the quantization is performed and how quantum mechanical properties such as the quantum mechanical Purcell factor and system-bath correlation functions are obtained for this setup. We will also derive the correlation functions for a similar device with two coupled cavities.
}

\red{
\section{Quantization of quasinormal modes in structured environments}\label{sec:quantQNMs}
}
\subsection{QNM quantization}\label{sec:quanti}
\red{From macroscopic QED, the electric field operator for a dissipative, dispersive, non-magnetic, and isotropic medium is given by \cite{dung1998three, suttorp2004field, philbin2010canonical, scheel2009macroscopic}:}
\begin{align}\label{eq:welschquanti}
    \opvec{E}(\mathbf{r},\omega)= i\sqrt{\frac{\hbar}{\pi\epsilon_0}}\int\mathrm{d}^3r' \sqrt{\epsilon_I(\mathbf{r}',\omega)}\mathbf{G}(\mathbf{r},\mathbf{r}',\omega)\cdot\opvec{b}(\mathbf{r}',\omega) ,
\end{align}
where \(\epsilon_I(\mathbf{r},\omega)\) is the imaginary part of the permittivity, and \(\opvec{b}(\mathbf{r},\omega)\) are bosonic photon operators with continuous spatial and frequency indices, i.e.,
\begin{align}\label{eq:bcomm}
    \left[\hat{b}_p(\mathbf{r},\omega),\hat{b}^{\dagger}_q(\mathbf{r}',\omega')\right] = \delta_{pq}\delta(\mathbf{r}-\mathbf{r}')\delta(\omega-\omega').
\end{align}
\red{The macroscopic QED formalism has been extended in the literature to magnetic or anisotropic media  (see, e.g., Ref.~\cite{knoll2000qed}). Hence, the QNM quantization discussed here can be extended to these types of media, provided an (approximate) expansion of the full Green's function in terms of QNMs is available.}

The full electromagnetic field Hamiltonian can be diagonalized to read \cite{dung1998three, suttorp2004field, philbin2010canonical}, 
\begin{align} \label{eq:bHam}
    H = \hbar\int_0^{\infty}\mathrm{d}\omega\int\mathrm{d}^3r \omega \opvec{b}^{\dagger}(\mathbf{r},\omega)\cdot\opvec{b}(\mathbf{r},\omega),
\end{align}
and the frequency-independent field operator is obtained via integration over all frequencies \(\opvec{E}(\mathbf{r}) = \int_0^{\infty}\mathrm{d}\omega \opvec{E}(\mathbf{r},\omega) {+}\mathrm{H.a.}\)

The bosonic photon operators \(\opvec{b}(\mathbf{r},\omega)\) quantize the entire electromagnetic field. To obtain operators for the discrete QNMs of a single cavity only, we perform a projection onto the QNM subspace via \cite{franke2020quantized, fuchs2024quantization, fuchs2026quantum}
\begin{align}\label{eq:QNMops}
    \hat{a}_{i_{\mu}} = \int_0^{\infty}\mathrm{d}\omega\int\mathrm{d}^3r \mathbf{L}_{i_{\mu}}(\mathbf{r},\omega)\cdot\opvec{b}(\mathbf{r},\omega),
\end{align}
where the QNM projector kernels \(\mathbf{L}_{i_{\mu}}(\mathbf{r},\omega)\) read,
\begin{align}\label{eq:QNMproj}
    \mathbf{L}_{i_{\mu}}(\mathbf{r},\omega) = \sum_{\nu}\left(S^{-1/2}_{\rm \red{intra}}\right)_{i_{\mu}i_{\nu}}\sqrt{\frac{2}{\pi\omega_{i_{\nu}}}} A_{i_{\nu}}(\omega)\Big[\chi_{V_i}(\mathbf{r})\sqrt{\epsilon_I(\mathbf{r},\omega)}\qnm{f}{i_{\nu}}(\mathbf{r})\nonumber\\
    +\chi_{\overline{V_i}}(\mathbf{r})\sqrt{\epsilon_{{\rm back},I}(\mathbf{r},\omega)}\qnm{F}{i_{\nu}}(\mathbf{r},\omega)\Big].
\end{align}
Here, \(A_{i_{\nu}}(\omega) = \omega/[2(\tilde{\omega}_{i_{\nu}}-\omega)]\), and \(\chi_V(\mathbf{r})\) is unity if \(\mathbf{r}\in V\) and zero elsewhere. Hence, the QNMs \(\qnm{f}{i_{\nu}}(\mathbf{r})\) are replaced with the regularized QNM fields \(\qnm{F}{i_{\nu}}(\mathbf{r},\omega)\) for positions outside the cavity, and the imaginary part of the full permittivity \(\epsilon_I(\mathbf{r},\omega)\) is replaced with the imaginary part of the background permittivity \(\epsilon_{{\rm back},I}(\mathbf{r},\omega)\) so that only sources from within the cavity enter the projector, while sources from other cavities are removed. Furthermore, \(\left(S^{-1/2}_{\rm \red{intra}}\right)_{i_{\mu}i_{\nu}}\) is the inverse square root of the intracavity overlap matrix defined in Appendix~\ref{appsec:overlap}.

From the definition of the QNM operators \(\hat{a}_{i_{\mu}}\) in Eq.~\eqref{eq:QNMops} together with the commutator of the operators \(\opvec{b}(\mathbf{r},\omega)\) from Eq.~\eqref{eq:bcomm}, it follows that 
\begin{align}\label{eq:QNMcomm}
    [\hat{a}_{i_{\mu}},\hat{a}^{\dagger}_{j_{\nu}}] &= \sum_{\mu'\nu'}\left(S^{-1/2}_{\rm \red{intra}}\right)_{i_{\mu}i_{\mu'}}S_{i_{\mu'}j_{\nu'}}\left(S^{-1/2}_{\rm \red{intra}}\right)_{j_{\nu'}j_{\nu}}= \delta_{ij}\delta_{\mu\nu}+(1-\delta_{ij})\tilde{S}^{\rm inter}_{i_{\mu}j_{\nu}},
\end{align}
where \(S_{i_{\mu}j_{\nu}} = \delta_{ij}S^{\rm intra}_{i_{\mu}i_{\nu}} + (1-\delta_{ij})S^{\rm inter}_{i_{\mu}j_{\nu}}\) is the QNM overlap matrix, which separates into intra- and intercavity overlap. The overlap matrices are defined in Appendix~\ref{appsec:overlap}.

Due to the spatial separation of the cavities, the relative intercavity overlap \(\tilde{S}^{\rm inter}_{i_{\mu}j_{\nu}}\) in Eq.~\eqref{eq:QNMcomm} is exponentially small \cite{fuchs2024quantization}
\begin{align}\label{eq:expscale}
    \tilde{S}^{\rm inter}_{i_{\mu}j_{\nu}}= \sum_{\mu'\nu'} \left(S^{-1/2}_{\rm \red{intra}}\right)_{i_{\mu}i_{\mu'}}S^{\rm inter}_{i_{\mu'}j_{\nu'}}\left(S^{-1/2}_{\rm \red{intra}}\right)_{j_{\nu'}j_{\nu}} \sim \mathrm{e}^{-P_{i_{\mu}j_{\nu}}},
\end{align}
where we defined a quantitative measure for the separation, the \textit{cavity separation parameter} \cite{fuchs2024quantization} (see Appendix~\ref{appsec:cavsepparam} for the derivation of the generalization)
\begin{align}\label{eq:cavsepparam}
    P_{i_{\mu}j_{\nu}} = |S^{\rm imag, min}_{i_{\mu}j_{\nu}}| +T^{\min}_{i_{\mu}j_{\nu}}.
\end{align}

The terms contributing to the cavity separation parameter reflect distinct properties of the QNMs. In the first term, \red{\(S^{\rm imag, min}_{i_{\mu}j_{\nu}} = \min(\mathrm{Im}\{[S_{ij}(\omega)]_{\omega=\tilde{\omega}^*_{j_{\eta}}}\},\mathrm{Im}\{[S_{ij}(\omega)]^*_{\omega=\tilde{\omega}_{i_{\mu}}}\})\), where \cite{landau1975classical}
\begin{align}\label{eq:Eikonaldef}
    S_{ij}(\omega) = \int_{\mathcal{C}_{ij}}\mathrm{d}\mathbf{s}\cdot\mathbf{k}(\mathbf{s},\omega)
\end{align}
is the \textit{Eikonal} along the path \(\mathcal{C}_{ij}\) between the cavities \(i\) and \(j\) (e.g. through the center of a waveguide). 

The formulation via the Eikonal is very general to account for non-homogeneous media or guided propagation. In homogeneous media \(S_{ij}(\omega) = n_B \omega R_{ij}/c\) \cite{fuchs2024quantization}, while in inhomogeneous media \(S_{ij}(\omega) = \omega \Lambda_{ij}/c\), where \(\Lambda_{ij} = \int_{\mathcal{C}_{ij}}\mathrm{d}s\,n(s) = \overline{n}|\mathcal{C}_{ij}|\) \cite{Born_Wolf_2019} is the \textit{optical path length} along the path of geometric length \(|\mathcal{C}_{ij}|\) between the cavities, and \(\overline{n}\) is the average refractive index along this path. If the propagating waves are confined in some direction (e.g., for the 2D waveguide-coupled cavity in Fig.~\ref{fig:figure}) only a wavevector in the unconfined direction (i.e. the propagation direction) is guaranteed to exist and contributes to the Eikonal (e.g., \(S_{ij}(\omega) = k_x(\omega)R_{ij}\) for the 2D waveguide-coupled cavity from Sec.~\ref{sec:quantexample}). Thus, the Eikonal connects to basic properties of the outward propagating QNM field, and the intercavity overlap scales exponentially with the spatial separation of the cavities.
}

In the second term, \(T^{\rm min}_{i_{\mu}j_{\nu}} = \min[T_{i_{\mu},j_{\nu}},T_{j_{\nu},i_{\mu}}]\), where \( T_{i_{\mu},j_{\nu}} = -\mathrm{ln}[U^i_{j_{\nu}}/U^{\rm tot}_{j_{\nu}}]\) is the transfer factor, which describes the transferred power \(U^i_{j_{\nu}}\) to cavity \(i\) from mode \(j_{\nu}\) as a fraction of the total power \(U^{\rm tot}_{j_{\nu}}\) radiated from mode \(j_{\nu}\).

For certain setups and geometries, additional aspects of the mode may enter the cavity separation parameter. For example, the overlap also vanishes in the limit of very high-quality modes (cf.~Appendix~\ref{appsec:cavsepparam}). However, we do not consider these aspects in the following general discussion.

Therefore, for a large separation, i.e., \(P_{i_{\mu}j_{\nu}} \gg 0\), the operators of the different cavities commute [cf.~Eq.~\eqref{eq:QNMcomm}], allowing the construction of Fock states for the individual cavities. For small separations, hybridized modes or coupled QNM theory must be used instead \cite{Tao2020coupling, franke2022quantized, ren2022connecting}.

\subsection{The QNM Hamiltonian}
For well-separated cavities, the construction of QNM operators from Eq.~\eqref{eq:QNMops} yields bosonic operators for the quantized QNMs of the individual cavities. However, the integration over all frequencies in Eq.~\eqref{eq:QNMops} fixes the phase of the quantized QNMs, leading to modes that are quasi-bound, with propagation effects largely removed. 

From the operator projection in Eq.~\eqref{eq:QNMops} together with the commutator in Eq.~\eqref{eq:QNMcomm}, it follows that the operators \(\opvec{b}(\mathbf{r},\omega)\) can be decomposed into QNM and non-QNM contributions via
\begin{align} \label{eq:bdecomp}
    \opvec{b}(\mathbf{r},\omega) = \sum_{i,\mu}\mathbf{L}^*_{i_{\mu}}(\mathbf{r},\omega)\hat{a}_{i_{\mu}}+\opvec{c}(\mathbf{r},\omega).
\end{align}
Here, we introduced bath operators \(\opvec{c}(\mathbf{r},\omega)\) that are orthogonal to the QNM projector kernels \(\mathbf{L}_{i_{\mu}}(\mathbf{r},\omega)\) so that QNM and bath operators commute \cite{franke2020quantized, fuchs2024quantization, fuchs2026quantum}. Thus, the bath operators are mainly responsible for the propagation of photons through the background medium (the bath), and are generally non-bosonic [see Eq.~\eqref{eq:ccomm}].

Using this decomposition, we bring the electromagnetic field Hamiltonian from Eq.~\eqref{eq:bHam} into system-bath form \cite{franke2020quantized, fuchs2024quantization}
\begin{align}\label{eq:qnmham}
    H = H_S + H_B + H_{SB}&= \hbar\sum_{i,\mu\eta}\left(\int_0^{\infty}\mathrm{d}\omega\int\mathrm{d}^3r\, \omega\, \mathbf{L}_{i_{\mu}}(\mathbf{r},\omega)\cdot\mathbf{L}^*_{i_{\eta}}(\mathbf{r},\omega)\right)\hat{a}^{\dagger}_{i_{\mu}}\hat{a}_{i_{\eta}}\nonumber\\
    &\quad+ \hbar\int_0^{\infty}\mathrm{d}\omega\int\mathrm{d}^3r\, \omega\,\opvec{c}^{\dagger}(\mathbf{r},\omega)\cdot\opvec{c}(\mathbf{r},\omega)\nonumber\\
    &\quad + \hbar\sum_{i,\mu}\int_0^{\infty}\mathrm{d}\omega\int\mathrm{d}^3r\, \mathbf{g}_{i_{\mu}}(\mathbf{r},\omega)\cdot\opvec{c}(\mathbf{r},\omega)\,\hat{a}_{i_{\mu}}^{\dagger}{+}\mathrm{H.a.},
\end{align}
where \cite{franke2020quantized, fuchs2024quantization, fuchs2026quantum}
\begin{align}\label{eq:noisecoup}
    \mathbf{g}_{i_{\mu}}(\mathbf{r},\omega) = \sum_{\nu}(\delta_{\mu\nu}\omega-\chi_{i_{\mu}i_{\nu}})\mathbf{L}_{i_{\nu}}(\mathbf{r},\omega)
\end{align}
is a noise-coupling element associated with a broad frequency bandwidth, and \(\chi_{i_{\mu}i_{\nu}} = \sum_{\eta} (S^{-1/2}_{\rm red{intra}})_{i_{\mu}i_{\eta}}\tilde{\omega}_{i_{\eta}}(S^{1/2})_{i_{\eta}i_{\nu}}\) is the symmetrized QNM eigenfrequency.

Note that, due to the assumption of a negligible overlap (large separation \(P_{i_{\mu}j_{\eta}}\gg 0\)), there is no direct coupling between the QNMs of different cavities in the system part. Instead, the QNMs of all cavities couple to the same bath of propagating photons. In a time-dependent theory, we thus obtain effective, time-delayed coupling between the different cavities via photon propagation through the background medium (cf.~Sec.~\ref{sec:corrfunc}). While such coupling is generally weak in free space, the use of guiding structures, such as waveguides or plasmonic surfaces, can lead to significant coupling between separated systems in quantum technologies.

\subsection{Bath operators for separable background structures}
The bath operators are defined implicitly from Eq.~\eqref{eq:bdecomp}, and therefore obey the non-bosonic commutation relation
\cite{franke2020quantized},
\begin{align} \label{eq:ccomm}
    \left[\hat{c}_p(\mathbf{r},\omega),\hat{c}^{\dagger}_q(\mathbf{r}',\omega')\right]_- &= \delta_{pq}\delta(\mathbf{r}-\mathbf{r}')\delta(\omega-\omega')-\sum_{k,\mu}L^*_{k_{\mu},p}(\mathbf{r},\omega)L_{k_{\mu},q}(\mathbf{r}',\omega'),
\end{align}
where the second term on the right-hand side represents the completeness relation on the quantized QNM subspace, which is here removed from the full-space \(\delta\)-function.

In a complex structure consisting of multiple emitters and different kinds of background structures (cf.~Sec.~\ref{sec:combinedstructures}), the bath operators may be separated into a sum of operators
\begin{align}
    \opvec{c}(\mathbf{r},\omega) = \sum_{\alpha} \chi_{V_{\rm out}^{\alpha}}(\mathbf{r}) \opvec{c}^{\alpha}(\mathbf{r},\omega),
\end{align}
where the index \(\alpha\) refers to the \(\alpha\)-th background structure as in Sec.~\ref{sec:combinedstructures}. Then, the commutator from Eq.~\eqref{eq:ccomm} decomposes into two different cases. Firstly \(\mathbf{r},\mathbf{r}'\in V_{\rm out}^{\alpha}\), where:
\begin{align}
     \left[\hat{c}_p(\mathbf{r},\omega),\hat{c}^{\dagger}_q(\mathbf{r}',\omega')\right]_-\big|_{\mathbf{r},\mathbf{r}'\in V_{\rm out}^{\alpha}}
     &=  \left[\hat{c}_p^{\alpha}(\mathbf{r},\omega),\hat{c}^{\alpha,\dagger}_{q}(\mathbf{r}',\omega')\right]_-\nonumber\\
     &= \delta_{pq}\delta(\mathbf{r}-\mathbf{r}')\delta(\omega-\omega')-\sum_{k,\mu}L^*_{k_{\mu},p}(\mathbf{r},\omega)L_{k_{\mu},q}(\mathbf{r}',\omega').
\end{align}

Alternatively, if \(\mathbf{r}\in V_{\rm out}^{\alpha},\mathbf{r}'\in V_{\rm out}^{\beta},\, \alpha\neq \beta\), we find:
\begin{align}\label{eq:sepbathcomm}
    \left[\hat{c}_p(\mathbf{r},\omega),\hat{c}^{\dagger}_q(\mathbf{r}',\omega')\right]_-\big|_{\mathbf{r}\in V_{\rm out}^{\alpha},\mathbf{r}'\in V_{\rm out}^{\beta}}^{\alpha\neq \beta} =  \left[\hat{c}_{p}^{\alpha}(\mathbf{r},\omega),\hat{c}^{\beta,\dagger}_{q}(\mathbf{r}',\omega')\right]_- = -\sum_{k,\mu}L^*_{k_{\mu},p}(\mathbf{r},\omega)L_{k_{\mu},q}(\mathbf{r}',\omega').
\end{align}

Since the quasi-bound QNMs are (mostly) confined to their cavity, the commutator in Eq.~\eqref{eq:sepbathcomm} (approximately) vanishes if the background structures are separated, e.g., two non-intersecting waveguides. In this case, the bath operators separate into a sum of commuting operators for the different substructures. In the following, we use the full bath operators \(\opvec{c}(\mathbf{r},\omega)\) instead of the separated bath operators \(\opvec{c}^{\alpha}(\mathbf{r},\omega)\) for generality. 

\subsection{Coupling of quantized QNMs to quantum emitters}\label{sec:qnmtlscoup}
Many quantum technologies are based on quantum emitters coupled to electromagnetic modes, e.g., an atom in an optical microcavity or near a plasmonic nanoparticle. We use two-level systems (TLSs) to model the quantum emitters. The extension to multi-level emitters is straightforward.

\red{We model the TLSs as point dipoles that couple to the electric field operator via dipole coupling. Higher order multipole coupling or spatially extended structures such as quantum dots could also be included in the theory for more general emitters. Within the rotating-wave approximation, the TLS contributions to the Hamiltonian reads \cite{knoll2000qed, franke2019quantization}:}
\begin{align} \label{eq:TLSHam}
    H^{\rm TLS} &= \hbar\sum_m \omega_m \hat{\sigma}_m^+\hat{\sigma}_m-\sum_m \hat{\sigma}_m^+ \int_0^{\infty}\mathrm{d}\omega \mathbf{d}_m\cdot\opvec{E}(\mathbf{r}_m,\omega){-}\mathrm{H.a.},
\end{align}
where \(\hat{\sigma}_m^{+}\) creates an excitation with energy \(\hbar\omega_m\) in TLS \(m\), and \(\mathbf{d}_m\) is the dipole moment of TLS \(m\) coupled to the electric field operator from Eq.~\eqref{eq:welschquanti} at the location \(\mathrm{r}_m\). Accordingly, \(\hat{\sigma}_m^-\) annihilates an excitation in TLS \(m\).

Using Eq.~\eqref{eq:welschquanti} together with the decomposition of the operators \(\opvec{b}(\mathbf{r},\omega)\) from Eq.~\eqref{eq:bdecomp}, we bring the electric field operator into system-bath form:
\begin{align}\label{eq:Edecomp}
    \opvec{E}(\mathbf{r},\omega) = \sum_{i,\mu}\mathbf{E}_{i_{\mu}}(\mathbf{r},\omega)\hat{a}_{i_{\mu}} + i\sqrt{\frac{\hbar}{\epsilon_0\pi}}\int\mathrm{d}^3r'\sqrt{\epsilon_I(\mathbf{r}',\omega)}\mathbf{G}(\mathbf{r},\mathbf{r}',\omega)\cdot\opvec{c}(\mathbf{r}',\omega),
\end{align}
where we defined the \textit{QNM-generated electric field} \cite{fuchs2026quantum}
\begin{align}\label{eq:qnmgenefeld}
    \mathbf{E}_{i_{\mu}}(\mathbf{r},\omega) = \frac{i}{\omega\epsilon_0}\int\mathrm{d}^3r'\mathbf{G}(\mathbf{r},\mathbf{r}',\omega)\cdot\mathbf{j}_{i_{\mu}}(\mathbf{r}',\omega)
\end{align}
with the (classical) current density generated by the QNM, \(\mathbf{j}_{i_{\mu}}(\mathbf{r}, \omega) = \omega\sqrt{\frac{\hbar \epsilon_0}{\pi}\epsilon_I(\mathbf{r}, \omega)}\mathbf{L}^*_{i_{\mu}}(\mathbf{r}, \omega)\). The physical interpretation of the QNM-generated electric field is straightforward: The presence of the dissipative QNM creates a noise current that acts as a source of an electric field that a TLS interacts with. The QNM-generated electric field can be calculated numerically from a Helmholtz equation [cf.~Eq.~\eqref{eq:ehelm}] using \(\mathbf{j}_{i_{\mu}}(\mathbf{r}, \omega)\) as a source term. A more efficient approach uses the expansion of the Green's function in terms of the single-cavity QNMs to obtain an expression of the electric field in terms of few dominant QNMs \cite{fuchs2026greens}. 

Thus, with Eq.~\eqref{eq:Edecomp}, the Hamiltonian from Eq.~\eqref{eq:TLSHam} can be brought into system-bath form, where the TLSs and quasi-bound quantized QNMs act as the system and the propagating photons in the background lightguiding structure act as a bath.
Thus, with a system-bath Hamiltonian and coupling parameters that are rigorously defined in terms of numerically calculable QNM parameters, we can perform quantum dynamics calculations for multi-cavity systems and quantum emitters in structured environments using this QNM framework.

\begin{figure}
    \centering
    \includegraphics[width=0.95\linewidth]{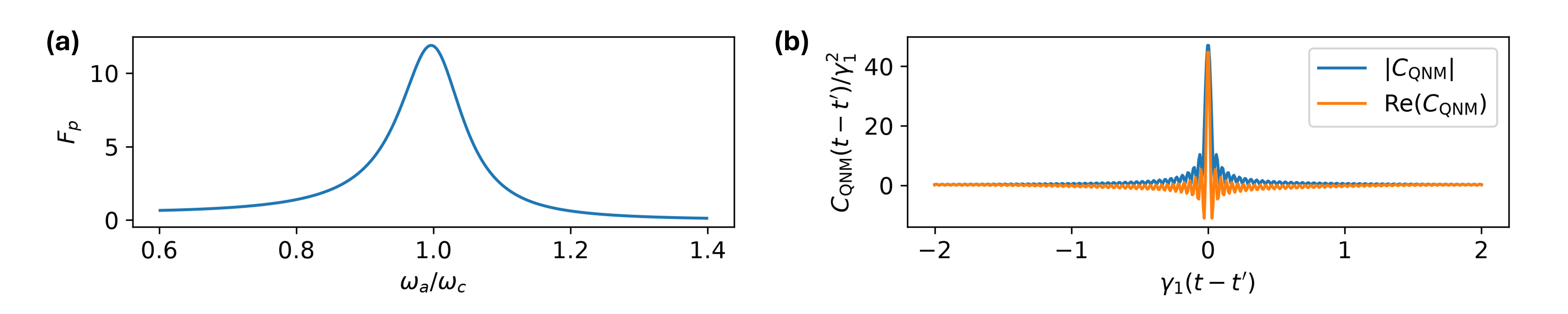}
    \caption{\red{(a) Purcell factor \(F_p\) of a \(z\)-polarized dipole located at the center of the 2D waveguide-coupled cavity from Fig.~\ref{fig:figure}(a) at \(\mathbf{r}_a = (0,L/2)\). (b) QNM Correlation function from Eq.~\eqref{eq:2DQNMcorrfunc} for the same 2D waveguide coupled cavity structure, assuming a single dominant QNM. The high-frequency oscillations stem from the implementation at a finite bandwidth.}}
    \label{fig:quantumQNM}
\end{figure}

\red{
\subsection{Example: QNMs and quantum emitters in a 2D waveguide-coupled cavity structure}\label{sec:quantexample}
We return to the example from Fig.~\ref{fig:figure} of a square silicon cavity coupled to two different waveguides (air and silicon dioxide, respectively).

For quantized QNMs in this structure, we use the regularized mode \(\tilde{F}_c(x,y)\) derived in Sec.~\ref{sec:classexample} to calculate the overlap matrix \(S_{cc}\). The 2D equivalent of Eq.~\eqref{appeq:Sintradef} reads:
\begin{align}\label{eq:2DSccintegral}
    S_{cc} = -\frac{c^2}{4\pi i\omega_c}\int_0^{\infty}\mathrm{d}\omega\frac{1}{(\omega-\tilde{\omega}_c)(\omega-\tilde{\omega}^*_c)}\int_0^L\mathrm{d}y \Big\{[\nabla\times \tilde{F}_c(x,y,\omega)\opvec{e}_z]\cdot[\opvec{e}_x\times\tilde{F}^*_c(x,y,\omega)\opvec{e}_z]\nonumber\\
    -[\opvec{e}_x\times \tilde{F}_c(x,y,\omega)\opvec{e}_z]\cdot[\nabla\times\tilde{F}^*_c(x,y,\omega)\opvec{e}_z]\Big\}_{x = -d/2}\nonumber\\
    +\frac{c^2}{4\pi i\omega_c}\int_0^{\infty}\mathrm{d}\omega\frac{1}{(\omega-\tilde{\omega}_c)(\omega-\tilde{\omega}^*_c)}\int_0^L\mathrm{d}y \Big\{[\nabla\times \tilde{F}_c(x,y,\omega)\opvec{e}_z]\cdot[\opvec{e}_x\times\tilde{F}^*_c(x,y,\omega)\opvec{e}_z]\nonumber\\
    -[\opvec{e}_x\times \tilde{F}_c(x,y,\omega)\opvec{e}_z]\cdot[\nabla\times\tilde{F}^*_c(x,y,\omega)\opvec{e}_z]\Big\}_{x=d/2},
\end{align}
where again, \(z\)-polarised fields are assumed. 
We now use the explicit form of \(\tilde{F}_c\) from Eq.~\eqref{eq:2Dreg}, focusing on frequencies above the air waveguide cutoff, so that \(k_{jx}\) is assumed to be real in both waveguides. Hence, we find:
\begin{align}\label{eq:2DS}
    S_{cc} = \frac{Lc^2}{8\gamma_c}\sum_{j=1,3}\Big(\sqrt{\epsilon_j\frac{\tilde{\omega}_c^2}{c^2}-k_y^2}+\sqrt{\epsilon_j\frac{\tilde{\omega}^{*2}_c}{c^2}-k_y^2}\Big),
\end{align}
where we (approximately) applied the residue theorem at the poles at \(\omega = \tilde{\omega}^{(*)}_c\) to solve the frequency integral in Eq.~\eqref{eq:2DSccintegral}, and \(\gamma_c = -\mathrm{Im}(\tilde{\omega}_c)\). For the present structure, we find \(S_{cc}\approx 5.9524\). 

To study the coupling to a quantum emitter, we consider a TLS point dipole located in the center of the cavity at \(\mathbf{r}_a = (0,L/2)\). The spontaneous emission rate of the TLS coupled to the quantized QNM of the cavity reads \cite{franke2019quantization}:
\begin{align}
    \Gamma_{\rm QNM} = |g_{c,a}|^2 \frac{\gamma_c}{\Delta_{ca}^2+\gamma_c^2},
\end{align}
where \(g_{c,a} = \sqrt{\omega_c S_{cc}/(2\hbar\epsilon_0)}\mathbf{d}_a\cdot\opvec{e}_z\tilde{f}_c(\mathbf{r}_a)\) is the QNM-TLS coupling strength \cite{franke2019quantization}, and \(\Delta_{ca} = \omega_c-\omega_a\) is the detuning between the QNM resonance and TLS transitions frequencies. For a \(z\)-polarized dipole, we plot the quantum Purcell Factor \(F_p = \Gamma_{\rm QNM}/\Gamma_0\) in Fig.~\ref{fig:quantumQNM}(a), where \(\Gamma_0\) is the emission rate in free space. The enhancement of the spontaneous emission near the resonance frequency is clearly visible. Below, we will also derive system-bath correlation functions for this structure, which can be implemented into common quantum dynamics schemes.

}

\section{System-bath correlation functions}\label{sec:corrfunc}
System-bath correlation functions play a central role in many established open quantum system dynamics methods, such as Nakajima-Zwanzig equations \cite{nakajima1958quantum, zwanzig1960ensemble, breuer2002theory} or the time-convolutionless approach (TCL) \cite{shibata1977generalized, breuer2002theory} as well as in path integral approaches \cite{caldeira1983path, grabert1988quantum, chernyak1996collective, garraway1997nonperturbative}, including the hierarchical equations of motion (HEOM) \cite{kubo1989time, tanimura2020numerically, fuchs2023hierarchical}. Therefore, in the following sections, we derive multicavity correlation functions with rigorously defined coupling parameters to characterize QNM-QNM, QNM-TLS, and TLS-TLS interactions in structured media. 

The correlation functions considered here are of the form
\begin{align} \label{eq:gencorrfunc}
    C_{\mathbb{x}\mathbb{y}}(t,t')= \sum_{pq}\int\mathrm{d}^3r\int_0^{\infty}\mathrm{d}\omega\int\mathrm{d}^3r'\int_0^{\infty}\mathrm{d}\omega' g_{\mathbb{x},p}(\mathbf{r},\omega)g^*_{\mathbb{y},q}(\mathbf{r}',\omega') \mathrm{tr}_B\left[\hat{c}_p(\mathbf{r},\omega,t)\hat{c}^{\dagger}_q(\mathbf{r}',\omega',t')\rho_B\right].
\end{align}
Here, \(\mathbb{x},\mathbb{y}\) are generic placeholder indices for the QNM or TLS indices \cite{fuchs2026quantum}, while \(\mathrm{tr}_B[...]\) is the trace over the bath degrees of freedom, and \(\rho_B = |{\rm vac}\rangle\langle {\rm vac}|\) is the initial vacuum bath state. Under the assumption of an initial vacuum bath state, it follows that
\begin{align*}
    \mathrm{tr}_B\left[\hat{c}_p(\mathbf{r},\omega,t)\hat{c}^{\dagger}_q(\mathbf{r}',\omega',t')\rho_B\right]= \left[\hat{c}_p(\mathbf{r},\omega,t),\hat{c}^{\dagger}_q(\mathbf{r}',\omega',t')\right]_-,
\end{align*}
which we employ in the derivation of the specific correlation functions below.

\subsection{Bath dynamics}\label{subsec:bathdyn}
To derive the correlation function from Eq.~\eqref{eq:gencorrfunc}, we need the dynamics of the bath operators with respect to the bath Hamiltonian [cf.~Eq.~\eqref{eq:qnmham}]
\begin{align}\label{eq:HBdef}
    H_B = \hbar\int_0^{\infty}\mathrm{d}\omega\int\mathrm{d}^3r\,\omega \opvec{c}^{\dagger}(\mathbf{r},\omega)\cdot\opvec{c}(\mathbf{r},\omega).
\end{align}

A detailed derivation of the bath dynamics was given in Ref.~\cite{fuchs2026quantum} in the context of cavities in a homogeneous three-dimensional background medium. The derivation of the time evolution of the bath operators for other environments follows the same steps, but with a different permittivity function and a different definition of the QNM projector kernels [Eq.~\eqref{eq:QNMproj}]. The time-dependent bath operators thus read, 
\begin{align}\label{eq:fullbath}
    \opvec{c}(\mathbf{r},\omega,t) &= \opvec{c}(\mathbf{r},\omega,t_0)\mathrm{e}^{-i\omega(t-t_0)}\nonumber\\
    &\quad+\sum_{ij,\mu\nu}\sum_{N=0}^{\infty}\mathbf{L}^*_{i_{\mu}}(\mathbf{r},\omega)\int_{t_0}^t\mathrm{d}t_1\mathrm{e}^{-i\omega(t-t_1)}\int_{t_0}^{t_1}\mathrm{d}t_2 \mathbb{K}^N_{i_{\mu}j_{\nu}}(t_1-t_2)\hat{C}_{j_{\nu}}(t_2),
\end{align}
where
\begin{align}\label{eq:initscat}
    &\hat{C}_{i_{\mu}}(t)=i\int\mathrm{d}^3r\int_0^{\infty}\mathrm{d}\omega \mathbf{g}_{i_{\mu}}(\mathbf{r},\omega)\cdot\opvec{c}(\mathbf{r},\omega,t_0)\mathrm{e}^{-i\omega(t-t_0)}
\end{align}
describes the annihilation of a noise quantum in the bath due to absorption by the QNM \(i_{\mu}\). Furthermore (leaving \(s>0\) implicit), 
\begin{align}\label{eq:KNdef}
    \mathbb{K}^1_{i_{\mu}j_{\nu}}(s) &= i\int\mathrm{d}^3r\int_0^{\infty}\mathrm{d}\omega \mathbf{g}_{i_{\mu}}(\mathbf{r},\omega)\cdot\mathbf{L}^*_{j_{\nu}}(\mathbf{r},\omega)\mathrm{e}^{-i\omega s},\nonumber\\
    \mathbb{K}^N_{i_{\mu}j_{\nu}}(s) &= \sum_{l,\eta}\int_0^s\mathrm{d}t_1 \mathbb{K}^{N-1}_{i_{\mu}l_{\eta}}(s-t_1)\mathbb{K}^1_{l_{\eta}j_{\nu}}(t_1)= \sum_{l,\eta}\int_0^s\mathrm{d}t_1 \mathbb{K}^1_{i_{\mu}l_{\eta}}(s-t_1)\mathbb{K}^{N-1}_{l_{\eta}j_{\nu}}(t_1),
\end{align} 
describes the \(N\)-th order \textit{intercavity scattering}, where the emission from one QNM cavity scatters at the quasi-bound QNM or outward propagating emission of another cavity. This is a retarded process, which only yields significant contributions if \(s\approx \tau\), where \(\tau\) is the retardation time. Intracavity terms \(i=j\) \red{to \(\mathbb{K}^1_{i_{\mu}j_{\nu}}\)} vanish. \red{This follows mathematically from causality, as shown in Refs.~\cite{fuchs2026quantum, gustin2026dissipation}. It can also be understood intuitively: From Eq.~\eqref{eq:bdecomp}, \(\mathbf{L}^*_{j_{\nu}}\) is related to the annihilation of a quantum in mode \(j_{\nu}\). Meanwhile, from Eq.~\eqref{eq:qnmham}, \(\mathbf{g}_{i_{\mu}}\) is associated with the creation of a quantum in mode \(i_{\mu}\) via coupling to a bath photon. Hence, \(\mathbb{K}^1_{i_{\mu}j_{\nu}}(s)\) can be understood as the annihilation of a photon in mode \(j_{\nu}\) which causes an outward propagating emission through the bath, which couples to mode \(i_{\mu}\) as an inward propagation at some time \(s\) later. For \(i=j\), the outward propagating emission cannot propagate back inward to the same cavity unless it is reflected (i.e., at another cavity), which requires an additional scattering process and is not included in the single-scattering term \(\mathbb{K}^1\). The higher-order \(\mathbb{K}^N\) can contain effective intracavity scattering if they consist of multiple intercavity scattering processes, e.g., a photon is emitted, scattered back at another cavity, and returns to the original cavity.
}
Similarly, \(\mathbb{K}^0_{i_{\mu}j_{\nu}}(s) = 2\delta_{ij}\delta_{\mu\nu}\Theta(s)\delta(s)\) is the zeroth-order term, where no scattering occurs. Thus, the first term in Eq.~\eqref{eq:fullbath} describes the unperturbed propagation of a bath photon through the background structure, while the second term corrects for scattering at the QNMs. 

\subsection{QNM correlation functions} \label{subsec:QNMcorr}
In the QNM Hamiltonian from Eq.~\eqref{eq:qnmham}, the QNMs of well-separated cavities (\(P_{i_{\mu}j_{\nu}}\gg 0\)) do not couple directly. Instead, different QNMs couple to the same bath of propagating photons, yielding an effective, time-delayed intercavity interaction. Following the derivation in Ref.~\cite{fuchs2026quantum}, the QNM correlation function reads (\(s = t-t'\)):
\begin{align}\label{eq:QNMcorrfunc}
    C^{\rm QNM}_{i_{\mu}j_{\nu}}(s) = \delta_{ij}C^{\rm bos}_{i_{\mu}j_{\nu}}(s)&+ \Theta(s)\sum_{N=1}^{\infty}\sum_{\eta}(\delta_{\eta\nu}\partial_{s}{+}i\chi_{j_{\eta}j_{\nu}})\mathbb{K}^N_{i_{\mu}j_{\eta}}(s) \nonumber\\
    & + \Theta(-s)\sum_{N=1}^{\infty}\sum_{\eta}\Big[(\delta_{\eta\mu}\partial_{-s}{+}i\chi_{i_{\eta}i_{\mu}})\mathbb{K}^N_{j_{\nu}i_{\eta}}(-s)\Big]^*,
\end{align}
where
\begin{align}\label{eq:Cbos}
    &C^{\rm bos}_{i_{\mu}j_{\nu}}(s) =\int_0^{\infty}\mathrm{d}\omega\int\mathrm{d}^3r \mathbf{g}_{i_{\mu}}(\mathbf{r},\omega)\cdot\mathbf{g}^*_{j_{\nu}}(\mathbf{r},\omega)\mathrm{e}^{-i\omega s}
\end{align}
is the QNM correlation function for a \textit{bosonic} bath \cite{franke2020quantized, fuchs2023hierarchical, fuchs2026quantum}. It is connected to the intercavity scattering \(\mathbb{K}^1_{i_{\mu}j_{\nu}}(s)\) via \cite{fuchs2026quantum}
\begin{align}\label{appeq:CbosK1}
     &\sum_{\nu'}(\delta_{\nu\nu'}\partial_{s}+i\chi^*_{j_{\nu}j_{\nu'}})\mathbb{K}^1_{i_{\mu}j_{\nu'}}(s)=(1-\delta_{ij})\Theta(s)C^{\rm bos}_{i_{\mu}j_{\nu}}(s),
\end{align}

A single intercavity scattering \(\mathbb{K}^1_{i_{\mu}j_{\nu}}(s)\) yields a significant contribution only if \(s\approx \tau_{ij}\), where \(\tau_{ij}\) is the photon retardation time between cavities \(i\) and \(j\). From Eq.~\eqref{eq:KNdef}, it follows that \(\mathbb{K}^N_{i_{\mu}j_{\nu}}(s)\) only yields a significant result if \(s\sim N\tau\). Equation~\eqref{eq:QNMcorrfunc} can therefore be truncated at finite \(N\) for a finite time difference \(s=t-t'\). 

\begin{figure}
    \centering
    \includegraphics[width=0.95\linewidth]{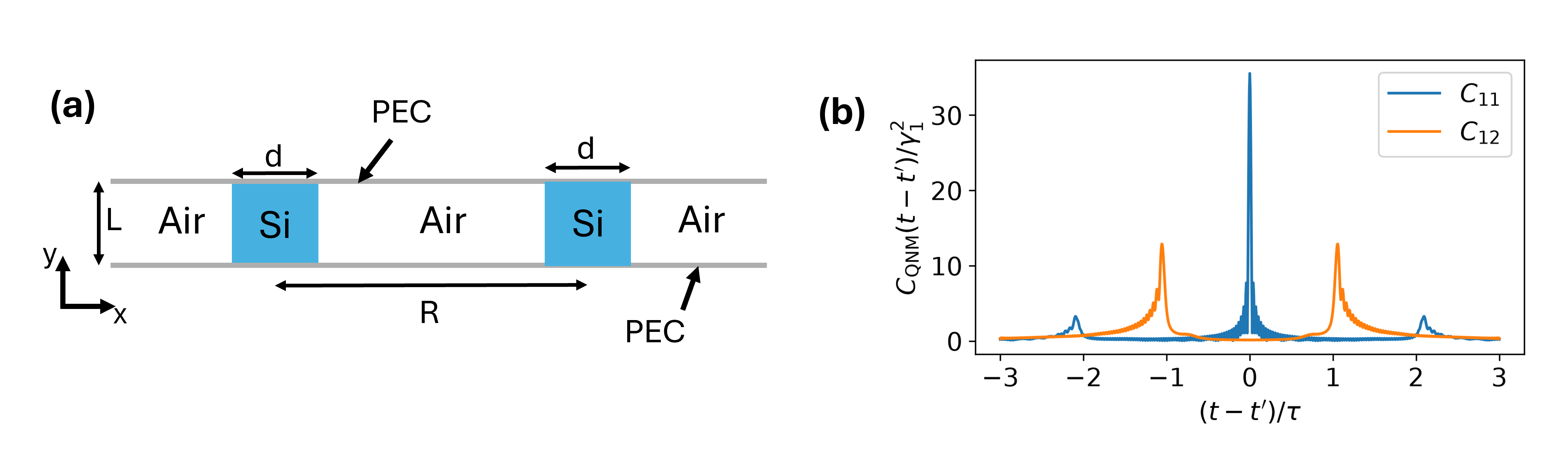}
    \caption{\red{(a) Sketch of two 2D square silicon cavities (\(d=L=300\,{\rm nm})\), separated by distance \(R=20d\), and located in an infinite, lossless 2D air waveguide made up of perfect electric conductors at \(y=0\) and \(y=L\). (b) Absolute value of the diagonal and off-diagonal terms of the QNM correlation function [Eq.~\eqref{eq:QNMcorrfunc}] for the coupled-cavity structure from (a). The diagonal component shows a peak at \(t-t'=0\) related to spontaneous decay of the single-cavity mode into the waveguide, and a small peak at \(|t-t'|=2\tau = 2R/c\) related to assisted emission via photon exchange with the other cavity. The off-diagonal term has dominant peaks at \(|t-t'|=\tau\) related to radiative energy exchange between the two cavities. The high-frequency oscillations arise due to an implementation at a finite bandwidth.}}
    \label{fig:2QNMcorr}
\end{figure}

\red{
\subsection{Example: Correlation function for a 2D waveguide-coupled cavity}
Now, we calculate the QNM correlation function for the 2D waveguide-coupled cavity from Fig.~\ref{fig:figure}(a). Since we consider a system with one dominant QNM inside one cavity, the QNM correlation function reads,
\begin{align}\label{eq:2DQNMcorrfunc}
    C^{\rm QNM}(t-t') = \int_0^{\infty}\mathrm{d}\omega\int\mathrm{d}^2r \mathbf{g}_c(\mathbf{r},\omega)\cdot\mathbf{g}_c^*(\mathbf{r},\omega)\mathrm{e}^{-i\omega(t-t')},
\end{align}
with the QNM noise coupling elements from Eq.~\eqref{eq:noisecoup}. Now, we employ the Markov approximation from Ref.~\cite{franke2020quantized}, where we approximate \(\int\mathrm{d}^2r \mathbf{g}_c(\mathbf{r},\omega)\cdot\mathbf{g}_c^*(\mathbf{r},\omega)\approx (1/\pi)\gamma_c\), and then extend the lower limit of the frequency integral to \(-\infty\) on the assumption that all relevant contributions are dominant near the QNM resonance frequency and are negligible for other frequencies. Then, we can approximate
\begin{align}
    C^{\rm QNM}(t-t') \approx 2\gamma_c \delta(t-t'),
\end{align}
which was also the result in Refs.~\cite{franke2020quantized, fuchs2023hierarchical, fuchs2026quantum}. We plot this result in Fig.~\ref{fig:quantumQNM}(b), where the high-frequency oscillations appear as a consequence of the implementation of the \(\delta\)-function at a finite bandwidth. 

For a coupled-cavity example, we use the slightly modified model from Fig~\ref{fig:2QNMcorr}(a) of an infinite waveguide filled with air and two square silicon cavities located at \(x=0\) and \(x=R\). We again take \(d=L = 300\,{\rm nm}\). The single-cavity QNMs are the same as in Eq.~\eqref{eq:2DQNM}, except that we now have only air waveguides. We again focus on the lowest-energy QNM above the waveguide cutoff frequency. The identical cavities have the same dominant mode with eigenfrequency: \(\tilde{\omega}_1 = \tilde{\omega}_2 = \tilde{\omega}_c \approx (3.779-0.0963i)(c/d)\), which corresponds to \(Q = 19.611\). 

From Eq.~\eqref{eq:2DS}, we obtain the intracavity overlap \((\epsilon_1 = \epsilon_3 = 1\)):
\begin{align}
    S_{11}=\frac{Lc^2}{2\gamma_c}\mathrm{Re}(k_{Bx}^{\mu}),
\end{align}
where \(k_{Bx}^{\mu} = \sqrt{\frac{\tilde{\omega}_c^2}{c^2}-k_y^2}\) is the \(x\)-component of the QNM wavevector inside the air waveguides.

For identical cavities, \(\tilde{f}_2(x,y) = \tilde{f}_1(x-R,y)\) holds, and results in the intercavity overlap [cf.~Eq.~\eqref{appeq:Sinterdef})  \footnote{\red{see also Ref.~\cite{fuchs2023hierarchical}, where the intercavity overlap is derived for two coupled 1D dielectric slabs, which is a very similar derivation to this 2D example}}:
\begin{align}
    S_{12} = \frac{Lc^2}{2\gamma_c}\mathrm{Re}\Big(k_{Bx}^{\mu}\mathrm{e}^{-ik_{Bx}^{\mu} R} \Big).
\end{align}

As a consequence, the relative intercavity overlap \(|S_{12}/S_{11}|\) scales as \(\mathrm{e}^{\mathrm{Im}(k_{Bx}^{\mu}) R}\). For the model from Fig.~\ref{fig:2QNMcorr}(a), we have \(k_{Bx}^{\mu}\approx (0.00703-0.00058i){\rm nm}^{-1}\), so that for \(R = 20 d\), the relative intercavity overlap is \(|S_{12}/S_{11}|\sim\mathrm{e}^{-3}<5\,\%\), and thus, our cavities are sufficiently well-separated. 

To calculate the multi-cavity correlation function from Eq.~\eqref{eq:QNMcorrfunc}, we use the results from Ref.~\cite{fuchs2026quantum}, which show that
\begin{align}\label{eq:Cbosapprox}
    &C^{\rm bos}_{ij}(t-t') \approx \gamma_c\int_0^{\infty}\mathrm{d}\omega \frac{\mathrm{e}^{-i\omega(t-t')}}{2\pi}\Big(\mathrm{e}^{-ik_{Bx}R_{ij}}+\mathrm{e}^{ik_{Bx}R_{ij}}\Big),
\end{align}
and
\begin{align}\label{eq:singleKapprox}
    &\mathbb{K}^1_{12}(t-t')\approx \frac{i\gamma_c}{2\pi }\int_0^{\infty} \mathrm{d}\omega \Big(\mathrm{e}^{-ik_{Bx}R_{ij}}+\mathrm{e}^{ik_{Bx}R_{ij}}\Big)\frac{\mathrm{e}^{-i\omega(t-t')}}{\omega-\tilde{\omega}^*_c}.
\end{align}

Here, \(i,j\in \{1,2\}\) are cavity indices, \(k_{Bx} = \sqrt{(\omega^2/c^2-k_y^2)}\) is the \(x\)-component of the wavevector in the waveguide , and \(R_{ij} = (1-\delta_{ij})R\) is the distance between cavities \(i\) and \(j\). The higher-order \(\mathbb{K}^N\) are obtained from Eq.~\eqref{eq:KNdef}. The full correlation function is calculated from Eq.~\eqref{eq:QNMcorrfunc}. 
In Fig.~\ref{fig:2QNMcorr}(b) we show the diagonal and off-diagonal terms of the correlation function. 
Since \(\mathbb{K}^N\) only significantly differs from \(0\) for \(|t-t'|\approx N\tau = N R/c\) (since it contains \(N\) intercavity scattering processes, requiring the photon to propagate between the cavities \(N\) times), we truncate the calculation at \(N_{\rm max} = 3\) in Fig.~\ref{fig:2QNMcorr}(b). For the two identical cavities \(C_{22} = C_{11}\) and \(C_{21} = C_{12}^*\). }

\red{As expected, the diagonal term of the correlation function \(C_{11}\) shows a peak at \(t-t'=0\) that is related to the sponanetous decay of the single-cavity mode into the waveguide. There are additional peaks visible at \(|t-t'| \approx 2\tau\) that are related to the assisted emission via photon exchange with the other cavity (requiring one round trip to the other cavity and back). In the off-diagonal term, the correlation function peaks at \(|t-t'| = \tau\) which is related to photon exchange between the cavities. The high-frequency oscillations of the correlation functions appear due to the implementation at a finite bandwidth. We stress again that the present study is limited to a small frequency regime near the lowest QNM resonance, and additional aspects, such as multi-mode coupling and the influence of the branch cut (which could lead to evanescent coupling between the cavities through the connecting waveguide) are neglected. Still, the results show how the correlation functions for multiple, waveguide-coupled cavities can be implemented. The correlation functions from Fig.~\ref{fig:quantumQNM}(b) or Fig.~\ref{fig:2QNMcorr}(b) could now be straight-forwardly applied to common open-quantum system methods to calculate the evolution of the system density matrix of the waveguide-coupled cavities \cite{nakajima1958quantum, zwanzig1960ensemble, breuer2002theory, richter2022enhanced, strathearn2017efficient, fuchs2023hierarchical}.
}

\subsection{QNM-TLS correlation functions}\label{subsec:QTcorr}
In Sec.~\ref{sec:qnmtlscoup}, we discussed the direct (non-bath mediated) coupling between a TLS inside (or near) a cavity. The TLSs also couple to the bath modes, leading to an effective, time-delayed interaction between the TLSs and the QNMs, mediated by the bath. This interaction is especially important for spatially separated TLSs and QNMs, i.e., a TLS inside another cavity or in the background medium. The interaction is characterized by the correlation function \cite{fuchs2026quantum}
\begin{align}\label{eq:QTcorr}
    C^{\rm Q-T}_{i_{\mu}a}(s)&=\frac{\Theta(s)}{i\hbar}\sum_{j,\eta\nu}\sum_{N=0}^{\infty}(\delta_{\eta\nu}\partial_s{+}i\chi_{j_{\eta}j_{\nu}})\int_0^s\mathrm{d}t_1\mathbb{K}^N_{i_{\mu}j_{\eta}}(s-t_1)\mathbf{d}^*_a\cdot\mathbf{E}^*_{j_{\nu}}(\mathbf{r}_a,-t_1)\nonumber\\
    &+\frac{\Theta(-s)}{i\hbar}\sum_{j,\eta\nu}\sum_{N=0}^{\infty}(\delta_{\eta\mu}\partial_s{+}i\chi^*_{i_{\eta}i_{\mu}})\int_s^0\mathrm{d}t_1\Big[\mathbb{K}^N_{j_{\nu}i_{\eta}}(t_1-s)\Big]^*\mathbf{d}^*_a\cdot\mathbf{E}^*_{j_{\nu}}(\mathbf{r}_a,-t_1),
\end{align}
where \(C^{\rm T-Q}_{a i_{\mu}}(s)=[C^{\rm Q-T}_{i_{\mu}a}(-s)]^*\), and
\begin{align}\label{eq:timedepQNMgenefield}
    \mathbf{E}_{i_{\mu}}(\mathbf{r},t)= \int_0^{\infty}\mathrm{d}\omega\mathbf{E}_{i_{\mu}}(\mathbf{r},\omega)\mathrm{e}^{-i\omega t},
\end{align}
is the time-dependent QNM-generated electric field, which is obtained from Eq.~\eqref{eq:qnmgenefeld}. Again, the correlation function can be truncated at finite \(N\) for a practical implementations.

\subsection{TLS correlation functions}\label{subsec:TLScorr}
A photon emitted from a TLS into the bath can either re-excite the same TLS or propagate through the bath to excite another TLS in a cavity or the background medium. A TLS inside a cavity will mainly couple to the quasibound quantized QNMs of that cavity, whereas a TLS in the outside medium will primarily couple to the bath, which is a crucial driver for the interactions with other systems. The correlation function for the bath-mediated interactions between two TLSs read:
\begin{align}\label{eq:tlscorrfunc}
    C_{ab}^{\rm TLS}(s)&= \frac{1}{\pi\hbar\epsilon_0}\int_0^{\infty}\mathrm{d}\omega \mathbf{d}_{a}\cdot{\rm Im}\Big[\mathbf{G}(\mathbf{r}_{a},\mathbf{r}_{b},\omega)\Big]\cdot\mathbf{d}^*_{b}\mathrm{e}^{-i\omega s}\nonumber\\
    &\;-\frac{\Theta(s)}{\hbar^2}\sum_{ij,\mu\nu\eta}\sum_{N=0}^{\infty}\int_{0}^s\mathrm{d}t_1\int_0^{t_1}\mathrm{d}t_2 \mathbf{d}_{a}\cdot\mathbf{E}_{i_{\mu}}(\mathbf{r}_{a},s-t_1)\nonumber\\
    &\qquad\qquad\qquad\times\Big[(\delta_{\eta\nu}\partial_{t_1}{+}i\chi_{j_{\eta}j_{\nu}})\mathbb{K}^N_{i_{\mu}j_{\nu}}(t_1-t_2)\Big]\mathbf{d}^*_{b}\cdot\mathbf{E}^*_{j_{\nu}}(\mathbf{r}_{b},-t_2)\nonumber\\
    &\;-\frac{\Theta(-s)}{\hbar^2}\sum_{ij,\mu\nu\eta}\sum_{N=0}^{\infty}\int_s^0\mathrm{d}t_1\int_s^{t_1}\mathrm{d}t_2 \mathbf{d}_{a}\cdot\mathbf{E}_{i_{\mu}}(\mathbf{r}_{a},s-t_2)\nonumber\\
    &\qquad\qquad\qquad\times\Big[\mathbf{d}_{b}\cdot\mathbf{E}_{j_{\nu}}(\mathbf{r}_{b},-t_1)(\delta_{\eta\mu}\partial_{t_1}{+}i\chi_{i_{\eta}i_{\mu}})\mathbb{K}^N_{j_{\nu}i_{\mu}}(t_1-t_2)\Big]^*.
\end{align}
Here, the first line on the right-hand side is the full dipole coupling between the TLSs via the electromagnetic field \cite{hughes2005extrinsic, franke2019quantization, fuchs2026quantum}, while the other terms are corrections due to scattering at the quasibound QNMs, which are here removed from the full dipole coupling. Again, the correlation function can be truncated at finite \(N\) for a finite time difference \(s=t-t'\). 

\section{Conclusions}
In conclusion, we presented a quantization scheme for QNMs coupled to different lightguiding structures, such as waveguides or surfaces. We formulated a boundary integral, from which specific boundary conditions for the different background structures and combination of different types of lightguiding structures can be derived. Our results are summarized in Tab.~\ref{tab:summary}. 

These boundary conditions are used to formulate QNMs in various background structures. The QNMs of well-separated cavities can be quantized independently of each other, and we adapted the cavity separation parameters \(P_{i_{\mu}j_{\nu}}\) as a simple measure of the separation. The quantization naturally yields a non-bosonic bath of propagating photons that the QNMs of all cavities interact with. These bath operators can be separated into sets of commuting operators for the different lightguiding structures, e.g., for a network of non-intersecting waveguides.

In addition to the QNM quantization, we included quantum emitters modeled as TLSs that couple to both the QNMs and the bath. From the system-bath Hamiltonian, we derived correlation functions for the time-delayed interactions between the spatially separated QNMs and TLSs. All decay rates and coupling elements are \red{systematically} defined in terms of QNM parameters that can be obtained numerically.

The QNM approach allows for a \red{systematic} treatment of coupled-cavity systems with losses. Thus, we extend the previous theory of cavities in a homogeneous background to setups with complex lightguiding structures, which are the basis for realistic quantum devices, such as on-chip quantum information devices.

\red{We illustrated the approach on the simple example of 2D waveguide-coupled cavities. For a single cavity coupled to two different waveguides, we derived the lowest energy QNM above the waveguide cutoff frequency, and performed the regularization. We then quantized the QNM and calculated the quantum Purcell Factor of a point-dipole two-level system in the center of the cavity. We furthermore calculated the QNM correlation functions for the single-cavity model as well as for a similar mode of two identical 2D cavities embedded in a 2D waveguide and separated by a finite distance \(R\). The modeling of more complex and realistic devices is subject for future work.}

\ack{R.M.F. and M.R. acknowledge support from the Deutsche Forschungsgemeinschaft (Project number 525575745). We also thank Stephen Hughes for insightful discussions and comments to this manuscript.}

\data{The data supporting this study can be made available upon reasonable request.}

\appendix
\section{Electrodynamics in one and two dimensions}\label{appsec:1D2D}
Here, we derive the 1D and 2D Helmholtz equations used as the basis for the derivations in Sec.~\ref{sec:BChomogeneous}.
\subsection{1D}
We assume a setup that depends only on one spatial variable \(x\) and is homogeneous in the \(y,z\)-directions. Furthermore, we assume linearly polarized waves,
\begin{align}
    \mathbf{E}(x,\omega) = E(x,\omega)\opvec{e}_z,\quad \mathbf{J}(x,\omega) = J(x,\omega)\mathbf{e}_z,
\end{align}
so that the Helmholtz equation~\eqref{eq:ehelm} reads,
\begin{align}\label{appeq:1Dehelm}
    \partial_x^2 E(x,\omega) +\frac{\omega^2}{c^2}\epsilon(x,\omega)E(x,\omega) = -i\omega\mu_0 J(x,\omega).
\end{align}

Accordingly, the Helmholtz equation for the scalar Green's function reads,
\begin{align}\label{appeq:1Dgreenhelm}
    \Big[\partial_x^2+\frac{\omega^2}{c^2}\epsilon(x,\omega)\Big]G(x,x',\omega) = -\frac{\omega^2}{c^2}\delta(x-x').
\end{align}

By integrating Eq.~\eqref{appeq:1Dehelm} and Eq.~\eqref{appeq:1Dgreenhelm} with respect to \(x\) and employing the fundamental theorem of calculus (i.e. the one dimensional equivalent to Green's theorem), we obtain the expression from Eq.~\eqref{eq:BCintegral1D}. 

\subsection{2D}
For a two-dimensional problem, Maxwell's equations read
\begin{align}
    &\nabla\cdot[\epsilon(\mathbf{r},\omega)\mathbf{E}(\mathbf{r},\omega)] = \frac{\sigma(\mathbf{r},\omega)}{\epsilon_0}\nonumber\\
    &\nabla_{\perp}\cdot\mathbf{E}(\mathbf{r},\omega) = -i\omega B(\mathbf{r},\omega)\nonumber\\
    &\nabla_{\perp}B(\mathbf{r},\omega) = \mu_0\mathbf{J}(\mathbf{r},\omega) -\frac{\omega^2}{c^2}\epsilon(\mathbf{r},\omega)\mathbf{E}(\mathbf{r},\omega),
\end{align}
where \(\mathbf{r} = (x,y)\), \(\nabla = (\partial_x,\partial_y)\), \(\nabla_{\perp} = (\partial_y,-\partial_x)\), and \(B=B_z\) is the z-component of the magnetic field vector. Furthermore, \(\sigma\) are surface charges and \(\mathbf{J}\) are surface currents. Next, we employ \(\nabla_{\perp}(\nabla_{\perp}\cdot\mathbf{E}) = \Delta\mathbf{E}-\nabla(\nabla\cdot\mathbf{E})\), to obtain the Helmholtz equation analog in 2D:
\begin{align}\label{appeq:2Dehelm}
    \Big[\Delta+\frac{\omega^2}{c^2}\epsilon(\mathbf{r},\omega)\Big]\mathbf{E}(\mathbf{r},\omega) = -i\omega\mu_0\mathbf{J}_T(\mathbf{r},\omega),
\end{align}
where 
\begin{align}
    \mathbf{J}_T(\mathbf{r},\omega)&= \mathbf{J}(\mathbf{r},\omega)-\frac{1}{i\omega\mu_0}\nabla\left[\frac{\sigma(\mathbf{r},\omega)}{\epsilon_0\epsilon(\mathbf{r},\omega)}-\frac{\mathbf{E}(\mathbf{r},\omega)\cdot\nabla\epsilon(\mathbf{r},\omega)}{\epsilon(\mathbf{r},\omega)}\right]
\end{align}
is the transverse surface current. Note that, since we assume a homogeneous background medium \(\epsilon(\mathbf{r},\omega)\big|_{\mathbf{r}\notin V_{\rm in}} = \epsilon_B\), the transverse current vanishes everywhere outside the scattering volume, and hence, it does not play a role in the derivation of boundary conditions for the electric field.

Thus the Green's function in 2D solves the Helmholtz equation
\begin{align}\label{appeq:2Dgreenhelm}
    \Big[\Delta+\frac{\omega^2}{c^2}\epsilon(\mathbf{r},\omega)\Big] \mathbf{G}(\mathbf{r},\mathbf{r}',\omega) = -\frac{\omega^2}{c^2}\delta(\mathbf{r}-\mathbf{r}').
\end{align}

From Eq.~\eqref{appeq:2Dehelm} and Eq.~\eqref{appeq:2Dgreenhelm}, we then obtain the condition from Eq.~\eqref{eq:BCintegral2D} via Gauss's integral theorem. 

\section{Boundary conditions at a surface}\label{appsec:greensurface}
\subsection{Green's function}
Here, we provide additional information on the Green's function near a surface from Sec.~\ref{sec:surfaceBC}. The Green's function is given via the Sommerfeld integral from Eq.~\eqref{eq:sommerfeldG}, which depends on the Kernel (cf.~Refs.~\cite{sipe1987new, novotny2012principles})
\begin{align}\label{appeq:Kkappaspace}
    \mathbf{K}(\rho,\kappa,\omega) = \frac{i}{8\pi^2}\int_0^{2\pi}\mathrm{d}\varphi \left\{[1+r_s(\kappa)]\opvec{e}_s\opvec{e}_s + \opvec{e}_{p_+}\opvec{e}_{p_+} - r_p(\kappa) \opvec{e}_{p_-}\opvec{e}_{p_+}\right\}\mathrm{e}^{i\kappa\rho \cos\varphi},
\end{align}
where \(\opvec{e}_{p_\pm} = (\kappa\opvec{e}_z\pm q_1\opvec{e}_{\kappa})/(\sqrt{\epsilon_1}k)\), \(\opvec{e}_s = \opvec{e}_{\kappa}\times\opvec{e}_z\), and \(\opvec{e}_{\kappa}\) is the unit vector in the \(\kappa\)-direction \cite{sipe1987new}. \red{Note that in Sec.~~\ref{sec:surfaceBC} we consider a small cavity directly at the interface at \(\mathbf{r}\approx 0\) for a simple illustration and also focus only on the region above the interface \(z>0\). Therefore, only upward propagating waves \(\opvec{e}_{p_+}\opvec{e}_{p_+}\) appear in Eq.~\eqref{appeq:Kkappaspace}. But we stress that downward propagating waves \(\opvec{e}_{p_-}\opvec{e}_{p_-}\) could also be included for a more detailed discussion including the space below the interface or for a cavity located a finite distance above the interface.}
The integration is carried out over the angle \(\varphi\) between \(\opvec{e}_{\kappa}\) and \(\opvec{e}_{\rho}\). The unit vectors \(\opvec{e}_s,\opvec{e}_{\kappa}\) depend implicitly on \(\varphi\), and hence, the integration will yield an expression in terms of Bessel functions \(J_n(\kappa\rho)\), which we do not specify here, since they do not factor into the boundary conditions we derive in the main text (see e.g. Ref.~\cite{paulus2000accurate} for an example of the different entries of the Green's dyadic). 

Sufficiently far from the surface, pure surface contributions to the Green's function vanish, and we obtain a far-field representation of the Green's function purely in terms of spherical waves. The reflected part is given in terms of the matrix \(\mathbf{A}(\theta)\) from Eq.~\eqref{eq:Gspace}. This matrix reads \cite{novotny2012principles}
\begin{align}\label{appeq:Adef}
    \mathbf{A}^{\rm ref}(\theta) = r_s(\theta)\opvec{e}_{\varphi}\opvec{e}_{\varphi} -r_p(\theta) \opvec{e}_-\opvec{e}_+,
\end{align}
where \(\opvec{e}_{\pm} = \cos(\theta)\opvec{e}_{\rho}\pm\sin(\theta)\opvec{e}_z\), and \(\cos(\theta)= z/r,\, \sin(\theta) = \rho/r\).

\subsection{Boundary conditions from the free space contributions}
Here, we derive the boundary conditions from the free space contributions, given in Eq.~\eqref{eq:surfaceBCfree}. The Green's function \(\mathbf{G}_{\rm free}\) is given in Eq.~\eqref{eq:Gspace} with \(\mathbf{A}^{\rm ref}(\theta)\) from Eq.~\eqref{appeq:Adef}. We write the Green's function as a dyadic product \(\mathbf{G}_{\rm free}(\mathbf{r}) = \sum_j \mathbf{G}_j(\mathbf{r})\opvec{e}_j\), where \(\opvec{e}_j\) is the \(j\)-th unit vector, and \(\mathbf{G}_j\) is the \(j\)-th column vector of the dyadic. Here and in the following, we neglect the frequency dependence of the Green's function for notational brevity. 

Now, the boundary condition integral from Eq.~\eqref{eq:BCintegral} takes the form:
\begin{align}\label{appeq:BCintegralj}
    0 = \sum_j \opvec{e}_j \lim_{R\to \infty}\oint_{\mathcal{S}(R)}\mathrm{d}A_R \opvec{e}_r \cdot \Big\{ \mathbf{E}(\mathbf{R})\times\big[\nabla\times\mathbf{G}_j(\mathbf{R})\big]-\mathbf{G}_j(\mathbf{R})\times\big[\nabla\times\mathbf{E}(\mathbf{R})\big]\Big\},
\end{align}
where we have chosen a half-sphere with radius \(R\to\infty\) as our far-field surface, as sketched in Fig.~\ref{fig:surfacecav}(a). Since the \(\opvec{e}_j\) are linearly independent, the contributions for different \(j\) have to vanish independently. Unlike in the case of a homogeneous background medium, the different polarization directions have to be treated separately to account for the different reflection of \(s\)- and \(p\)-polarized waves. From Eq.~\eqref{eq:Gspace} together with Eq.~\eqref{appeq:Adef}, we find
\begin{align}
    &\mathbf{G}_{\varphi}(\mathbf{R}) = \frac{\mathrm{e}^{ikR}}{4\pi R}[1+r_s(\theta)]\opvec{e}_{\varphi},\quad \nabla\times\mathbf{G}_{\varphi}(\mathbf{R}) = \frac{\mathrm{e}^{ikR}}{4\pi R}[\nabla r_s(\theta)]\times\opvec{e}_{\varphi} - ik\frac{\mathrm{e}^{ikR}}{4\pi R}[1+r_s(\theta)] \opvec{e}_{\theta},\nonumber\\
    &\mathbf{G}_{\theta}(\mathbf{R}) = \frac{\mathrm{e}^{ikR}}{4\pi R}[1-\cos(2\theta)r_p(\theta)]\opvec{e}_{\theta} + \frac{\mathrm{e}^{ikR}}{4\pi R}\sin(2\theta)r_p(\theta)\opvec{e}_r,\nonumber\\
    &\nabla\times \mathbf{G}_{\theta}(\mathbf{R}) = \frac{\mathrm{e}^{ikR}}{4\pi R}\Big\{ ik[1-\cos(2\theta)r_p(\theta)]\opvec{e}_{\varphi}+[\nabla \sin(2\theta)r_p(\theta)]\times\opvec{e}_r\Big\},
\end{align}
where we used spherical coordinates to write, from Eq.~\eqref{appeq:Adef}, \(\opvec{e}_- = \opvec{e}_{\theta}\) and \(\opvec{e}_+ = \cos(2\theta)\opvec{e}_{\theta}+\sin(2\theta)\opvec{e}_r\). 

Now, we insert these results into Eq.~\eqref{appeq:BCintegralj}, which yields two boundary conditions for the \(\theta\)- and \(\varphi\)-contributions. Writing the integral over \(\mathcal{S}(R)\) as an integral over the solid angle \(\Omega\) of the half sphere (with \(\mathrm{d}A_R = R^2 \mathrm{d}\Omega\)), we obtain, from the \(\mathbf{G}_{\varphi}\)-contributions:
\begin{align}\label{appeq:phicontributions}
    0 = \lim_{R\to \infty} \oint_{\Omega}\mathrm{d}\Omega \mathrm{e}^{ikr} [1+r_s(\theta)] R\opvec{e}_r \cdot \Big\{ -ik\mathbf{E}(\mathbf{R})\times\opvec{e}_{\theta}-\opvec{e}_{\varphi}\times\Big[\nabla\times\mathbf{E}(\mathbf{R})-\frac{\nabla r_s(\theta)}{1+r_s(\theta)}\times \mathbf{E}(\mathbf{R}) \Big]\Big\},
\end{align}
which, after reordering the terms, yields the boundary conditions from Eq.~\eqref{eq:surfaceBCfree}. Note that, we implicitly assumed \(1+r_s \neq 0\) in Eq.~\eqref{appeq:phicontributions}, which does not hold for \(\theta\to \pi/2\). In this limit, all terms in the curly brackets in Eq.~\eqref{appeq:phicontributions} vanish except the last term. So, for the integral to vanish, it follows that \(\mathbf{E}(\mathbf{R})\) must go to zero in this limit as fast as \(1+r_s\). Notably, the boundary conditions from Eq.~\eqref{eq:surfaceBCfree} already imply a form for the electric field \(\mathbf{E}\sim (1+r_s)\mathrm{e}^{ikr}/r\), which satisfies this condition.

For the \(\mathbf{G}_{\theta}\)-contributions, we have
\begin{align}
    0 &= \lim_{R\to \infty} \oint_{\Omega}\mathrm{d}\Omega \mathrm{e}^{ikr}  R\opvec{e}_r \cdot \Big\{  ik[1-\cos(2\theta)r_p(\theta)]\mathbf{E}(\mathbf{R})\times\opvec{e}_{\varphi}\nonumber\\
    &+\mathbf{E}(\mathbf{R}) \times\big[ \{\nabla \sin(2\theta)r_p(\theta)\}\times\opvec{e}_r\big]-\big[1-\cos(2\theta)r_p(\theta)\big]\opvec{e}_{\theta}\times \big[\nabla\times\mathbf{E}(\mathbf{R})\big]\Big\}.
\end{align}

To obtain the boundary conditions from Eq.~\eqref{eq:surfaceBCfree}, we make use of the fact that \(\nabla\sin(2\theta)r_p(\theta)\) is parallel to \(\opvec{e}_{\theta}\), and hence \(\{[\nabla\sin(2\theta)r_p(\theta)]\times\opvec{e}_r\}\times\opvec{e}_r = -\nabla\sin(2\theta)r_p(\theta) = -\opvec{e}_{\theta}[\opvec{e}_{\theta}\cdot\nabla\sin(2\theta)r_p(\theta)]\).

\subsection{Boundary conditions from the bound surface wave contributions}
For the bound surface wave contributions to the Green's function from Eq.~\eqref{eq:surfaceG}, we use a cylindrical far-field surface as sketched in Fig.~\ref{fig:surfacecav}(b). We again represent the Green's function via the dyadic product \(\mathbf{G}_{\rm sur}(\rho,z) = \sum_j \mathbf{G}_j(\rho,z)\opvec{e}_j\), where \(\mathbf{G}_j(\rho,z) = \mathbf{B}_j(\kappa_{\rm spp})\mathrm{e}^{i\kappa_{\rm spp}\rho-q_{\rm spp}z}/\sqrt{\rho}\), where \(\mathbf{B}_j(\kappa_{\rm spp})\) are the column vectors of the dyadic \(\mathbf{B}(\kappa_{\rm spp})\) from Eq.~\eqref{eq:surfaceG}, which we leave general here, since they do not factor into the boundary conditions below. We also again leave the \(\omega\)-dependence implicit for brevity. Hence, the boundary conditions integral reads,
\begin{align}
    0 = \sum_j\opvec{e}_j\lim_{\rho\to\infty}\lim_{h\to\infty}\int_0^{2\pi}\mathrm{d}\varphi\int_0^{h}\mathrm{d}z \sqrt{\rho} \mathrm{e}^{i\kappa_{\rm spp}\rho-q_{\rm spp}z}\opvec{e}_{\rho}\cdot \Big\{ \mathbf{E}(\mathbf{R})\times\big[i\kappa_{\rm spp}\opvec{e}_{\rho}\times\mathbf{B}_j(\kappa_{\rm spp})\nonumber\\
    -q_{\rm spp}\opvec{e}_z\times\mathbf{B}_j(\kappa_{\rm spp})\big]-\mathbf{E}(\mathbf{R})\times \big[(1/\rho)\opvec{e}_{\rho}\times\mathbf{B}_j(\kappa_{\rm spp})\big]-\mathbf{B}_j(\kappa_{\rm spp})\times[\nabla\times\mathbf{E}(\mathbf{R})]\Big\},
\end{align}
where \(\mathbf{R} = \rho\opvec{e}_{\rho}+z\opvec{e}_z\). Using the Graßmann identity, we can reorder the terms in the curly brackets into the form from Eq.~\eqref{eq:surfaceBCz0}. 

\red{
\subsection{Branch point contribution}
For finite \(z\) and \(\rho\to\infty\), there is an additional singularity in the integral from Eq.~\eqref{eq:sommerfeldG}, namely the branch point \(\kappa = \sqrt{\epsilon_1} k\), where \(q_1 = 0\). Near this branch point, \(\opvec{e}_{p\pm} \approx \opvec{e}_z\) and \(r_s\approx r_p \approx -1\), so that from Eq.~\eqref{appeq:Kkappaspace}:
\begin{align}
    \mathbf{K}(\kappa,\rho,\omega) \sim J_0(\kappa\rho)\opvec{e}_z\opvec{e}_z, 
\end{align}
and therefore \(\mathbf{G}(\rho,z,\omega) \sim \opvec{e}_z\opvec{e}_z I(\rho,z,\omega)\), where
\begin{align}
    I(\rho,z,\omega) = \int_0^{\infty}\mathrm{d}\kappa J_0(\kappa\rho)\frac{\kappa}{q_1}\mathrm{e}^{iq_1 z}. 
\end{align}

Now, we perform a change of variables \(\kappa\to s=k_1-\kappa\), where \(k_1 = \sqrt{\epsilon_1}\omega/c\). Near the branch point \(q_1\approx \sqrt{2k_1(k_1-\kappa)} =\sqrt{2k_1 s}\), and hence:
\begin{align}\label{appeq:Isintegral}
    I(\rho,z,\omega) \approx \sqrt{\frac{k_1}{2}}J_0(k_1 \rho) \int_{-\infty}^{k_1}\mathrm{d}s \frac{\mathrm{e}^{-i\rho s + i\sqrt{2k_1 s}z}}{\sqrt{s}},
\end{align}
where we used \(J_0(\kappa\rho) = J_0(k_1 \rho)\mathrm{e}^{-i\rho s}\) near the branch point.

Furthermore, since we assume \(\rho\to\infty\), the exponential will oscillate fast, leading to destructive interference, unless \(s \sim 1/\rho \to 0\). Hence, we can focus on the regime \(s\approx 0\). This means that locally near the branch point, Eq.~\eqref{appeq:Isintegral} is approximately equal to
\begin{align}
    I(\rho,z,\omega) &\approx \sqrt{\frac{k_1}{2}}J_0(k_1 \rho) \int_0^{\infty}\mathrm{d}s  \frac{\mathrm{e}^{-i\rho s + i\sqrt{2k_1 s}z}}{\sqrt{s}}\nonumber\\
    &\approx \sqrt{\frac{k_1}{2}}J_0(k_1\rho) \int_0^{\infty}\mathrm{d}u \mathrm{e}^{-i\rho u^2/(2k_1)}\mathrm{e}^{iuz} \approx \frac{J_0(k_1 \rho)}{2}\sqrt{\frac{2\pi k_1}{\rho}}\sim \frac{\mathrm{e}^{ik_1 \rho}}{\rho},
\end{align}
where we performed another coordinate transformation \(u = \sqrt{2k_1 s}\) and used the fact that near the branch point \(u\approx 0\) and for finite \(z\) but \(\rho\to\infty\), the exponential \(\mathrm{e}^{iuz}\) is negligible against \(\mathrm{e}^{-i\rho u^2/(2k_1)}\) in the \(u\)-integral. 

Thus, the branch point contribution to the asymptotic Green's function behaves as \(\mathbf{G}(\rho,z,\omega) \sim \opvec{e}_z\opvec{e}_z \mathrm{e}^{i\sqrt{\epsilon_1} \omega \rho/c}/\rho\). This is a spherical wave propagating outwards parallel to the interface. The boundary conditions derived from this contribution read
\begin{align}
    \opvec{e}_{\rho} \times\nabla\times\mathbf{E}\to -ik\mathbf{E},\rho\to\infty,
\end{align}
which are identical to the free-space boundary conditions from Eq.~\eqref{eq:surfaceBCfree} for the limit \(\theta\to\pi/2\). 

For the example of a small cavity located at the surface at \(\mathbf{r}\approx 0\) considered in the main text, emission parallel to the surface propagates at \(z\approx 0\), where the SPP contributions dominate the field far away from the cavity. The SPP contributions decay as \(1/\sqrt{\rho}\), while the branch point contributions decay as \(1/\rho\). Hence, we neglected the branch-point contributions near the interface in the simple example in the main text. If the interface does not support bound SPPs, the branch point contributions are in fact the dominant contributions in this regime.
}

\section{QNM regularization}\label{appsec:regularization}
We derive the regularized QNM fields \(\qnm{F}{i_{\mu}}(\mathbf{r},\omega)\) for an individual cavity, using sources on the cavity surface, as in Refs.~\cite{franke2020quantized, ren2020near}. The full Green's function and electric field for a system of coupled cavities are then constructed from the single-cavity QNMs by using the scheme from Ref.~\cite{fuchs2026greens}. 

Therefore, let \(\mathbf{G}^{(i)}(\mathbf{r},\mathbf{r}')\) be the full Green's function for a single QNM cavity \(i\) in an arbitrary background medium (we leave the \(\omega\)-dependence implicit in the Green's functions and permittivities for notational convenience), which solves Eq.~\eqref{eq:greenhelm} for \(\epsilon(\mathbf{r},\omega) = \epsilon^{(i)}(\mathbf{r},\omega)\). We assume that, for positions \(\mathbf{r},\mathbf{r}',\in V_i\) inside the cavity, the full Green's function is approximated by the QNMs,
\begin{align}\label{appeq:GQNMinside}
    \mathbf{G}^{(i)}(\mathbf{r},\mathbf{r}')\big|_{\mathbf{r},\mathbf{r}'\in V_i} = \sum_{\mu}A_{i_{\mu}}(\omega)\qnm{f}{i_{\mu}}(\mathbf{r})\qnm{f}{i_{\mu}}(\mathbf{r}'). 
\end{align}

From the Helmholtz equations for \(\mathbf{G}^{(i)}(\mathbf{r},\mathbf{r}')\) and the background Green's function, we then obtain 
\begin{align}\label{appeq:dysonderiv}
    \mathbf{G}^{(i)}(\mathbf{r},\mathbf{r}')-\mathbf{G}_{\rm back}(\mathbf{r},\mathbf{r}') &= \int_V\mathrm{d}^3s \Big[\delta(\mathbf{r}-\mathbf{s})\mathbf{G}^{(i)}(\mathbf{s},\mathbf{r}')-\mathbf{G}_{\rm back}(\mathbf{r},\mathbf{s})\delta(\mathbf{s}-\mathbf{r}')\Big]\nonumber\\
    &=\frac{c^2}{\omega^2}\int_V\mathrm{d}^3s \Big\{\Big[\nabla_s\times\nabla_s\times\mathbf{G}_{\rm back}(\mathbf{s},\mathbf{r})\Big]^T\cdot\mathbf{G}^{(i)}(\mathbf{s},\mathbf{r}')\nonumber\\
    &\qquad\qquad -\mathbf{G}_{\rm back}(\mathbf{r},\mathbf{s})\cdot\Big[\nabla_s\times\nabla_s\times\mathbf{G}^{(i)}(\mathbf{s},\mathbf{r}')\Big]\Big\}\nonumber\\
    &\qquad+\int_V\mathrm{d}^3s \Delta\epsilon(\mathbf{s}) \mathbf{G}_{\rm back}(\mathbf{r},\mathbf{s})\cdot \mathbf{G}^{(i)}(\mathbf{s},\mathbf{r}'),
\end{align}
where we used the Green's function identity \([\mathbf{G}(\mathbf{s},\mathbf{r})]^T = \mathbf{G}(\mathbf{r},\mathbf{s})\), and \(V\) is the integration volume. Since we are interested in the fields outside the cavity, we have in Eq.~\eqref{appeq:dysonderiv} either \(\mathbf{r}\notin V_i\) or \(\mathbf{r}'\notin V_i\), or both. Without loss of generality, we assume \(\mathbf{r}\notin V_i\), and leave \(\mathbf{r}'\) general. For the case \(\mathbf{r}\in V_i, \mathbf{r}'\notin V_i\), the transpose of Eq.~\eqref{appeq:dysonderiv} is used instead. Hence, we can set in Eq.~\eqref{appeq:dysonderiv} \(V \to \overline{V}_i\), where \(\overline{V}_i\) denotes the complement of the cavity volume \(V_i\). Then, the last term on the right-hand-side of Eq.~\eqref{appeq:dysonderiv} vanishes, since \(\Delta\epsilon(\mathbf{r}) = \epsilon(\mathbf{r})-\epsilon_{\rm back}(\mathbf{r})\) is zero everywhere except inside the cavity volume \(V_i\). On the first term on the right-hand-side of Eq.~\eqref{appeq:dysonderiv}, we employ the second Green's identity [cf.~Eq.~\eqref{eq:secondgreen}] to obtain a surface integrals over the two surfaces of \(\overline{V}_i\), which are the cavity surface \(\mathcal{S}_i\), and the infinite far-field surface \(\mathcal{S}_{\infty}\). The far-field surface integral vanishes as a consequence of the outgoing boundary conditions [cf.~Eq.~\eqref{eq:BCintegral}]. Hence Eq.~\eqref{appeq:dysonderiv} becomes
\cite{franke2020quantized, fuchs2026greens},
\begin{align}\label{appeq:surfacedyson}
    \mathbf{G}^{(i)}(\mathbf{r},\mathbf{r}') = \chi_{\overline{V}_i}(\mathbf{r}')\mathbf{G}_{\rm back}(\mathbf{r},\mathbf{r}')+\frac{c^2}{\omega^2}\oint_{\mathcal{S}_i} \mathrm{d}A_s \Big\{ \Big[\opvec{n}_s\times\mathbf{G}_{\rm back}(\mathbf{s},\mathbf{r})\Big]^T\cdot\Big[\nabla_s\times\mathbf{G}^{(i)}(\mathbf{s},\mathbf{r}')\Big]\nonumber\\
    -\Big[\nabla_s\times\mathbf{G}_{\rm back}(\mathbf{s},\mathbf{r})\Big]^T\cdot\Big[\opvec{n}_s\times\mathbf{G}^{(i)}(\mathbf{s},\mathbf{r}')\Big]\Big\},
\end{align}
where \(\opvec{n}_s\) is the surface normal vector on \(\mathcal{S}_i\), pointing outwards with respect to \(V_i\), and \(\chi_{\overline{V}_i}(\mathbf{r}')\) is unity if \(\mathbf{r}'\in \overline{V}_i\), and zero if \(\mathbf{r}'\in V_i\). 

Now, we consider \(\mathbf{r}'\in V_i\). Inserting Eq.~\eqref{appeq:GQNMinside} on the right-hand side of Eq.~\eqref{appeq:surfacedyson} yields
\begin{align}\label{appeq:greenssurface}
    \mathbf{G}^{(i)}(\mathbf{r},\mathbf{r}')\big|_{\mathbf{r}\notin V_i,\mathbf{r}'\in V_i} = \sum_{\mu}A_{i_{\mu}}(\omega)\qnm{F}{i_{\mu}}(\mathbf{r},\omega) \qnm{f}{i_{\mu}}(\mathbf{r}')
\end{align}
with the regularized QNM fields from Eq.~\eqref{eq:surfacereg}. For \(\mathbf{r}'\notin V_i\), the transpose of Eq.~\eqref{appeq:surfacedyson} is inserted on the right-hand-side 
of Eq.~\eqref{appeq:surfacedyson}, to yield:
\begin{align}
    \mathbf{G}^{(i)}(\mathbf{r},\mathbf{r}')\big|_{\mathbf{r},\mathbf{r}'\notin V_i} =\mathbf{G}_{\rm back}(\mathbf{r},\mathbf{r}')+ \sum_{\mu}A_{i_{\mu}}(\omega)\qnm{F}{i_{\mu}}(\mathbf{r},\omega) \qnm{F}{i_{\mu}}(\mathbf{r}',\omega).
\end{align}

An alternative approach uses a Dysons scattering equation to obtain the regularized QNM fields \cite{ge2014quasinormal}. Towards this, we take Eq.~\eqref{appeq:dysonderiv} and set \(V\) to be the entire space (cavity+background). Then, the first term on the right-hand-side of Eq.~\eqref{appeq:dysonderiv} is turned into a single surface integral over \(\mathcal{S}_{\infty}\), which vanishes due to the outgoing boundary conditions [cf.~Eq.~\eqref{eq:BCintegral}]. The second term on the right-hand-side of Eq.~\eqref{appeq:dysonderiv} vanishes everywhere except inside \(V_i\), so that
\begin{align}\label{appeq:dyson}
    \mathbf{G}^{(i)}(\mathbf{r},\mathbf{r}')=\mathbf{G}_{\rm back}(\mathbf{r},\mathbf{r}') +\int_{V_i}\mathrm{d}^3s \Delta\epsilon(\mathbf{s}) \mathbf{G}_{\rm back}(\mathbf{r},\mathbf{s})\cdot \mathbf{G}^{(i)}(\mathbf{s},\mathbf{r}').
\end{align}

For \(\mathbf{r}\notin V_i, \mathbf{r}'\in V_i\), the Green's function then reads,
\begin{align}
    \mathbf{G}^{(i)}(\mathbf{r},\mathbf{r}')\big|_{\mathbf{r}\notin V_i, \mathbf{r}'\in V_i} = \mathbf{G}_{\rm back}(\mathbf{r},\mathbf{r}')+\sum_{\mu}A_{i_{\mu}}(\omega)\qnm{F}{i_{\mu}}(\mathbf{r},\omega) \qnm{f}{i_{\mu}}(\mathbf{r}'),
\end{align}
where now, the regularized QNM fields read,
\begin{align}\label{appeq:regQNM}
    \qnm{F}{i_{\mu}}(\mathbf{r},\omega) = \int_{V_i}\mathrm{d}^3s \Delta\epsilon(\mathbf{s}) \mathbf{G}_{\rm back}(\mathbf{r},\mathbf{s})\cdot \qnm{f}{i_{\mu}}(\mathbf{s}).
\end{align}

Note that different parts of the QNMs contribute on the surface vs. inside the cavity, and hence Eq.~\eqref{appeq:surfacedyson} and Eq.~\eqref{appeq:dyson} only agree if all QNMs are included \cite{franke2023impact}. 

\section{QNM overlap integrals}\label{appsec:overlap}
From the definition of the QNM operators \(\hat{a}_{i_{\mu}}\) in Eq.~\eqref{eq:QNMops} together with the commutator of the operators \(\opvec{b}(\mathbf{r},\omega)\) from Eq.~\eqref{eq:bcomm} follows that
\begin{align}
    [\hat{a}_{i_{\mu}},\hat{a}^{\dagger}_{j_{\nu}}]_- &=
    \int_0^{\infty}\mathrm{d}\omega\int\mathrm{d}^3r \mathbf{L}_{i_{\mu}}(\mathbf{r},\omega)\cdot\mathbf{L}^*_{j_{\nu}}(\mathbf{r},\omega)\nonumber\\
    &=\sum_{\mu'\nu'}\left(S^{-1/2}_{\rm \red{intra}}\right)_{i_{\mu}i_{\mu'}}\Big[\delta_{ij} S^{\rm intra}_{i_{\mu'}i_{\nu'}}+(1-\delta_{ij})S^{\rm inter}_{i_{\mu'}j_{\nu'}}\Big]\left(S^{-1/2}_{\rm \red{intra}}\right)_{j_{\nu'}j_{\nu}},
\end{align}
where we used the definition of the QNM projector kernels \(\mathbf{L}_{i_{\mu}}(\mathbf{r},\omega)\) from Eq.~\eqref{eq:QNMproj} to obtain the intracavity overlap matrix \(S^{\rm intra}_{i_{\mu}i_{\nu}}\) and the intercavity overlap matrix \(S^{\rm inter}_{i_{\mu}j_{\nu}}\). 

\subsection{Intracavity overlap}
From the definition of the projector kernels in Eq.~\eqref{eq:QNMproj}, it follows that the intracavity overlap reads
\begin{align}\label{appeq:Sintraformal}
    S^{\rm intra}_{i_{\mu}i_{\nu}} = \frac{2}{\pi \sqrt{\omega_{i_{\mu}}\omega_{i_{\nu}}}}\int_0^{\infty}\mathrm{d}\omega A_{i_{\mu}}(\omega)A_{j_{\nu}}^*(\omega)\Bigg[\int_{V_i}\mathrm{d}^3 r\epsilon_I(\mathbf{r},\omega)\qnm{f}{i_{\mu}}(\mathbf{r})\cdot\qnm{f}{i_{\nu}}^*(\mathbf{r})\nonumber\\
    +\int_{\overline{V}_i}\mathrm{d}^3 r\epsilon_{{\rm back},I}(\mathbf{r},\omega)\qnm{F}{i_{\mu}}(\mathbf{r},\omega)\cdot\qnm{F}{i_{\nu}}^*(\mathbf{r},\omega)\Bigg],
\end{align}
where \(\overline{V}_i\) is the complement of the cavity volume \(V_i\). As discussed in Ref.~\cite{franke2020fluctuation} for cavities in three-dimensional homogeneous media, the \(\overline{V}_i\)-integral does not vanish in the limit of a non-absorptive background medium. This can also be seen from the representation of \(S^{\rm intra}_{i_{\mu}i_{\nu}}\) from Eq.~\eqref{appeq:Sintradef}, which we derive below, and which does not vanish for \(\epsilon_{{\rm back},I}\to 0\).

Towards this, we decompose \(\epsilon_{{\rm back},I}(\mathbf{r},\omega) = [\epsilon_{{\rm back}}(\mathbf{r},\omega)-\epsilon^*_{{\rm back}}(\mathbf{r},\omega)]/(2i)\) and use the definition of the regularized QNM fields \(\qnm{F}{i_{\mu}}(\mathbf{r},\omega)\) from Eq.~\eqref{appeq:regQNM} to obtain (for a brief notation, we omit the \(\omega\)-dependence in the permittivities and Green's functions) \cite{franke2020fluctuation, fuchs2024quantization}
\begin{align}\label{appeq:overlineVintra}
    \int_{\overline{V}_i}\mathrm{d}^3 r\epsilon_{{\rm back},I}(\mathbf{r},\omega)\qnm{F}{i_{\mu}}(\mathbf{r},\omega)\cdot\qnm{F}{i_{\nu}}^*(\mathbf{r},\omega)=\int_{V_i}\mathrm{d}^3r \Delta\epsilon(\mathbf{r})\int_{V_i}\mathrm{d}^3r'\Delta\epsilon(\mathbf{r}') \qnm{f}{i_{\mu}}(\mathbf{r})\cdot \mathbf{M}(\mathbf{r},\mathbf{r}',\omega)\cdot\qnm{f}{i_{\nu}}^*(\mathbf{r}').
\end{align}

Here [using the Helmholtz equation~\eqref{eq:greenhelm} for the background Green's function \(\mathbf{G}_{\rm back}\)],
\begin{align}\label{appeq:Mintradef}
    2i \mathbf{M}(\mathbf{r},\mathbf{r}',\omega) &= \frac{c^2}{\omega^2}\int_{\overline{V}_i}\mathrm{d}^3s \Big\{ \Big[\nabla_s\times\nabla_s\times\mathbf{G}_{\rm back}(\mathbf{s},\mathbf{r})\Big]^T\cdot\mathbf{G}_{\rm back}^*(\mathbf{s},\mathbf{r}')\nonumber\\
    &\qquad\qquad\qquad-\mathbf{G}_{\rm back}(\mathbf{r},\mathbf{s})\cdot\Big[\nabla_s\times\nabla_s\times\mathbf{G}_{\rm back}^*(\mathbf{s},\mathbf{r}')\Big]\Big\}\nonumber\\
    &+\int_{\overline{V}_i}\mathrm{d}^3s \Big\{ \mathbf{G}_{\rm back}(\mathbf{s},\mathbf{r})\delta(\mathbf{s}-\mathbf{r}')
   -\delta(\mathbf{s}-\mathbf{r})\mathbf{G}_{\rm back}^*(\mathbf{s},\mathbf{r}')\Big\},
\end{align}
where we used \([\mathbf{G}_{\rm back}(\mathbf{s},\mathbf{r})]^T = \mathbf{G}_{\rm back}(\mathbf{r},\mathbf{s})\).

We then utilize the second Green's identity [cf.~Eq.~\eqref{eq:secondgreen}] on the first integral on the right-hand-side of Eq.~\eqref{appeq:Mintradef} to obtain an integral over the surface \(\mathcal{S}= \mathcal{S}_{\infty}\cup \mathcal{S}_i\) of \(\overline{V}_i\). Here, \(\mathcal{S}_i\) is a closed, near-field surface surrounding cavity \(i\), while \(\mathcal{S}_{\infty}\) is an infinite far-field surface. The integral over \(\mathcal{S}_{\infty}\) vanishes due to the outgoing boundary conditions [cf.~Eq.~\eqref{eq:BCintegral}, see also Refs.~\cite{franke2020fluctuation, fuchs2024quantization}]. Furthermore, the integrals with the \(\delta\)-functions in Eq.~\eqref{appeq:Mintradef} vanish because \(\mathbf{r},\mathbf{r}'\in V_i\) and \(\mathbf{s}\notin V_i\). Thus, we obtain,
\begin{align}
    2i \mathbf{M}(\mathbf{r},\mathbf{r}',\omega)=\frac{c^2}{\omega^2}\oint_{\mathcal{S}_i}\mathrm{d}A_s \Big\{\Big[\opvec{n}_s\times\mathbf{G}_{\rm back}(\mathbf{s},\mathbf{r})\Big]^T\cdot\Big[\nabla_s\times\mathbf{G}_{\rm back}^*(\mathbf{s},\mathbf{r}')\Big]\nonumber\\
    -\Big[\nabla_s\times\mathbf{G}_{\rm back}(\mathbf{s},\mathbf{r})\Big]^T\cdot\Big[\opvec{n}_s\times\mathbf{G}_{\rm back}^*(\mathbf{s},\mathbf{r}')\Big]\Big\}.
\end{align}
We reinsert this result into Eq.~\eqref{appeq:overlineVintra}, so that
\begin{align}
    \int_{\overline{V}_i}\mathrm{d}^3 r\epsilon_{{\rm back},I}(\mathbf{r},\omega)\qnm{F}{i_{\mu}}(\mathbf{r},\omega)\cdot\qnm{F}{i_{\nu}}^*(\mathbf{r},\omega)=\frac{c^2}{2i\omega^2}\oint_{\mathcal{S}_i}\mathrm{d}A_s \Big\{\Big[\opvec{n}_s\times\qnm{F}{i_{\mu}}(\mathbf{s},\omega)\Big]\cdot\Big[\nabla_s\times\qnm{F}{i_{\nu}}^*(\mathbf{s},\omega)\Big]\nonumber\\
    -\Big[\nabla_s\times\qnm{F}{i_{\mu}}(\mathbf{s},\omega)\Big]\cdot\Big[\opvec{n}_s\times\qnm{F}{i_{\nu}}^*(\mathbf{s},\omega)\Big]\Big\}.
\end{align}

Together with the definition of the regularized magnetic QNM fields \(\qnm{H}{i_{\mu}}(\mathbf{s},\omega) = \nabla_s\times\qnm{F}{i_{\mu}}(\mathbf{s},\omega)/(i\mu_0\omega)\), \(S^{\rm intra}_{i_{\mu}i_{\nu}}\) from Eq.~\eqref{appeq:Sintraformal} becomes
\begin{align}  \label{appeq:Sintradef}
    S^{\rm intra}_{i_{\mu}i_{\eta}} =& \frac{2}{\pi\sqrt{\omega_{i_{\mu}}\omega_{i_{\eta}}}}\int_0^{\infty}\mathrm{d}\omega\, A_{i_{\mu}}(\omega)A^*_{i_{\eta}}(\omega)\Big\{\int_{V_i}\mathrm{d}^3r\epsilon_I(\mathbf{r},\omega)\qnm{f}{i_{\mu}}(\mathbf{r})\cdot\qnm{f}{i_{\eta}}^*(\mathbf{r})\nonumber\\
    &+\frac{1}{2\omega\epsilon_0}\oint_{\mathcal{S}_i}\mathrm{d}A_s\Big[\left(\qnm{H}{i_{\mu}}(\mathbf{s},\omega)\times\opvec{n}_s\right)\cdot\qnm{F}{i_{\eta}}^*(\mathbf{s},\omega)+\mathrm{c.c.}(\mu\leftrightarrow \eta)\Big]\Big\},
\end{align}

As discussed above, the surface integral in \(S^{\rm intra}_{i_{\mu}i_{\nu}}\) does not vanish, even in the limit of a non-absorptive background medium \(\epsilon_{{\rm back},I}\to 0\).

The matrix \(S^{\rm intra}_{i_{\mu}i_{\nu}}\) is a Gramian matrix and therefore invertible, so that by definition, 
\begin{align}
    \delta_{ij}\sum_{\mu'\nu'}\left(S^{-1/2}_{\rm \red{intra}}\right)_{i_{\mu}i_{\mu'}}S^{\rm intra}_{i_{\mu'}i_{\nu'}}\left(S^{-1/2}_{\rm \red{intra}}\right)_{i_{\nu'}i_{\nu}} = \delta_{ij}\delta_{\mu\nu},
\end{align}
as in Eq.~\eqref{eq:QNMcomm}. 

\subsection{Intercavity overlap}
Again using the definition of the projector kernels \(\mathbf{L}_{i_{\mu}}(\mathbf{r},\omega)\) from Eq.~\eqref{eq:QNMproj}, the intercavity overlap reads (leaving \(i\neq j\) implicit),
\begin{align}\label{appeq:Sinterformal}
    S^{\rm inter}_{i_{\mu}j_{\nu}} = \frac{2}{\pi\sqrt{\omega_{i_{\mu}}\omega_{j_{\nu}}}}\int_0^{\infty}\mathrm{d}\omega A_{i_{\mu}}(\omega)A^*_{j_{\nu}}(\omega)\Bigg[\int_{\overline{V}_{ij}}\mathrm{d}^3r \epsilon_{{\rm back},I}(\mathbf{r},\omega)\qnm{F}{i_{\mu}}(\mathbf{r},\omega)\cdot\qnm{F}{j_{\nu}}^*(\mathbf{r},\omega)\nonumber\\
    +\int_{V_i}\mathrm{d}^3r \sqrt{\epsilon_I(\mathbf{r},\omega)\epsilon_{{\rm back},I}(\mathbf{r},\omega)}\qnm{f}{i_{\mu}}(\mathbf{r})\cdot\qnm{F}{j_{\nu}}^*(\mathbf{r},\omega)\nonumber\\
    +\int_{V_j}\mathrm{d}^3r \sqrt{\epsilon_I(\mathbf{r},\omega)\epsilon_{{\rm back},I}(\mathbf{r},\omega)}\qnm{F}{i_{\mu}}(\mathbf{r},\omega)\cdot\qnm{f}{j_{\nu}}^*(\mathbf{r})\Bigg].
\end{align}

Here, \(\overline{V}_{ij}\) is the complement of \(V_i\cup V_j\), where \(V_i,V_j\) are the cavity volumes. For a non-absorptive background medium \(\epsilon_{{\rm back},I}\to 0\), the integrals over \(V_i\) and \(V_j\) vanish. The \(\overline{V}_{ij}\)-integral is transformed into a surface integral using the same steps as in the intracavity case. The surface of \(\overline{V}_{ij}\) consists of the separate cavity surfaces \(\mathcal{S}_i,\mathcal{S}_j\) and the infinite far-field surface \(\mathcal{S}_{\infty}\). The integral over \(\mathcal{S}_{\infty}\) vanishes due to the outgoing boundary conditions, as discussed in the intracavity case, leaving only the integrals over the cavity surfaces. Thus, \(S^{\rm inter}_{i_{\mu}j_{\nu}}\) from Eq.~\eqref{appeq:Sinterformal} becomes 
\begin{align} \label{appeq:Sinterdef}
    S^{\rm inter}_{i_{\mu}j_{\eta}}\Big|_{i\neq j} =& \frac{2}{\pi\sqrt{\omega_{i_{\mu}}\omega_{j_{\eta}}}}\int_0^{\infty}\mathrm{d}\omega\,A_{i_{\mu}}(\omega)A^*_{j_{\eta}}(\omega)\nonumber\\
    &\quad\times\Bigg\{\frac{1}{2\omega \epsilon_0}\oint_{\mathcal{S}_i}\mathrm{d}A_s\Big[\left(\qnm{H}{i_{\mu}}(\mathbf{s},\omega)\times\opvec{n}_s\right)\cdot\qnm{F}{j_{\eta}}^*(\mathbf{s},\omega){+}\mathrm{c.c.}(i_{\mu}\leftrightarrow j_{\eta})\Big]\nonumber\\
    &\qquad+\frac{1}{2\omega \epsilon_0}\oint_{\mathcal{S}_j}\mathrm{d}A_s\Big[\left(\qnm{H}{i_{\mu}}(\mathbf{s},\omega)\times\opvec{n}_s\right)\cdot\qnm{F}{j_{\eta}}^*(\mathbf{s},\omega){+}\mathrm{c.c.}(i_{\mu}\leftrightarrow j_{\eta})\Big]\nonumber\\
    &\qquad+\int_{V_i}\mathrm{d}^3r \sqrt{\epsilon_I(\mathbf{r},\omega)\epsilon_{{\rm back},I}(\mathbf{r},\omega)}\qnm{f}{i_{\mu}}(\mathbf{r})\cdot\qnm{F}{j_{\nu}}^*(\mathbf{r},\omega)\nonumber\\
    &\qquad+\int_{V_j}\mathrm{d}^3r \sqrt{\epsilon_I(\mathbf{r},\omega)\epsilon_{{\rm back},I}(\mathbf{r},\omega)}\qnm{F}{i_{\mu}}(\mathbf{r},\omega)\cdot\qnm{f}{j_{\nu}}^*(\mathbf{r})\Bigg\}.
\end{align}

As pointed out for the intracavity overlap above, the intercavity overlap also does not vanish in the limit of a non-absoprtive background medium \(\epsilon_{{\rm back},I}\to 0\).

\section{Cavity separation parameter}\label{appsec:cavsepparam}
The intercavity overlap \(S^{\rm inter}_{i_{\mu}j_{\nu}}\) (with \(i\neq j\) implicit) from Eq.~\eqref{appeq:Sinterdef} decreases for an increasing separation between the cavities, where separation is meant in the sense of optical path. To quantify this separation, we derive the \textit{cavity separation parameter}, following the steps from Ref.~\cite{fuchs2024quantization}. 
For brevity, we illustrate the derivation of the cavity separation parameter on the term
\begin{align}\label{appeq:Iterm}
    \mathcal{I}_{i_{\mu}j_{\nu}} = \frac{2}{\pi\sqrt{\omega_{i_{\mu}}\omega_{j_{\nu}}}}\int_0^{\infty}\mathrm{d}\omega A_{i_{\mu}}(\omega)A^*_{j_{\nu}}(\omega)\int_{V_i}\mathrm{d}^3r \sqrt{\epsilon_I(\mathbf{r},\omega)\epsilon_{{\rm back},I}(\mathbf{r},\omega)}\qnm{f}{i_{\mu}}(\mathbf{r})\cdot\qnm{F}{j_{\nu}}^*(\mathbf{r},\omega)
\end{align}
which appears in the intercavity overlap from Eq.~\eqref{appeq:Sinterdef}. The other terms are treated in the same way.

\subsection{Effect of retardation}
Retardation is the first aspect entering the cavity separation parameter. In the construction of the QNM operators \(\hat{a}_{i_{\mu}}\), we fix the phase by integrating over all frequencies [cf.~Eq.~\eqref{eq:QNMops}]. As a consequence, the quantized QNMs are \textit{quasi-bound} in the sense that they are mostly concentrated in their cavity. However, due to the open boundary conditions, the QNMs extend into the exterior medium, resulting in finite overlap even in the absence of a time delay. To investigate this aspect, we note that the spatial integral in Eq.~\eqref{appeq:Iterm} is restricted to the cavity volume \(V_i\). The regularized QNM field \(\qnm{F}{j_{\nu}}^*(\mathbf{r},\omega)\), meanwhile, is created by sources inside the cavity \(V_j\). 
In three-dimensional homogeneous media, this spatial separation was treated by decomposing \(\qnm{F}{j_{\nu}}^*(\mathbf{r},\omega)\) into a slow-varying envelope and fast-rotating plane wave \(\mathrm{e}^{-i\omega r/c}\) \cite{fuchs2024quantization}. In structured environments such as waveguides or surfaces, the spatial dependence of the wave is more complicated. Therefore, we perform the factorization (\(i\neq j\))
\red{
\begin{align}\label{appeq:Fdecomp}
    \qnm{F}{j_{\nu}}^*(\mathbf{r},\omega)\big|_{\mathbf{r}\in V_i} = [\qnm{F}{j_{\nu}}'(\mathbf{r},\omega)]^* \mathrm{e}^{-i[S_{ij}(\omega)]^*},
\end{align}
where
\begin{align}\label{appeq:Eikonaldef}
    S_{ij}(\omega) = \int_{\mathcal{C}_{ij}}\mathrm{d}\mathbf{s}\cdot\mathbf{k}(\mathbf{s},\omega)
\end{align}
is the \textit{Eikonal} for the propagation of the wave along the optical path \(\mathcal{C}_{ij}\) (e.g., through the core of a waveguide) between the cavities \(i\) and \(j\) \cite{landau1975classical}. Here, \(\mathbf{k}\) is the generalized wavevector along the path, which depends on the frequency via a dispersion relation. If a wavenumber can be defined for all spatial directions, the dispersion relation reads \cite{landau1975classical},
\begin{align}\label{appeq:generaldispersion}
    |\mathbf{k}(\mathbf{r},\omega)|^2 = \epsilon(\mathbf{r},\omega)\frac{\omega^2}{c^2}.
\end{align}

The definition of the Eikonal from Eq.~\eqref{appeq:Eikonaldef} is very general to also account for non-homogeneous media and propagation in media where the light is confined in one direction, while propagating in another direction (as is the case for the 2D waveguide from Fig.~\ref{fig:figure}). For homogeneous media \(S_{ij}(\omega) = n_B \omega R_{ij}/c\) \cite{fuchs2024quantization}, while in structured media \(S_{ij}(\omega) = \omega \Lambda_{ij}/c\), where \(\Lambda_{ij} = \int_{\mathcal{C}_{ij}}\mathrm{d}s \,n(s) \approx \overline{n} |\mathcal{C}_{ij}|\) is the \textit{optical path length} with geometric length \(|\mathcal{C}_{ij}|\), and \(\overline{n}\) is the average refractive index along the path. Meanwhile, if the wave is confined in a direction perpendicular to the propagation direction, as in the waveguide from Fig.~\ref{fig:figure}, parts of the wavevector will be perpendicular to the propagation direction and not contribute to the Eikonal, because the Eikonal equation is only valid for distances larger than the wavelength and can not capture the non propagating confined part perpendicular to the propagation direction, so that \(S_{ij}(\omega) = k_{\parallel}(\omega) \Lambda_{ij}\), where \(k_{\parallel}(\omega)\) is the wavevector component parallel to the propagation direction, e.g., \(k_x\) in the case of the 2D waveguide. 

Note that in Eq.~\eqref{appeq:Fdecomp}, we have separated out only the Eikonal for the extremal path between the two cavity centers and neglected contributions from propagation within the cavities. For a wave propagating between two specific points \(\mathbf{r}_i,\mathbf{r}_j\) inside the two cavities \(i\) and \(j\) respectively, we approximated \(S_{\mathbf{r}_i\mathbf{r}_j}(\omega)\approx S_{ij}(\omega)\) [where \(S_{\mathbf{r}_i\mathbf{r}_j}(\omega)\) is the Eikonal along the full path between the two points], assuming the cavities are small compared to the distance between them. On this basis, we also assume that the envelope function \([\qnm{F}{j_{\nu}}'(\mathbf{r},\omega)]^*\) in Eq.~\eqref{appeq:Fdecomp} varies only slowly with frequency. In Eq.~\eqref{appeq:Iterm}, the frequency integral is dominated by the poles at the QNM eigenfrequencies contained in \(A_{i_{\mu}}(\omega)A^*_{j_{\nu}}(\omega)\). For practical applications, the frequency interval is limited to a narrow range around these resonances. 
We therefore assume that the envelope function can be treated in a pole approximation \cite{ren2020near}
\begin{align}
    \qnm{F}{j_{\nu}}'(\mathbf{r},\omega) \to \qnm{F}{j_{\nu}}'(\mathbf{r},\omega_{j_{\nu}}).
\end{align}
We perform the same approximation on all \(\omega\)-dependent terms in Eq.~\eqref{appeq:Iterm} except the poles and the fast-rotating exponential \(\mathrm{e}^{-i[S_{ij}(\omega)]^*}\). 
Thus, we obtain
\begin{align}\label{appeq:Ibeforeresidue}
    &\mathcal{I}_{i_{\mu}j_{\nu}} \approx \frac{1}{2\pi}\int_0^{\infty}\mathrm{d}\omega \frac{\sqrt{\omega_{i_{\mu}}\omega_{j_{\nu}}}\mathrm{e}^{-i[S_{ij}(\omega)]^*}}{(\omega-\tilde{\omega}_{i_{\mu}})(\omega-\tilde{\omega}^*_{j_{\nu}})}\nonumber\\
    &\qquad\qquad\times\int_{V_i}\mathrm{d}^3r \sqrt{\epsilon_I(\mathbf{r},\sqrt{\omega_{i_{\mu}}\omega_{j_{\nu}}})\epsilon_{{\rm back},I}(\mathbf{r},\sqrt{\omega_{i_{\mu}}\omega_{j_{\nu}}})}\qnm{f}{i_{\mu}}(\mathbf{r})\cdot[\qnm{F}{j_{\nu}}'(\mathbf{r},\omega_{j_{\nu}})]^*.
\end{align}

Since the generalization of the Eikonal to the partially confined case is a projection of the wavevector in the propagation direction and we consider an outward emission from cavity \(j\) to cavity \(i\), we can assume \(\mathrm{Re}\{[S_{ij}(\omega)]^*\}>0\). We thus solve the \(\omega\)-integral in Eq.~\eqref{appeq:Ibeforeresidue} by (approximately) applying the residue theorem at the pole \(\omega=\tilde{\omega}_{i_{\mu}}\). Thus, the integral scales as 
\begin{align}
    \mathcal{I}_{i_{\mu}j_{\nu}}\sim  \mathrm{e}^{-i[S_{ij}(\omega)]^*_{\omega = \tilde{\omega}_{i_{\mu}}}} \sim  \mathrm{e}^{-\mathrm{Im}\{[S_{ij}(\omega)]^*_{\omega=\tilde{\omega}_{i_{\mu}}}\}},
\end{align}
i.e., it decays exponentially as a function of the imaginary part of the Eikonal between the two cavity centers, evaluated at the QNM eigenfrequency. For homogeneous propagation, \(\mathrm{Im}\{[S_{ij}(\omega)]^*_{\omega=\tilde{\omega}_{i_{\mu}}}\} = \gamma_{i_{\mu}}\Lambda_{ij}/c\) \cite{fuchs2024quantization}, while for guided propagation, e.g., in the 2D waveguide from Fig.~\ref{fig:figure}, \(\mathrm{Im}\{[S_{ij}(\omega)]^*_{\omega=\tilde{\omega}_{i_{\mu}}}\} = \mathrm{Im}[k_x(\tilde{\omega}_{i_{\mu}})]\Lambda_{ij}\), i.e., \(\mathcal{I}_{i_{\mu}j_{\nu}}\) decreases exponentially as a function of the separation of the cavities. 

All other terms in \(S^{\rm inter}_{i_{\mu}j_{\nu}}\) show similar scaling, so that the intercavity overlap scales as
\begin{align}
    S^{\rm inter}_{i_{\mu}j_{\nu}} \sim \mathrm{e}^{-|S^{\rm imag, min}_{i_{\mu}j_{\nu}}|},
\end{align}
where \(S^{\rm imag, min}_{i_{\mu}j_{\nu}} = \min(\mathrm{Im}\{[S_{ij}(\omega)]_{\omega=\tilde{\omega}^*_{j_{\eta}}}\},\mathrm{Im}\{[S_{ij}(\omega)]^*_{\omega=\tilde{\omega}_{i_{\mu}}}\})\). 

We expect this recipe to be rough guidance that must be adapted to each type of nanostructure.
}

\subsection{Effect of geometry}
As mentioned above, the specific geometry of the background (vacuum, waveguides, surfaces, etc.) and the cavity profoundly affects the mode overlap between different cavities. A simple example is that of a structure of waveguide-coupled cavities. If two cavities are directly coupled via a waveguide, the overlap will be much stronger than if they are not directly coupled and are instead coupled only to a third cavity. 

Determinative for this aspect of the overlap is how much of the total radiated power from cavity \(j\) actually reaches cavity \(i\) to contribute to the overlap from Eq.~\eqref{appeq:Iterm}. We express this aspect via the general scaling:
\begin{align}\label{appeq:directionalityscaling}
    \mathcal{I}_{i_{\mu}j_{\nu}}\sim \frac{U^i_{j_{\nu}}}{U^{\rm tot}_{j_{\nu}}},
\end{align}
where \cite{kristensen2017theory, jackson2012classical}
\begin{align}\label{appeq:powerradtotal}
    U^{\rm tot}_{j_{\nu}} = \frac{1}{2}\int_{\mathcal{S}_j}\mathrm{d}A_s \opvec{n}_s\cdot\mathrm{Re}[\qnm{F}{j_{\nu}}(\mathbf{s},\omega_{j_{\nu}})\times\qnm{H}{j_{\nu}}^*(\mathbf{s},\omega_{j_{\nu}})]
\end{align}
is the total power radiated by mode \(j_{\nu}\) at frequency \(\omega_{j_{\nu}}\) through the surface \(\mathcal{S}_j\) surrounding cavity \(j\). Meanwhile, \(U^i_{j_{\nu}}\) is that part of the power radiated by mode \(j_{\nu}\) at frequency \(\omega_{j_{\nu}}\) which reaches cavity \(i\). To compute the power transfer, we separate the background volume \(V_{\rm out}\) into smaller volumes \(V_{\rm out}^{\alpha}\) to account for different propagation paths through the background: \(V_{\rm out} = \cup_{\alpha} V_{\rm out}^{\alpha}\), similar to what was done in Sec.~\ref{sec:combinedstructures}. Next, we denote by \(\alpha_i\) that part of the background structure which guides light towards cavity \(i\), so that:
\begin{align}\label{appeq:powerradtoi}
    U^i_{j_{\nu}} = \frac{1}{2}\int_{\mathcal{S}_j^{\alpha_i}}\mathrm{d}A_s \opvec{n}_s\cdot\mathrm{Re}[\qnm{F}{j_{\nu}}(\mathbf{s},\omega_{j_{\nu}})\times\qnm{H}{j_{\nu}}^*(\mathbf{s},\omega_{j_{\nu}})],
\end{align}
where \(\mathcal{S}_j^{\alpha_i}\) denotes that part of the surface \(\mathcal{S}_j\) which intersects with the part of the background which guides light towards cavity \(i\). This is similar to what we did in the QNM regularization for combined backgrounds in Eq.~\eqref{eq:combinedsurfacereg}. If no light is guided from cavity \(j\) to \(i\), then \(\mathcal{S}_j^{\alpha_i} = \emptyset\), and therefore \(U^i_{j_{\nu}}=0\). 

Thus, Eq.~\eqref{appeq:directionalityscaling} can be understood intuitively as the power transferred to cavity \(i\) from mode \(j_{\nu}\) relative to the total output power by mode \(j_{\nu}\) at frequency \(\omega_{j_{\nu}}\). 
We define the transfer factor 
\begin{align}\label{appeq:couplingfactor}
    T_{i_{\mu},j_{\nu}} = -\mathrm{ln}[U^i_{j_{\nu}}/U^{\rm tot}_{j_{\nu}}]
\end{align}
in analogy to the coupling factor in directional couplers (e.g., Ref.~\cite{sanna2018design}). Note that Eq.~\eqref{appeq:couplingfactor} does not depend on the mode index \(\mu\), since only the total input power to cavity \(i\) from mode \(j_{\nu}\) matters. However, we keep the index in our notation.
Therefore, from Eq.~\eqref{appeq:Sinterdef}, 
\begin{align}
    S^{\rm inter}_{i_{\mu}j_{\nu}}\sim \mathrm{e}^{-T^{\rm min}_{i_{\mu}j_{\nu}}},
\end{align}
where \(T^{\rm min}_{i_{\mu}j_{\nu}} = \min[T_{i_{\mu},j_{\nu}},T_{j_{\nu},i_{\mu}}]\). For a homogeneous background medium, \(\mathcal{S}_j^{\alpha_i}\) can be expressed via the solid angle \(\Omega_{ji}\) into which energy must be radiated to reach cavity \(i\) from cavity \(j\). Therefore, the transfer factor in this case is related to the directivity via \(T_{i_{\mu},j_{\nu}} = -\mathrm{ln}[D_{j_{\nu}}(\Omega_{ji})]\), which was the result in Ref.~\cite{fuchs2024quantization}.

\subsection{Additional aspects}
Additional contributions may enter into the cavity separation parameter. For example, a high-Q cavity mode loses energy at a very slow rate, thereby decreasing its overlap with modes from another cavity. In the limit of a closed cavity, the overlap with other cavities is zero. In Ref.~\cite{fuchs2024quantization}, it was shown that, for a high-Q mode,
\begin{align}
    \qnm{F}{j_{\nu}}(\mathbf{r},\omega)\sim 1/\sqrt{Q_{j_{\nu}}},
\end{align}
where \(Q_{j_{\nu}}=\omega_{j_{\nu}}/(2\gamma_{j_{\nu}})\). Thus, in certain setups, the overlap may be further reduced.

Another aspect is the detuning. When solving Eq.~\eqref{appeq:Ibeforeresidue} with the residue theorem, the result scales with \(1/(\tilde{\omega}_{i_{\mu}}-\tilde{\omega}^*_{j_{\nu}})\), which goes to zero if the detuning between the cavities is large. However, we usually assume cavities with resonances in the same frequency range, and hence, the detuning plays a negligible role in the separate quantization of the cavities. We therefore leave these additional aspects out of the cavity separation parameter from Eq.~\eqref{eq:cavsepparam}.

\printbibliography

\end{document}